\documentclass{aa} % for a referee version
\usepackage{graphicx}
\usepackage[draft]{hyperref}

\usepackage{txfonts}
\begin{document}

   \title{ARGOS at the LBT}

   \subtitle{Binocular laser guided ground-layer adaptive optics}

   \author{S.~Rabien \inst{1}  \and R.~Angel \inst{7}  \and L.~Barl \inst{1}  \and U.~Beckmann \inst{6}  \and L.~Busoni \inst{2}  \and S.~Belli \inst{1}  \and M.~Bonaglia \inst{2}  \and J.~Borelli \inst{3}  \and J.~Brynnel \inst{8,} \inst{5} \and P.~Buschkamp \inst{1} \and A.~Cardwell \inst{8} \and A.~Contursi \inst{1}\and C.~Connot \inst{6} \and R.~Davies \inst{1} \and M.~Deysenroth  \inst{1} \and O.~Durney \inst{7} \and F.~Eisenhauer \inst{1} \and M.~Elberich \inst{6}  \and S.~Esposito \inst{2} \and B.~Frye \inst{7} \and W.~Gaessler \inst{3} \and V.~Gasho \inst{7} \and H.~Gemperlein \inst{1} \and R.~Genzel \inst{1} \and I.~Y.~Georgiev \inst{3}\and R.~Green \inst{8,} \inst{7}\and M.~Hart\inst{7} \and C.~Kohlmann \inst{1} \and M.~Kulas \inst{3} \and M.~Lefebvre \inst{8} \and T.~Mazzoni \inst{2} \and J.~Noenickx \inst{7} \and G.~Orban~de~Xivry \inst{1,}\inst{10} \and T.~Ott \inst{1} \and D.~Peter \inst{3} \and A.~Puglisi \inst{2} \and Y.~Qin \inst{7} \and A.~Quirrenbach \inst{4} \and W.~Raab \inst{1,}\inst{9} \and M.~Rademacher \inst{7} \and G.~Rahmer \inst{8} \and M.~Rosensteiner \inst{1} \and H.W.~Rix \inst{3} \and P.~Salinari \inst{2} \and C.~Schwab \inst{4} \and A.~Sivitilli \inst{3}\and M.~Steinmetz \inst{5} \and J.~Storm \inst{5} \and C.~Veillet \inst{8} \and G.~Weigelt \inst{6} \and J.~Ziegleder \inst{1}
          }

   \institute{Max-Planck-Institut f\"ur extraterrestrische Physik, Giessenbachstr.
              Garching, Germany\\
              \email{srabien@mpe.mpg.de}
              \and
              INAF Osservatorio Astrofisico di Arcetri, Florence, Italy
              \and
              Max-Planck-Institut f\"ur Astronomie, Königstuhl 17, Heidelberg, Germany
              \and
              Landesternwarte, Königstuhl 12, Heidelberg, Germany
              \and
              Leibniz-Institut für Astrophysik Potsdam (AIP), An der Sternwarte 16, Potsdam, Germany
              \and
              Max Planck Institut f\"ur Radioastronomie, Auf dem H\"ugel 69, Bonn, Germany
              \and
              Steward Observatory, 933 North Cherry Avenue, University of Arizona, Tucson, Arizona,  USA
              \and
              Large Binocular Telescope Observatory, 933 North Cherry Avenue, Tucson, Arizona, USA
              \and
              European Space Agency - ESTEC, Keplerlaan 1, Noordwijk
              \and
              Space sciences, Technologies and Astrophysics Research (STAR) Institute Universit\'e de Li\` ege, Belgium
              }

   \date{Received June 25, 2018; accepted October 19, 2018}

% \abstract{}{}{}{}{}
% 5 {} token are mandatory
\abstract{ Having completed its commissioning phase, the Advanced Rayleigh guided Ground-layer adaptive Optics System (ARGOS)  facility is coming online for scientific observations at the Large Binocular Telescope (LBT). With six Rayleigh laser guide stars in two constellations and the corresponding wavefront sensing, ARGOS corrects the ground-layer distortions for both LBT 8.4\,m eyes with their adaptive secondary mirrors. Under regular observing conditions, this set-up delivers a point spread function (PSF) size reduction by a factor of ~2--3 compared to a seeing-limited  operation.  With the two LUCI infrared imaging and multi-object spectroscopy instruments receiving the corrected images, observations in the near-infrared  can be performed at high spatial and spectral resolution. We discuss the final ARGOS technical set-up  and the adaptive optics performance. We show that imaging cases with ground-layer adaptive optics (GLAO) are enhancing several scientific programmes, from cluster  colour magnitude diagrams and Milky Way embedded star formation, to nuclei of nearby galaxies or extragalactic lensing fields. In the unique combination of ARGOS with the multi-object near-infrared spectroscopy available in LUCI over a $4\times 4$ arcmin field of view, the first scientific observations have been performed on local and high-$z$ objects. Those high spatial and spectral resolution observations demonstrate the capabilities now at hand with ARGOS at the LBT.}
   \maketitle
-------------------------------------------------------------------
\section{Introduction}

\begin{figure*}
   \centering
   \includegraphics[width=18.3cm]{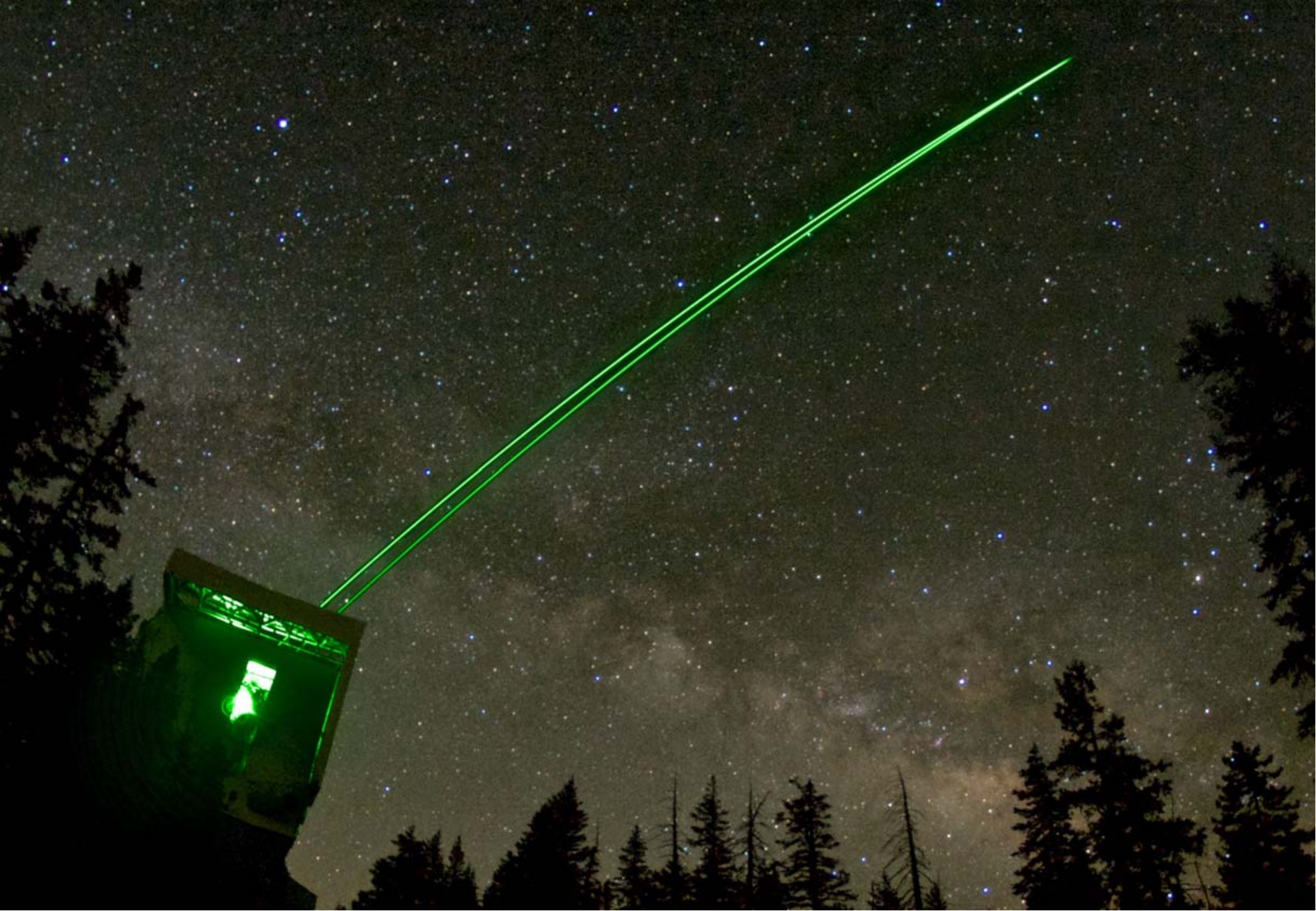}
   \caption{ The ARGOS system propagating a bundle of laser beams on each side of the large binocular telescope. This wide-angle photograph was taken in 2017 with a 25s exposure time. Each visible green beam in this image consists of three individual laser rays forming the wide-field constellations of guide stars in the atmosphere. The light pulses from the high-power lasers are subject to Rayleigh scattering by air molecules in the Earth's atmosphere. Having the wavefront sensors gated and adjusted to receive  the photons only from a distance of  12km,  a sharp reference beacon constellation is formed. Measuring the wavefronts and correcting the atmospheric ground-layer distortions with the two adaptive secondary mirrors yields a wide-field correction for imaging and spectroscopy.}
              \label{Fig_Lasers_on_sky}
    \end{figure*}

  The Advanced Rayleigh guided Ground-layer adaptive Optics System (ARGOS) has been implemented to deliver a ground-layer adaptive optics (GLAO) correction to both of the 8.4\,m `eyes' of the  Large Binocular Telescope (LBT). GLAO is a technique used to correct the atmospheric induced optical distortions over a large field of view, enhancing the image quality homogenously. To conduct this correction, ARGOS utilizes two constellations of multiple guide stars, generated artificially above the LBT with Rayleigh backscattering of high-power pulsed green lasers. Figure \ref{Fig_Lasers_on_sky} shows the ARGOS binocular laser beams when propagated to sky. With range-gated wavefront sensing systems for the laser beacons and the LBT's adaptive secondary mirrors, the correction yields an improved point spread function (PSF) for imaging and spectroscopic observations.
  Enhancing the image quality with GLAO  has several advantages.  Increasing the spatial resolution gives insights into the details of an object's structure. Additionally, the signal-to-noise ratio (S/N) in  spectroscopy benefits strongly from sharpening the image. With the required integration time to reach a given S/N being inversely proportional to the square of the PSF diameter, observations can be carried out in a much shorter time. Due to the smaller PSF size the spectroscopic slit width can be decreased accordingly, enhancing the spectral resolution and allowing spatially resolved spectroscopic observations. With GLAO delivering a wide field-of-view correction, science cases benefit from adaptive optics that cannot be done with single-conjugate systems or in seeing-limited mode. With a factor 2 to 3 PSF size reduction, ARGOS can be considered  a ‘seeing enhancer’ beneficial for the two facility instruments LUCI1 and LUCI2 offering imaging and multi-object spectroscopy (MOS) in the near-infrared (NIR) wavelength regime. A multitude of science cases will benefit from the enhanced resolution and encircled energy that ARGOS delivers. Amongst others, scientific topics that can be addressed with the aid of GLAO span a wide range, from extragalactic cases such as high-$z$ galaxy dynamics, active galactic nuclei, and Quasar host galaxies, to Galactic astrophysical questions about planets, Cepheids, or stellar clusters. To summarize briefly, ARGOS with LUCI at the LBT offers the following benefits:
   \begin{itemize}
\item Binocular observations, using the two 8.4\,m telescopes of LBT at once;
\item GLAO correction with a 0.2$\arcsec$ to 0.3$\arcsec$ resolution over a 4x4\,arcmin field of view at both telescopes;
\item A fairly homogeneous PSF shape over this full field (see Section \ref{sec_performance});
\item A large $2 \times 3$ arcmin field for the tilt star selection;
\item NIR imaging of the full field at the GLAO spatial resolution;
\item GLAO corrected NIR multi-object spectroscopy with custom cut slit masks and high spectral resolution.
   \end{itemize}
    The last  point  emphasizes one of the  unique capabilities that ARGOS provides. Currently this combination is only available at the LBT, enabling spectroscopic observations at high spatial and spectral resolution with slits cut to the object's shape. The high spatial resolution allows the spectroscopic slits to be cut as narrow as 0.25$\arcsec$ to 0.3$\arcsec$, pushing the spectral resolution up to R \textasciitilde 10000, enabling the detection of structures in the velocity distribution of  high-$z$ galaxies in great detail, and reducing the fraction of atmospheric bands for which the spectra are disturbed by the OH night sky emission lines.
    
    In  Section \ref{sec_system} we describe the ARGOS laser guide star (LGS) and adaptive optics system, now operational at the LBT. We give an overview of how the system is assembled, describe the laser beacon generation, the wavefront sensor units and its adaptive optics correction system. At the end of commissioning we can now show the resulting performance of the ARGOS binocular GLAO system. The analysis of the adaptive optics performance and the achieved image quality in the J, H, and K bands is discussed in section \ref{sec_performance}. Having targeted a variety of objects in imaging and spectroscopic mode over the commissioning period, we can show selected science cases which highlight the capabilities of the system. Section \ref{sec_observations} shows imaging programmes  that benefit from the PSF size reduction:  the globular cluster NGC2419; the nearby galaxy NGC6384, which  hosts a massive nuclear star cluster and a black hole; and three lensing galaxy clusters. In particular, PLCK G165.7+67.0 (G165) ARGOS-corrected K-band imaging complementing HST WFC3/IR data allows us to detect the counter-image of one of the gravitationally lensed sources. One of the most powerful properties of ARGOS at the LBT is the combination of the custom cut slit masks in the LUCI spectrographs with the light concentrating GLAO system. We present curved slit spectroscopic observations of gravitationally lensed high-redshift galaxies which follows the long axis of the arcs.  These spectroscopic programmes are focused  on the `8 o'clock arc' and SDSS 1038+4849, which reveal structures within the interstellar medium of these systems undetected by other methods.

--------------------------------------------------------------------

\section{The ARGOS GLAO system}\label{sec_system}

    \begin{figure}
   \centering
    \includegraphics[width=8cm]{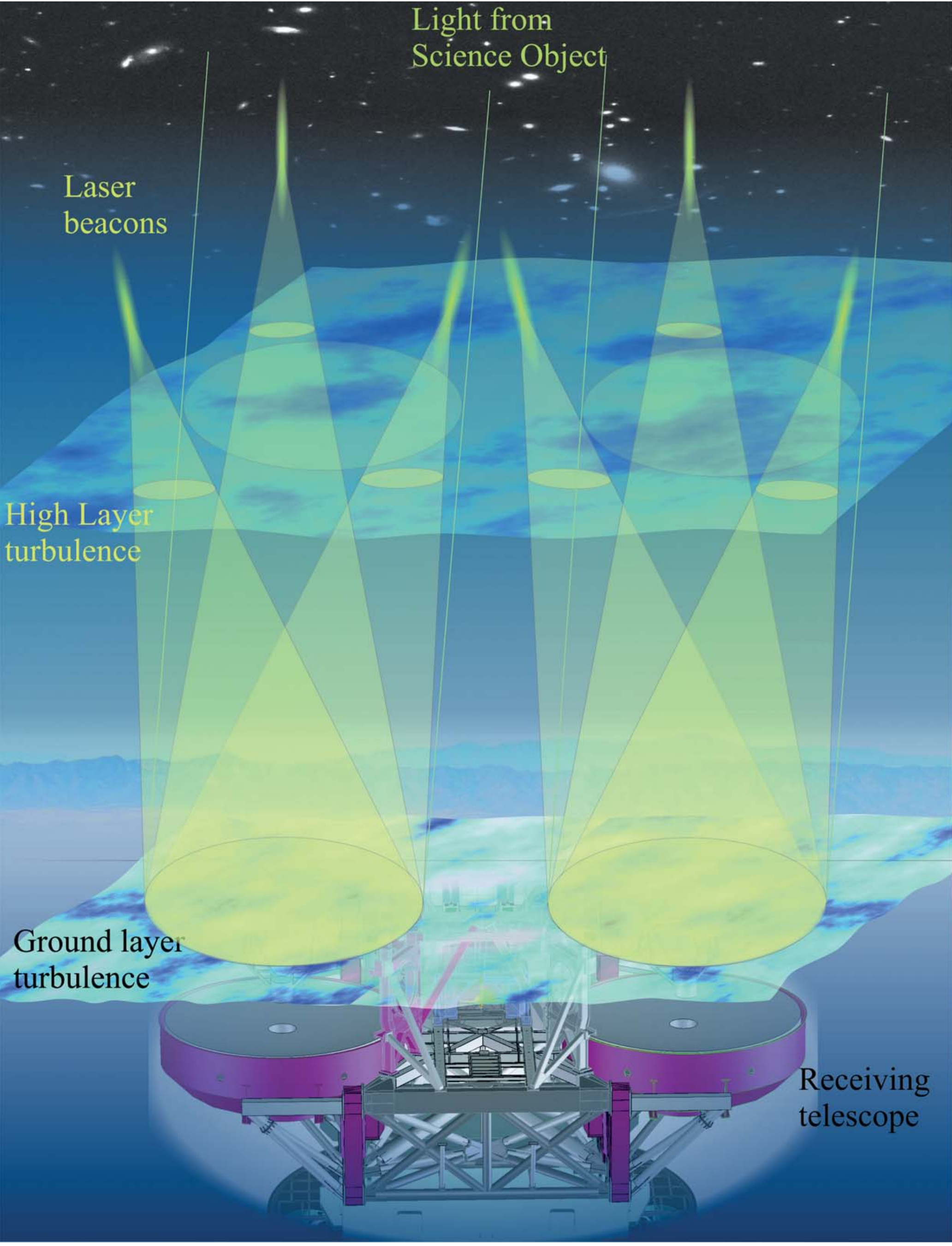}
      \caption{ Sketch of the basic geometry of the GLAO scheme with the six laser beacons serving as reference for the adaptive optics. Located at a finite height above the telescope, the light from the laser beacons travels down on a cone through the Earth's atmosphere. Turbulent layers located close to the ground will be sampled equally by all three beacons on each side, while on the high layers the footprints are completely separated and smaller than the illumination from the science object at infinity. The wavefront measurements from the high layers will average out over time, and the adaptive optics correction therefore is strongest for turbulent layers close to the ground.}
         \label{Fig_GLAO_show}
   \end{figure}
   
Within the multiple flavours of adaptive optics, the single-conjugate adaptive optics (SCAO) with sodium layer LGS is currently standard equipment at many large ground-based telescopes. The ESO UT4 \citep{Lewis2014, Rabien2003}, the Keck \citep{Dam2006, Wizinowich2006}, or Gemini telescopes \citep{boccas2006} offer facilities of this kind. Since SCAO mainly corrects the column of turbulence in the direction of the guide star the usable field of view is limited by the angular anisoplanatism to a $\sim$30$\arcsec$\ patch. In order to extend the suitable field of view, it is possible to  implement multi-conjugate adaptive optics (MCAO) with multiple guide stars and multiple deformable mirrors. Systems with this technique utilizing natural guide stars (NGSs) include  the MAD test system \citep{Marchetti2007} and LINC-NIRVANA \citep{Herbst2016} at the LBT. GEMS at the Gemini south telescope \citep{Rigaut2014, Neichel2014} is a MCAO system based on a sodium layer LGS constellation. Proposed by \citet{Rigaut02},  GLAO utilizes a single deformable mirror for the correction of the distortions located in the lower atmosphere, measured on multiple guide stars over a larger field. Natural guide stars can be used as reference sources, but many fields will not offer enough bright stars in a proper constellation to serve as adaptive optics probes. Laser guide stars are a natural choice for the wavefront measurement, offering a bright reference at the location of choice and constant positions for repeatable results. Successful realizations and tests of GLAO with Rayleigh lasers at smaller telescopes have been done at SOAR \citep{Tokovinin2008}, the WHT \citep{Morris2004}, and at the MMT \citep{Hart2010}. The basic geometry of a GLAO system is shown in Figure \ref{Fig_GLAO_show}. The constellation of laser beacons is placed at a finite distance above the telescope.  Therefore, the guide star light travels downwards through the atmosphere forming a `cone' on its path. Due to this geometry, the high-layer footprints of the beams are small and the constellations measurements will average over time, leaving the ground layer as the only common significant contribution for the AO correction. As has been shown in several studies \citep[e.g.][]{Avila1998, Garcia2007, Tokovinin2005, DaliAli2010}, and in measurements at Mt. Graham \citep{Masciadri2010}, the vertical atmospheric turbulence distribution often shows a prominent ground layer, allowing GLAO to remove most of the distortions. The favourable property originating from the ground-conjugate geometry is the wide field of view where the correction is effective. Placing the guide star constellation at a 4\,arcmin diameter circle corrects a similarly sized field for the science observation. The LUCI1 and LUCI2 instruments \citep{Seifert2003, Buschkamp12} cover this 4\,arcmin field for imaging and MOS spectroscopy. Having been designed as an adaptive telescope from the beginning, the LBT \citep{Hill2010} is equipped with adaptive secondary mirrors on both sides \citep{Riccardi2010}. Since the Gregorian telescope has the adaptive mirror conjugated close to the ground at \textasciitilde\,100m above the primary and the wide-field LUCI's as receiving instruments, the choice for GLAO was obvious. ARGOS saw the first laser light on  sky in November 2013 \citep{Rabien2014} and the adaptive optics loop running successfully in early 2015 \citep{Xivry2015}. Now that commissioning is coming to an end, the  8.4\,m apertures of the LBT are both equipped with an operational LGS-based GLAO system.

\subsection{System description}

     \begin{figure}
   \centering
   \includegraphics[width=9cm]{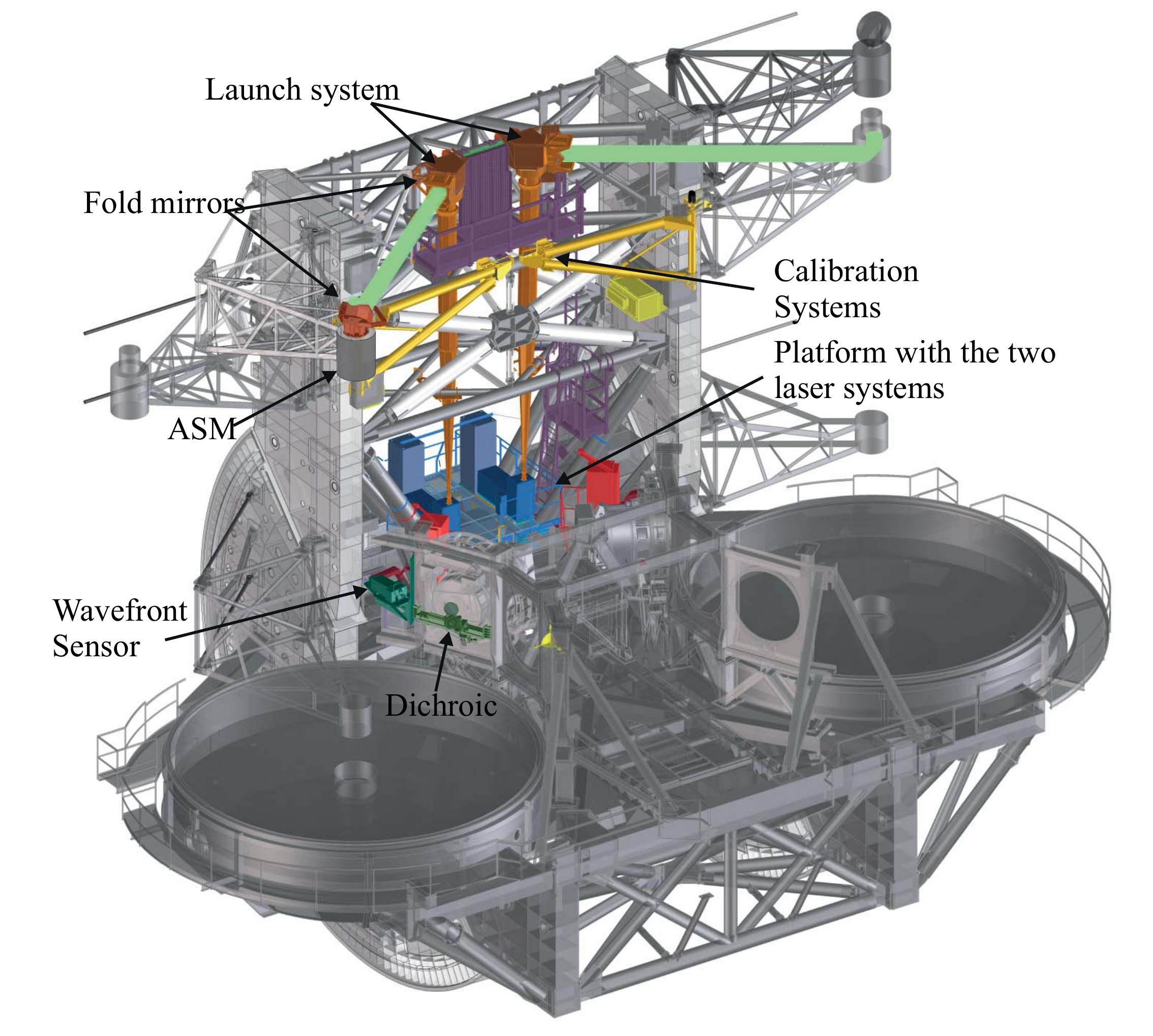}
   \caption{Overview of the ARGOS components as installed at the LBT. The laser units are located in the centre piece between the two mirrors on a dedicated platform. Leaving the laser systems, the beams are expanded with a refractive beam expander built into the LBT structure. The beams are directed  behind the secondary mirror and sent to the sky by large flat mirrors. On the return path the photons are split off with a large dichroic mirror, which sends the light to the wavefront sensor. Calibration light sources and optics mounted on swing arms can be moved into the Gregorian prime focus for the adaptive optics calibration. A total of eight electronics racks host the required infrastructure of drivers, readout, and controllers.}
              \label{Fig_argos_overview1}
    \end{figure}

ARGOS is based on constellations of laser beacons created by Rayleigh scattering from air molecules with multiple high-power pulsed green lasers. Upon a trigger command, all six laser heads fire synchronously the \textasciitilde40\,ns  light pulses. As they travel through the optical train of the laser systems and the launch telescope the beams are shaped, expanded, and focused at a distance of  12\,km. While propagating through the Earth's atmosphere, some photons out of each pulse are scattered backwards by air molecules, given the well-known Rayleigh cross section. After 80\,$\mu$s, photons scattered at a distance of  12\,km  arrive back again at the telescope and are detected by the wavefront sensor. In front of the instrument rotator structure we separate these photons with a dichroic beam splitter and direct the light towards the LGS wavefront sensors (LGSW). Inside these LGSWs the beams are first collimated before being sent through the Pockels cells gating units. These optical shutters are driven with a fast high-voltage switch to slice out exactly the photons scattered at 12\,km  within a selectable range, which is  usually set to 300\,m. The light out of this limited volume then propagates through the lenslet array onto the wavefront sensor detector. The detector itself is a fast, large frame PnCCD \citep{Hartmann2008} allowing all three LGSs to be imaged on a single frame. In terms of timing, the lasers and Pockels cells are triggered at a 10\,kHz repetition rate and the detector accumulates the photons from ten pulses before being read out, thus delivering a 1\,kHz frame rate. The frames from the CCD are then transferred to the ARGOS slope basic computational unit (BCU). This computer calculates the centroid position of all spots and sends the resulting slope vector to the adaptive secondary mirror where the reconstruction is performed in dedicated fast parallel computers, and the thin shell of the adaptive secondary mirror is set. Since the LGS position as measured on   sky does not reflect the atmospheric tilt properly, a separate NGS tip-tilt sensor is required. For this purpose ARGOS can use either its own avalanche photo-diode quad cell (APD-QC) set-up, or the First Light AO (FLAO) pyramid sensor \citep{Esposito2010a, Esposito2010b}. While the APD system can detect fainter stars, the usage of the FLAO pyramid has proved to be more convenient during operation. Indeed, the pyramid wavefront sensor is used to sense the ‘true’ values for 21 low-order wavefront modes and slowly offset the LGS wavefront slopes. Its usage as tilt sensor additionally avoids a complicated calibration of the APDs individual gains. The purpose of the truth sensing is to correct for non-common path aberrations, imperfect calibrations and the slight difference between elongated spots on sky and round calibration spots. ARGOS components are widely distributed over the telescope as can be seen in Figure \ref{Fig_argos_overview1}. Apart from the core units of the two laser boxes, the launch expanders and the wavefront sensors being described in this paper, the system relies on a multitude of auxiliary units to be functional:

\begin{itemize}
  \item A calibration system for each side mounted on swing arms \citep{Schwab2010}. This unit based on computer-generated holograms (CGHs)  delivers light sources mimicking the far off-axis LGSs and a single  NGS on sky in the Gregorian prime focus;
  \item A laser alignment telescope system to detect the initial locations of the lasers on sky and automatically adjust the position \citep{Sivitili2016};
  \item The high-speed slope computing units \citep{Biasi2004} based on a field programmable gate arrays system;
  \item A vibration compensation system based on accelerometers and fast laser uplink stabilization \citep{Peter2012};
  \item Transponder-Based Aircraft Avoidance (TBAD), the aircraft detection units for an operation without human aircraft spotters \citep{Rahmer2014};
  \item An infrastructure of drivers, controllers, cameras, computing units, safety systems, etc., filling eight electronics racks at the telescope.
\end{itemize}

The ARGOS main system parameters are listed in Table \ref{table_sys}.

\begin{table}
\caption{The parameters of the  ARGOS main system.}\label{table_sys}
  \centering
  \begin{tabular}{ll}
  \hline
  \noalign{\smallskip}
  Laser beacons per LBT side & 3  \\
  Laser beacon gating height & 12\,km\\
  Average power per laser & 18\,W  \\
  Laser repetition rate & 10\,kHz  \\
  Laser beam quality $M^2$ & $< 1.2$ \\
  Launch beam diameter at $\frac{1}{e^2}$ & 300\,mm\\
  Wavefront sensor frame rate & 1\,kHz \\
  Number of sub-apertures across primary & 15x15 \\
  Nominal range gating time & 2\,$\mu$s \\
  Nominal LGS photon flux on the WFS $^*$ & $5.8 \cdot 10^6 $ $ m^{-2} s^{-1}$ \\
  \noalign{\smallskip}
  \hline
  \noalign{\smallskip}
\end{tabular}
\raggedright ($^*$ this number includes an optical round trip efficiency of 25\%)
\end{table}

\subsection{Creating the laser beacons}

\begin{figure}
   \centering
    \includegraphics[width=8.3cm]{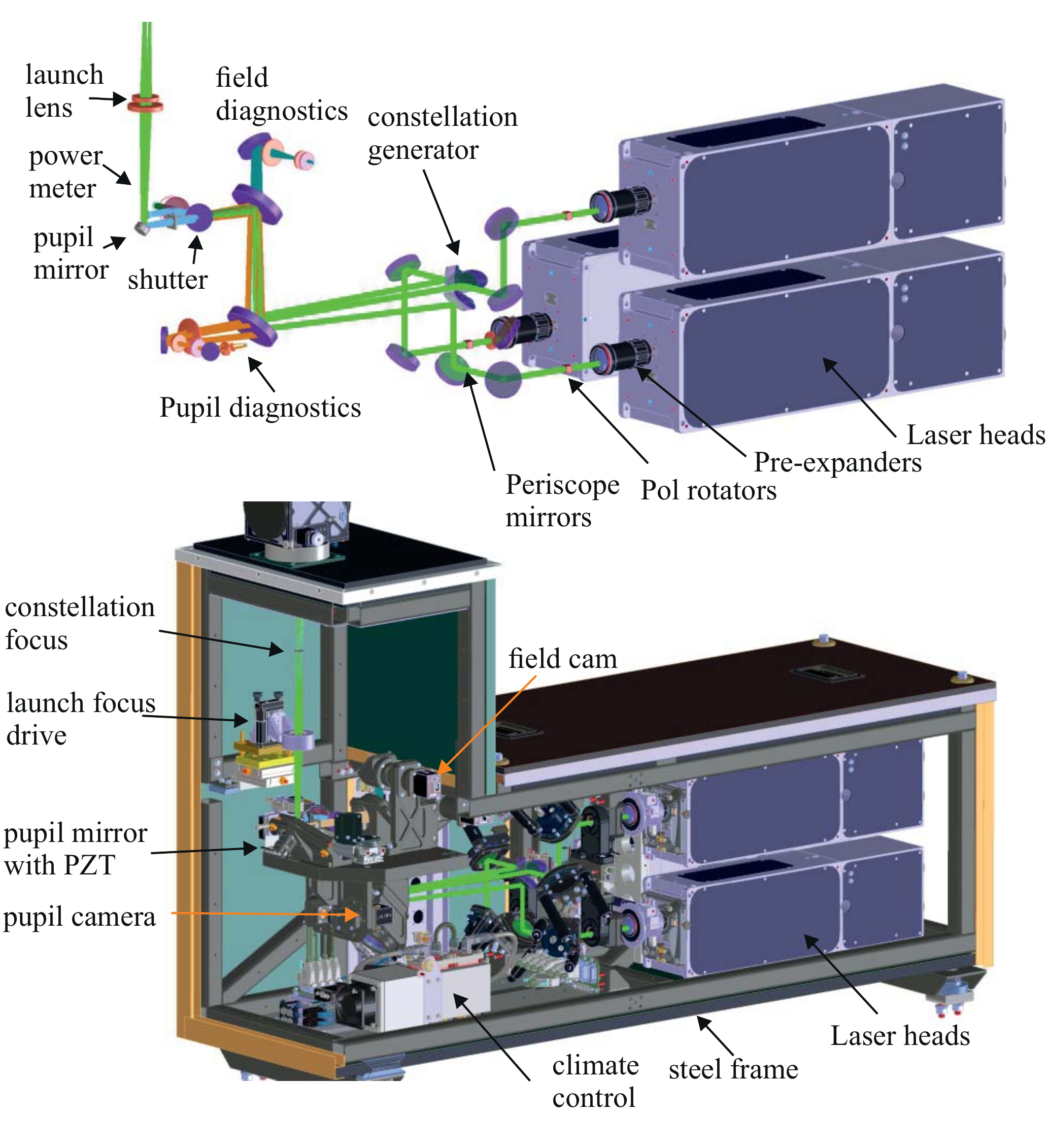}
      \caption{Optical layout (top) and the CAD model of one of the laser systems. Starting from the right the laser heads can be seen. In the beam path a pre-expander, polarization rotators, and a periscope assembly are installed before the constellation is generated and passed via fold mirrors towards the launch system. Diagnostics allow for an automatic pupil and field position control, stabilizing the constellation on sky. Average laser power and pulse shape measurements are built in. Finally, a safety shutter enables the beams to pass to the launch beam expander when opened.
              }
         \label{Fig_laser_box}
   \end{figure}

Although the very early LGS  systems already used Rayleigh scattering of high-power laser beams for the creation of the beacons \citep[see e.g.][]{Fugate91, Parenti92}, this method has become less popular since the adoption of sodium layer beacons. Beacons created in the sodium layer are at much higher altitude; therefore, the focal anisoplanatism is less problematic with this method. Nevertheless Rayleigh beacons offer several advantages over sodium beacons, making them very attractive:
\begin{itemize}
  \item  Rayleigh backscatter yields a high photon flux within the dense parts of the atmosphere;
  \item  Standard short-wavelength, visible, or UV pulsed lasers can be used. Lasers of this kind are currently available for industrial applications, with high power and beam quality. In contrast, sodium line lasers are still expensive and relatively bulky. This is a special advantage if multiple beacons are required, keeping the system compact and affordable;
  \item  The scattering height can be chosen freely, trading off altitude and volume coverage with backscattered photons;
  \item  By gating out all unwanted light in the wavefront sensors, Rayleigh guide star systems do not suffer from the `fratricide  effect', the pollution from scatter at low altitudes;
  \item  With slight adjustment of the gating time, an easy way of adjusting the wavefront sensor focus can be used. This is especially useful for a fast and stable way to close the adaptive optics loop;
  \item  In contrast to sodium beacons, Rayleigh beacons are not strongly sensitive to variations in particle abundance, nor distance variations of the backscattering layers.
  \item  The spot elongation is a free parameter in the design: longer gating times yield more return flux and more elongation in the outer sub-apertures.
\end{itemize}
Apart from the above mentioned focal anisoplanatism, drawbacks of Rayleigh beacons are that the gating height may be close to high-layer cirrus clouds. We have seen cases when the signal on the wavefront sensors increased massively due to clouds, disrupting the wavefront sensing. Fluorescence issues, as reported by the early Rayleigh guide star systems sharing the launch and receiving path, are not seen in systems with separate launch optics.
For ARGOS we use six frequency doubled Nd:YAG lasers from Innolas GmbH as beam sources. These lasers emit 40\,ns  pulses at 532\,nm with a nominal 18\,W average power each at 10\,kHz. With this pulse energy the calculated photon return from the beacons is of the order of 1800 photons per  sub-aperture and ms. At that high level of signal from the laser beacons, there is a factor \textasciitilde10 margin built into the system to account for  degrading laser power, optics, and atmospheric transmission, etc.
The lasers are built into two boxes, one per side of the LBT. One of the boxes is shown in Figure \ref{Fig_laser_box}. Inside each box we shape and direct the laser beams into the desired locations. Following the beam train after the exit from the laser heads, we collimate and pre-expand the beams to a 6\,mm $\frac{1}{e^2}$ diameter, rotate the polarization to match the Pockels cell gating units, steer the direction of each laser individually to position it on the constellation, send it to a common piezoelectric (PZT)  mirror in the launch pupil plane, and let it exit from the box into the main beam expander. To make the set-up work, additional diagnostics and controls are built-in:
\begin{itemize}
  \item  Two cameras measuring the location of the beams in the field and pupil plane, feeding a loop to adjust the positions;
  \item  A specially developed white-light shearing interferometer to control the beams collimation on demand;
  \item  A power meter to measure the laser's power and photo diodes to monitor the pulse shape and emission timing;
  \item  A safety shutter to enable or disable the laser propagation to sky;
  \item  The vibration compensating piezoelectric pupil mirror;
  \item  Several devices to control the temperature, humidity, and cleanliness of the unit.
  \end{itemize}

   \begin{figure}
   \centering
    \includegraphics[width=6.38cm]{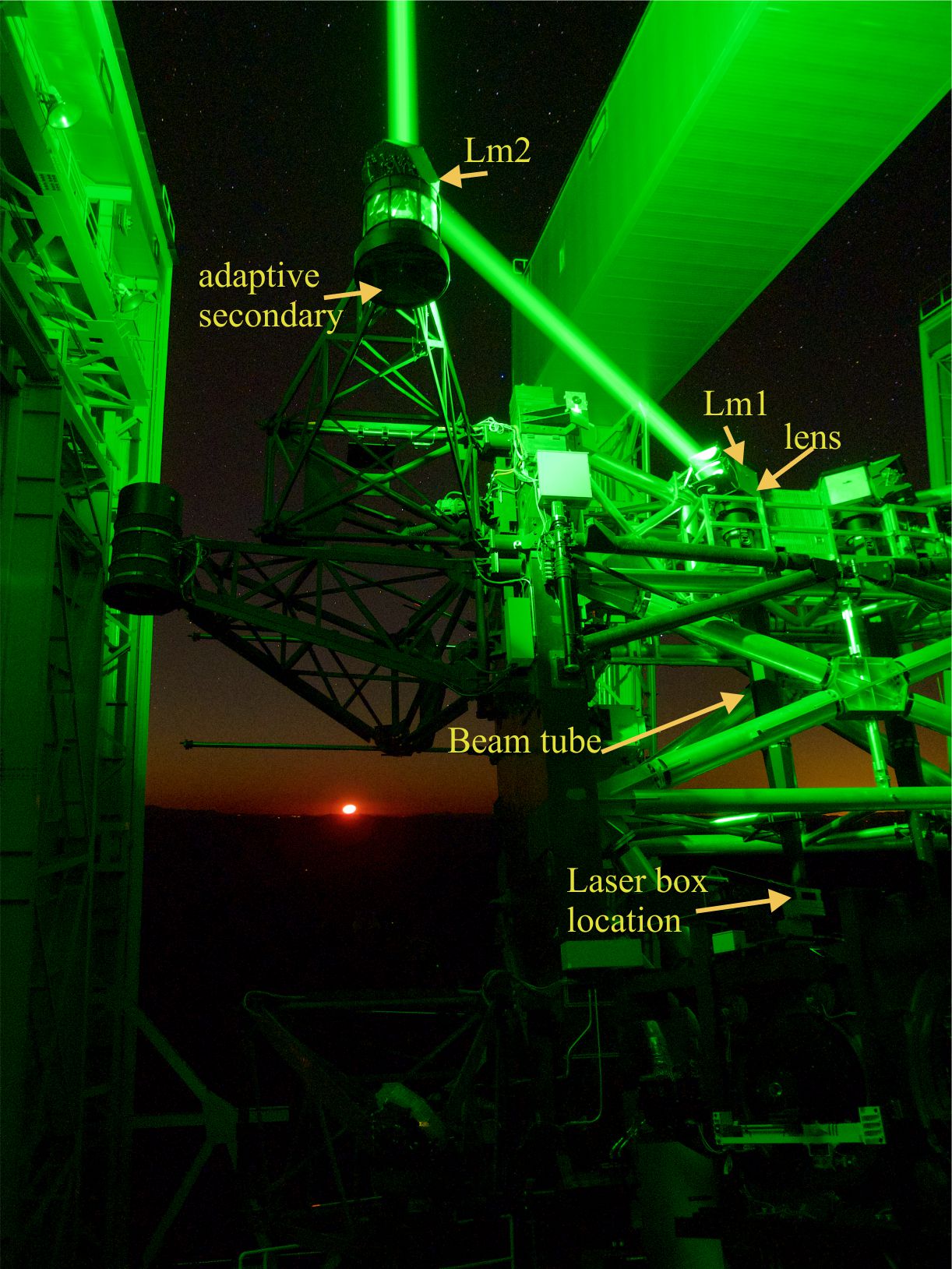}
      \caption{Photograph of the left side of the telescope with the lasers in operation. The launch beam path from the exit to the first fold mirror is enclosed in a beam tube for safety and light scattering reasons. Between the first fold mirror and the mirror behind the adaptive secondary the beam is propagated in free air above the primary mirrors. From the mirror behind the secondary the laser bundle is launched to sky, forming the beacon constellation at a distance of  12\,km.
              }
         \label{Fig_launch_image}
   \end{figure}

\begin{figure}
   \centering
    \includegraphics[width=5.6cm]{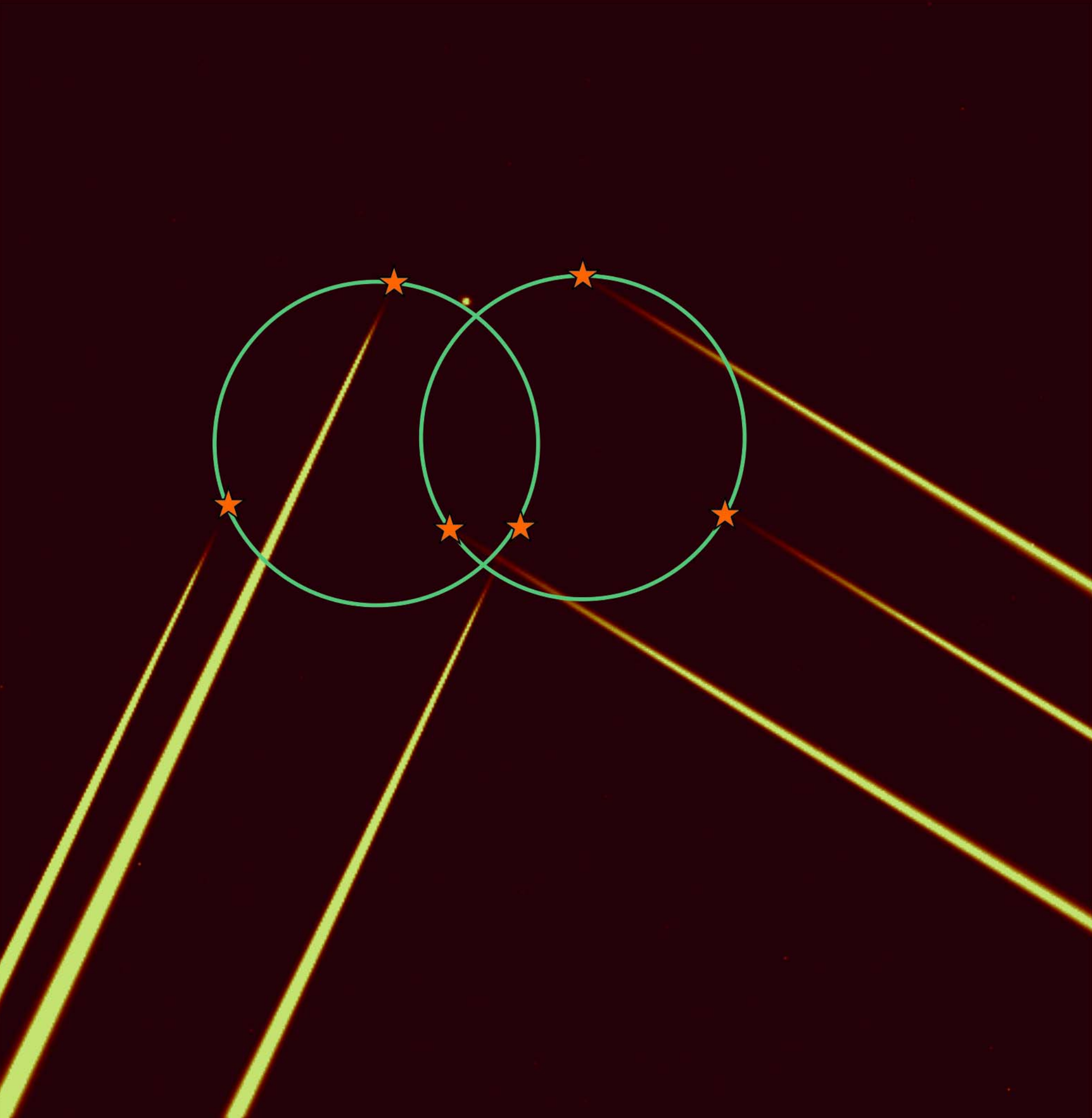}
      \caption{Image of the double laser constellation on sky, taken with a small auxiliary telescope. Each bundle of lasers belongs to one side of the binocular telescope. The circles guide the  eye, showing the constellation diameters of 4 arcmin on sky. The stars indicate the locations of the Rayleigh tip from where the scattered light is used for the wavefront sensing.}
         \label{Fig_constellation}
   \end{figure}

  Being able to create a small spot at a distance of 12\,km  from the telescope requires a sufficiently large beam to be launched. The enlargement of the 6\,mm laser beams and steering onto the axis of the telescopes are done by the launch telescope. A photograph of the launch path in operation is shown in Figure \ref{Fig_launch_image}. It consists of a refractive aspheric beam expander and two large fold mirrors. The expander has been designed with  \textasciitilde 8\,m focal length aspheric fused silica lenses, being held in the LBT's steel frame as a mount.
  The expanded 40\,cm aperture beam is reflected with a fold mirror across the primary and with a second fold mirror behind the adaptive secondary towards sky. These two fold mirrors are manufactured from Borofloat (a borosilicate glass), using an internal lightweight honeycomb rib structure with an optically polished front plate. During the commissioning phase we found this structure to bend slightly due to temperature differences between the front and back plate, mainly driven by the small amount of laser light that is absorbed and therefore heats the surface. This effect made it necessary to implement a twofold correction mechanism: a counter heater on the back surface of this mirror to keep the temperature difference at a constant level, and an astigmatic compensation system in the beam. Consisting of two tilted flat glass plates this compensation unit can be adjusted remotely to minimize the spot sizes on the wavefront sensor together with focusing the beams. The lasers focus is adjusted by moving the small launch entrance lenses in the expander units along the optical axis.
  
  Because the LBT is a binocular telescope with a large structure, significant flexure and differences in pointing can occur between the mount and the mirrors. The laser and launch system are built into the centre piece of the steel structure and follow its movement. When setting up at the beginning of the night, or moving to a new object, the constellation (shown in Figure \ref{Fig_constellation}) usually requires a new pointing correction. Finding the laser positions on sky is carried out with the help of the laser alignment telescope (LAT) and a routine calculating the required movements of the large fold mirror, as described in detail by \cite{Sivitili2016}. Once located in the field of view of the patrol cameras of the wavefront sensor, the laser beacons are placed by a `click-and-go' procedure in the central aperture.

\subsection{Sensing the laser beacons}

 \begin{figure*}
   \centering
    \includegraphics[width=16cm]{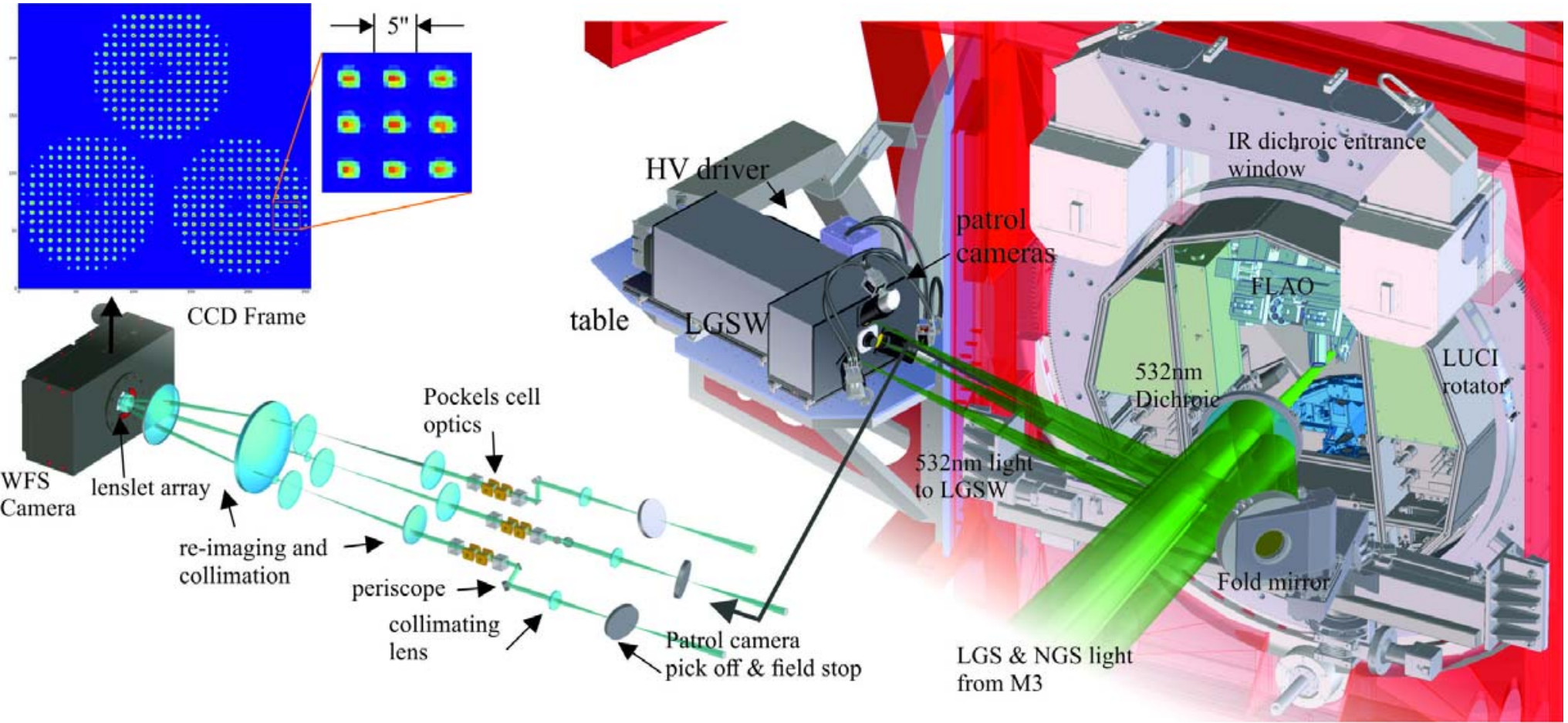}

      \caption{Optical path and CAD model of one of the wavefront sensors. The light from the LGSs enters from the right side in its 4 arcmin wide constellation. Patrol cameras enable a fast acquisition and position control of the guide stars. Inside the wavefront sensor units motorized mirrors stabilize the field and the pupil, driven by the signals from the WFS CCD. After that the gating units slice out the photon  bunches originating from the 12\,km distant scattering, letting a 300\,m range (i.e. 2$\mu$s) long part pass to the wavefront sensor detector. These Pockels cells units operate at the same 10\,kHz as the laser pulses are sent to sky. The wavefront sensor detector accumulates the charge originating from 10 consecutive pulses, being read out at a 1\,kHz frame rate. The wavefront sensor CCD frame of one telescope side with the three LGSs is shown on the left. Having passed a common microlens array in front of the detector, all three guide stars show a Shack-Hartmann spot pattern on the detector. From each spot the local wavefront slope can be measured, enabling the reconstruction and correction of the ground-layer turbulence. Because LGS spots are often larger than NGSs, the sub-aperture's field of view amounts to $\sim$5$\arcsec$ sampled with 8x8 pixels on the CCD frame.}
         \label{Fig_wfs_cad}
   \end{figure*}

\begin{figure}
   \centering
    \includegraphics[width=9cm]{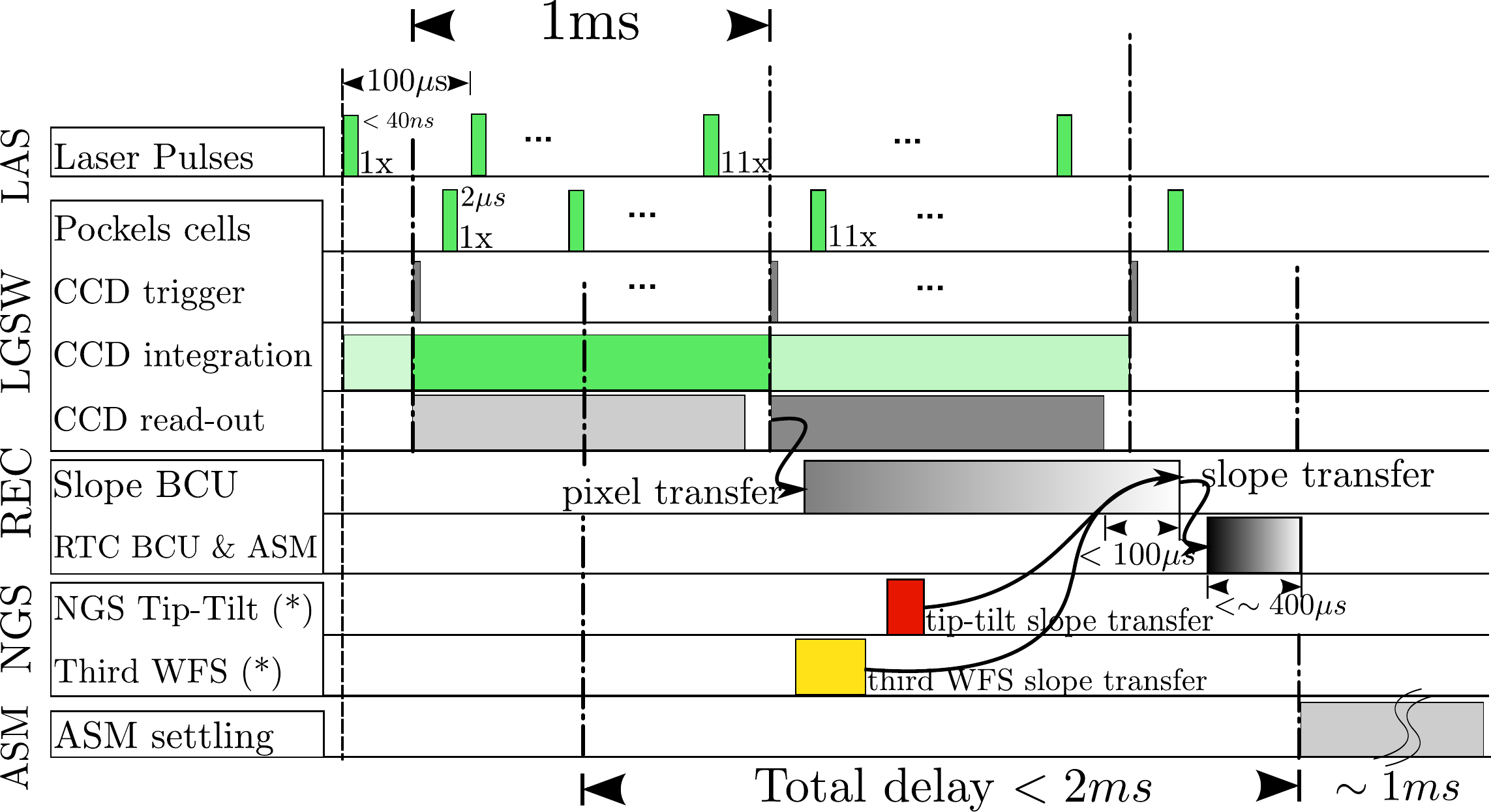}
      \caption{ Sketch of the complete ARGOS timing sequence from laser pulse launch down to the settling of the adaptive secondary mirror (ASM). The 40\,ns  laser pulses are generated at a 10\,kHz rate. The back and forth propagation in the atmosphere takes 80\,$\mu$s at which point the Pockels cells are opened for 2\,$\mu$s (corresponding to a 300\,m slab), thus exposing the PnCCD to the laser photons. Ten successive pulses are integrated on the detector, the readout is triggered taking just less than 1ms. The analogue outputs are directly sent to the slope BCU unit, which digitizes the analogue pixel signals and perform all the necessary operation to compute the Shack-Hartmann slopes in a parallel manner, thus minimizing the delay (est. < 100\,$\mu$s) after the end of the CCD readout. Asynchronous signals (indicated by a $\star$), a particular feature of ARGOS, such as tip-tilt signals from our APD or the FLAO pyramid slopes, can be concatenated to the LGS signals. The final slope vector is transmitted to the real-time computation BCU for reconstruction and control of the ASM. Disregarding the ASM settling time, the total delay is less than 2\,ms.}
         \label{Fig_WFS_timing}
   \end{figure}

 On the downwards path the LGS light is collected by the two primary mirrors of the LBT and sent via the secondary and the tertiary mirror towards the science instruments. Figure \ref{Fig_wfs_cad} shows a drawing of the optical path and the multiple wavefront sensors units as installed at the telescope. After the telescope's tertiary mirror, in front of the instrument rotator, a dichroic beam splitter separates off the 532\,nm laser light and sends it towards the LGS wavefront sensor (LGSW). Light from the  NGS passes the dichroic, is reflected by the LUCI entrance infrared-transmitting window, and is sent to the FLAO sensor. The infrared light passes the entrance window and enters the LUCI spectrographs.
 Both ARGOS LGS wavefront sensors are based on a Shack-Hartmann (SH) scheme to detect the LGSs. Details are described in \citet{Bonaglia2014}. Being located in a conjugated plane of the 12\,km focus, a field stop with $\sim$4$\arcsec$ diameter forms the entrance to the WFS optics. For each of the guide stars, a collimating lens re-images the pupil via a periscope onto the entrance of the Pockels cells. The periscope mirrors bring the beacon constellation closer together, and allow for field stabilization inside the sensor. The collimated beam – strictly collimated when the laser light comes from exactly 12\,km  – arrives at the Pockels cell assembly. This unit has been specially designed to ensure an excellent suppression over the large field of view of the Shack-Hartmann sub-apertures, 5$\arcsec$ on sky or the corresponding 1.6$^\circ$ ray angle at the level of the Pockels cells. Due to the 8.4\,m apertures of the primary mirrors, large ray angles are present at the Pockels cells assembly, prohibiting the use of standard commercial cells. The custom developed cells consist of double beta-barium borate (BBO) crystals in an optical arrangement ensuring the highest suppression over a large ray angle range.
 Applying a 9\,kV rectangular voltage pulse to the crystals electrodes, the cells act as the optical shutter. In the range gating sequence the opening is applied \textasciitilde 80\,$\mu$s after the laser pulse has been sent to sky and  then closed  2\,$\mu$s later. All light that is scattered in the atmosphere before or after the high-voltage pulse is suppressed by a factor of more than \textasciitilde 1000. This lets only those photons pass towards the CCD that have been scattered between 11.85 and 12.15\,km above the telescope. At that suppression rate we have calculated the resulting wavefront error in each affected radial mode to be less than 8\,nm RMS, sufficiently low to be of negligible influence. With the lasers repetition rate being set to 10\,kHz, we accumulate the charge of ten pulses on the sensor before the readout is triggered. The complete timing sequence is shown in Figure \ref{Fig_WFS_timing}. The readout of the CCD takes just less than 1\,ms during which the analogue signals are transmitted to the ARGOS slope BCU. This unit digitizes the analogue pixel signals, performs the necessary calibration and computes the slopes. The operations are performed in parallel to the arrival of the pixel charges and thus ensures a minimum latency, being estimated to be <100$\mu$s, with respect to the end of the readout. A particular feature of the ARGOS slope BCU is that it can asynchronously receive other WFS signals such as tip-tilt measurements from our APD or WFS slopes from the FLAO pyramid sensor. A final slope vector is concatenated and transmitted to the real-time computer (RTC) BCU that reconstructs and controls the adaptive secondary mirror (ASM). Without including the ASM settling time, the total latency in the timing sequence is less than 2\,ms.
 Optically, after the gating units in the LGSW, each beam passes a focusing lens and a common collimator that forms three pupil images on a single lenslet array directly before the CCD. The optics are adjusted such that the three SH patterns are distributed over the CCD as shown in the upper left inset of Figure~\ref{Fig_wfs_cad}. Each sub-aperture spans 8x8 pixels, corresponding to $\sim$5$\arcsec$$\times$5$\arcsec$ on sky,  and is aligned such that the split frame transfer does not cross the sub-apertures. More information on the wavefront sensor performance is given in \citet{Bonaglia2014}. The CCD itself is based on a deep-depletion technology, developed at the Max Planck semiconductor laboratory. It offers 264x264 pixels and 3.7\,e$^-$ readout noise at a 1\,kHz frame rate. Details can be found in \citet{Xivry2014}.

\subsection{Correcting the ground layer}\label{sec:correcting}

\begin{figure}
   \centering
    \includegraphics[width=9cm]{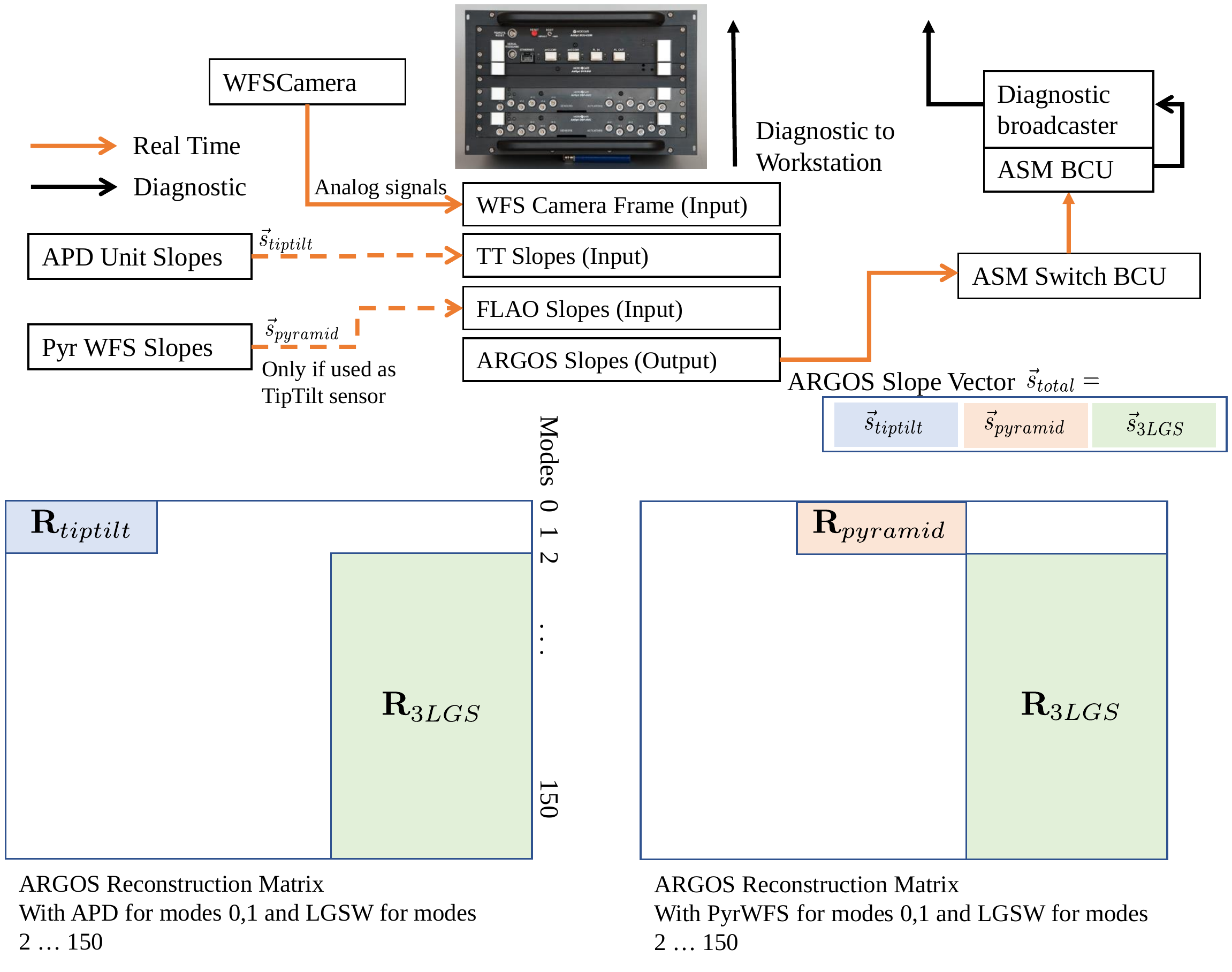}
     \caption{ Sketch of the ARGOS WFS computer and reconstruction matrix. First, the computer receives the analogue signals through 8 channels from the ARGOS pnCCD camera. It digitizes the data and sends the pixel values to the computing part of the board for LGS slope computation. In parallel, the computer can also receive the NGS tip-tilt slopes and the slopes from a third WFS, currently from the first light AO (FLAO). Finally, a slope vector is assembled and sent to the ASM switch BCU, selecting the working wavefront sensor, and then forwarded to the ASM control.}
          \label{Fig_WFS_computer}
   \end{figure}

The core of the adaptive optics computation with ARGOS is carried out by the wavefront sensor computer, as sketched in Figure \ref{Fig_WFS_computer}. It receives the PnCCD analogue signals, and finally transmits a slope vector to the LBTs adaptive secondary mirrors (ASM) where the wavefront reconstruction is performed. This computation architecture has to carry out several main tasks and fulfil tight constraints:

\begin{itemize}
    \item Digitizing the analogue signal from the ARGOS LGS wavefront sensor camera;
    \item Calibrating and background subtracting the frames;
    \item Computing the x and y local wavefront gradients for three SH pupils, each containing about 176 sub-apertures, hence 1056 displacements;
    \item Providing the interfaces to the other wavefront measurements, in particular to the NGS tip-tilt, mandatory for the AO operation, but also for the measurements of a third WFS providing an NGS or sodium guide star measurement. This allows a `hybrid' AO operation;
    \item Calculating the global tip and tilt signals for each LGS and controlling the overall field position correction on the CCD;
    \item Delivering a global tip-tilt signal to drive the uplink correction performed by the pupil mirror in the launch system;
    \item Transmitting the wavefront measurements to the LBT real-time reconstructor and deformable mirror controller.
\end{itemize}

    ARGOS has three different wavefront sensing measurements: the three combined  laser measurements, the NGS tip and tilt,  and a third wavefront sensor, that can be used to perform a `hybrid' correction scheme, combining GLAO with a SCAO mode. Currently the FLAO pyramid sensor can be concatenated into the slope vector. Those measurements are collected in a vector $s_f = [s_{3LGS}; s_{TT}; s_{FLAO}]$, for a total of 1600 slope measurements, and are sent to the ASM. At the ASM a reconstructor converts the slopes into the mirror Karhunen-Lo\`eve (KL) modal basis, using an integrator control. Once computed, the modes are then projected on the command space of the ASM by matrix multiplication, which is then directly used by the internal control of the ASM. An additional disturbance vector can be added to this command vector, which is typically used for calibration. 
    
\begin{figure}
   \centering
    \includegraphics[width=9cm]{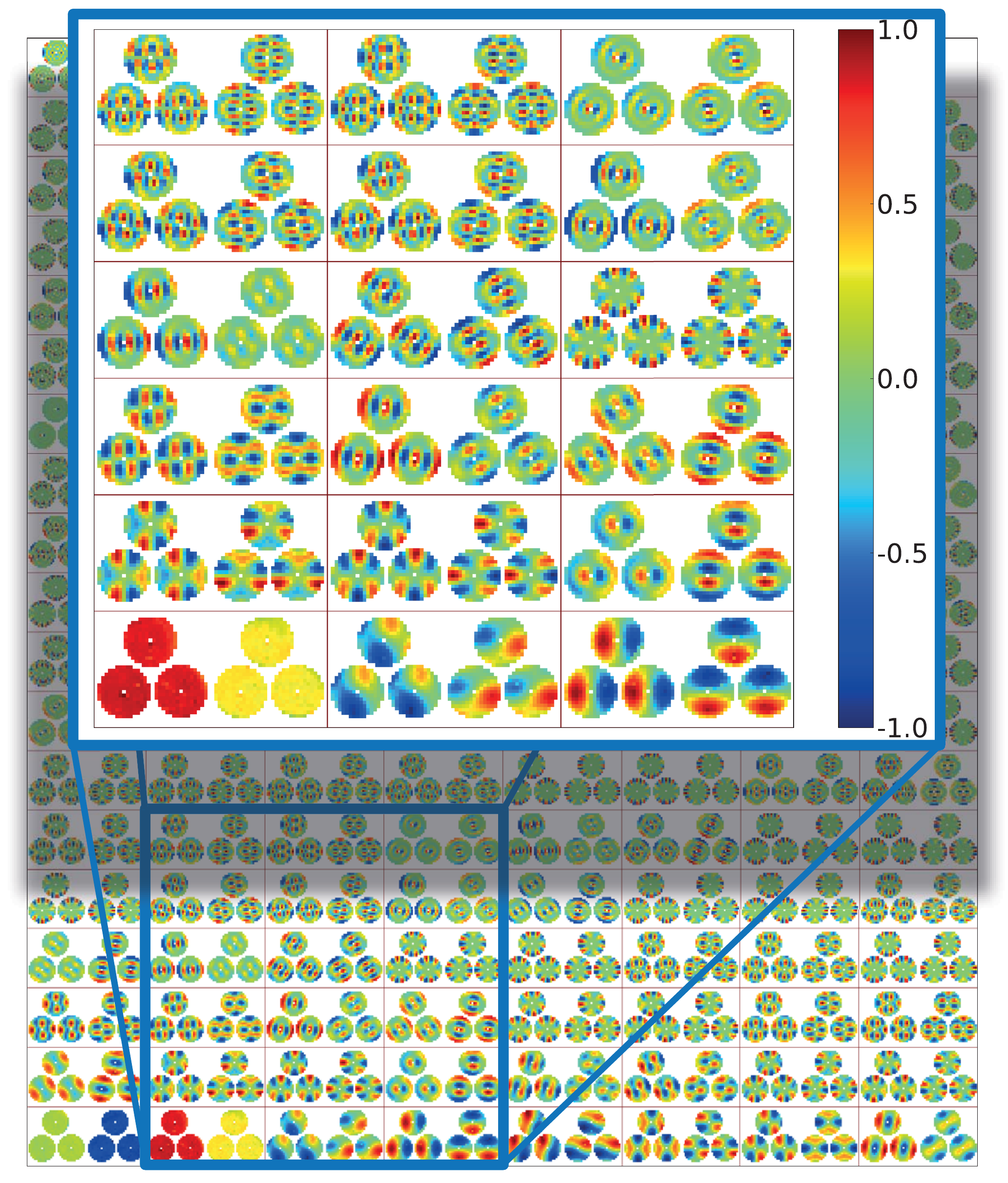}
      \caption{Image of the ARGOS interaction matrix showing the x- and y-direction slopes for each KL mode. Since the WFSs are SH, the image of the interaction matrix is the derivative of the wavefront in the x and y directions. The bottom line in the matrix displays the low-order modes: (from left to right) Tip, Tilt, Astigmatisms, Focus, and Coma. The inset  shows 18 modes for illustration. Values are in arbitrary units, scaled from -1 to 1, while the interaction matrix used for the computation is in units of RMS wavefront deviation divided by unity slope [m/arcsec].              }
         \label{Fig_IM}
   \end{figure}

    The main element to be calibrated from the ARGOS point of view is the reconstruction matrix used to perform the mapping, from the several wavefront measurements to the modal amplitudes. This matrix is obtained by pseudo-inversion of the interaction matrix, which  is part of the calibration of the ARGOS AO loop and is obtained as follows: utilizing a push-pull sequence, modes are successively applied to the ASM (i.e. by using the disturbance vector mentioned above). For each of them the wavefront measurements are recorded, thus constructing the matrix IM of dimension (Nslopes $\times$  Nmodes). The amplitudes of the push-pull of each mode is optimized to provide a uniform signal-to-noise ratio by re-scaling the applied amplitudes to have the same slope standard deviations \citep[for more details see][]{Esposito2010b}. An example of the full interaction matrix is shown in Figure \ref{Fig_IM}. The final ARGOS reconstructor is a block-wise matrix containing the
    GLAO reconstructor $R_{3LGSs}$ and $R_{TT}$ or $R_{FLAO}$ depending on which sensor is used for the tip-tilt correction.
    Considering the symmetry of the LGS constellation, the GLAO reconstructor that estimates the ground layer and averages out higher-layer altitudes is obtained by taking the pseudo-inverse of the three LGS interaction matrices at once: $R_{3LGSs} = IM_{3LGSs}^{\dagger}$, where $IM_{3LGS}$ is of dimension $n_{slopes} \times n_{modes}$ or typically $1057 \times 150 $.
    The block-wise reconstructor provides great flexibility regarding which wavefront sensor is used. In addition to these wavefront measurements and the modal reconstruction, ARGOS uses the NGS wavefront sensing unit (FLAO) to measure slowly changing non-common path aberrations between the LGS optical beam and the light on the pyramid sensor. This `truth' sensing is projected back to the LGS slopes and is added as new offsets in the ARGOS wavefront sensing computer.

\subsection{Operating ARGOS on sky}\label{sec:operating}

Getting the adaptive optics loop closed and operational on sky requires the whole chain of system items to be functional: from the telescope collimation itself to the focusing and shaping of the laser beams on sky, acquisition of the laser beams on the wavefront sensor, locking the LGS guiding loop, and the acquisition of the natural tilt star, and finally to a sequence of actions to close the adaptive optics loop. Having the loops closed then enables the science integrations to start. In the following, we briefly describe the acquisition and loop operations, and we provide insights into the way the laser star and tilt guide star operation is handled upon offsetting, sky frames pointing, and asynchronous interruptions such as aircraft passage or satellite closures. More details on the adaptive optics operation are given in \citet{Busoni2015}

\subsubsection{Acquiring the stars}

{\it Pre-acquisition:} After the telescope is properly collimated, a pre-set is executed to slew to the science field where the Satellite Avoidance System allows the laser operator to enable the propagation of the six lasers \citep[see e.g.][]{Rahmer2014}. To simultaneously align the LGS constellation on both of the telescope's optical axes, the laser alignment telescope (LAT) is used. This allows  the three LGSs to be acquired in the 40$\arcsec$ field of view of the patrol cameras of each LGSW.

{\it Beam sharpening:} The laser launch optics are equipped with two active devices to compensate for focus and astigmatism aberrations in the uplink path. While the need for focus compensation is obvious, astigmatism is introduced by the thermal bending of the launch mirrors LM1 and LM2 (see Figure \ref{Fig_launch_image}) in the launch system itself. To restore the laser beam quality the compensator position must be optimized by running an automatic procedure that scans the entire range of the focus stage while recording images of the laser spots on the LGSW patrol cameras in an intra-focus/extra-focus scheme. For each focus stage position the spot size is measured along two orthogonal directions. The distance between the two minima in the spot length curves is proportional to the total amount of astigmatism in the uplink propagation. The value is converted to a compensator adjustment by using a model of the launch optics.

\begin{figure}
   \centering
    \includegraphics[width=9cm]{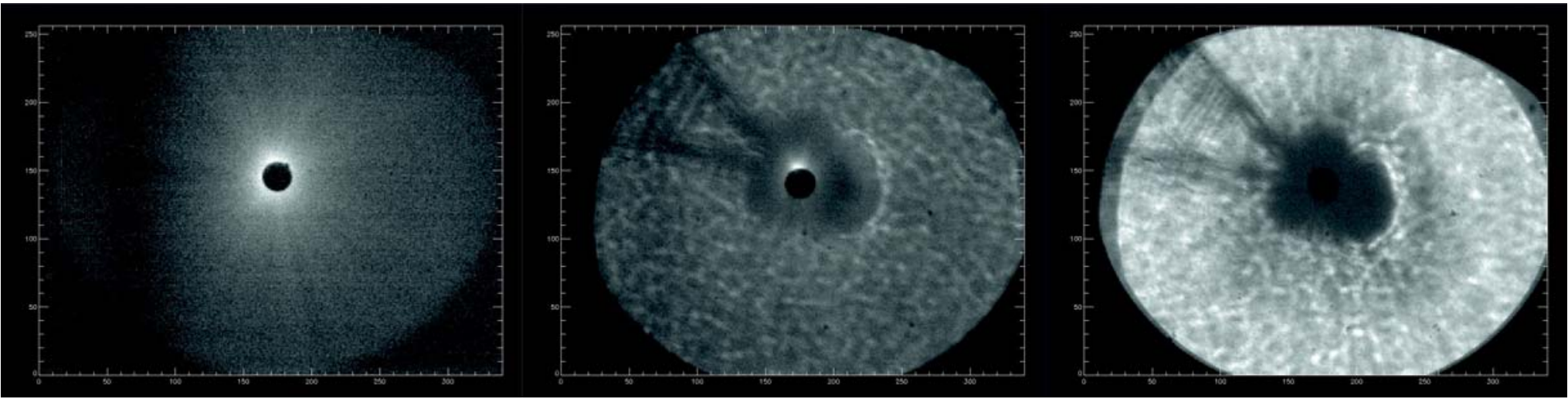}
      \caption{One of the LGSs as seen on the patrol camera. This camera has a field of view of 1 arcmin and is used to acquire the lasers in a click-and-go procedure. Shown in the left panel is a clear sky view of the laser as seen by this un-gated camera, focused to 12km. The laser spot vanishes in the central hole, being the entrance aperture of the WFS. In the middle panel scatter from lower altitude cirrus clouds can be seen, clearly distinguished from the LGS. In the right panel thick clouds make the star at 12km vanish completely. }
         \label{Fig_LGS_clouds}
   \end{figure}

{\it Final-acquisition:} The LGS acquisition requires human intervention: the ARGOS operator has to look at the LGSW patrol camera display and to click on the position of the LGS spots that are easily identifiable by eye. This has been shown to be much more robust than an automatic procedure, especially in the case of thin clouds creating scattered images on the LGSW patrol cameras. Images of the LGS on one patrol camera are shown in Figure \ref{Fig_LGS_clouds}.
As soon as the operator confirms the LGS acquisition, the LGS guiding loop is automatically closed. This algorithm uses the mean tip-tilt slopes recorded by the three SH sensors to evaluate the common drift of the LGS constellation on sky and to apply a proper tip-tilt correction to the launch system pupil mirror. This process is a real-time loop, running at 1 kHz implemented in the ARGOS slope BCU. The offset accumulated by the laser system pupil mirror is periodically offloaded to the LM1 that has the same optical effect on sky but with a larger stroke.

Together with the guiding loop the vibration compensation system is activated. This system uses the piezoelectric launch system pupil mirror compensating for the uplink laser jitter. The system relies on the measurement of eight accelerometers attached to the back of the 2 LMs in the launch optics and it implements an open-loop feed forward control and a Kalman recursive filter. \citep[Details in][]{Peter2012}.

In parallel with the process of LGS acquisition, the NGS sensor is configured and the natural star chosen by the LUCI script as the tip-tilt and truth sensing star is acquired. As ARGOS makes use of the FLAO hardware to implement the NGS WFS, it also uses its control procedures to move the acquisition stages on target, to configure the WFS hardware, and to activate the real-time communication with the slope computer.

\subsubsection{Enabling the adaptive correction}

Having both LGS and NGS acquired on the respective WFS automatically triggers the closure of the GLAO loop. This consists of several operations performed in sequence:

\begin{enumerate}

\item \textit{Configure the real-time computer (RTC)}. The reconstructor matrix is selected according to the chosen configuration (APD vs Pyramid, binning of pyramid WFS) and to the magnitude of the NGS. The reconstructor matrix is uploaded to the ASM RTC. The ARGOS slope computer BCU is re-configured, if needed.

\item \textit{Close the tip-tilt loop}. The real-time communication between the slope computer and the reconstructor on the ASM is enabled, with all modal gains set to zero. Then tip-tilt gains are ramped up from 0.01 to 0.2 in a few seconds.

\item \textit{Offload focus to time-of-flight}. The altitude to which the LGSW is conjugated can be varied by modifying the time-of-flight, i.e. the time passing between the trigger of the laser pulse and the signal that opens the gating units in the LGSW. This interval, being nominally 80\,$\mu$s and corresponding to the round-trip of the laser pulses to 12\,km, is adjusted to null the focus term measured by the SH WFS. This ensures a smooth operation when the LGS loop is closed.

\item \textit{Close the LGS loop}. The gain of the modes controlled by the LGS are increased in steps from 0.001 to 0.1 in a few seconds.

\item \textit{Start the Truth Sensing}. The FLAO control software computes autonomously the true wavefront error by projecting on the same modal basis that is used in ARGOS for the adaptive correction. The modal coefficients are read by the ARGOS control software and converted into LGSW signals through a multiplication by the LGS interaction matrix. This signal vector is integrated to the current LGSW slope-offset vector for all modes but the focus. Focus is offloaded to the time-of-flight, as described above, to retain the maximum dynamical range of the SH sensor.

\item \textit{Optimize modal gains}. As a last step, the modal gains can be optimized. An optimization script that scans a range of modal gains and searches for the values that minimize the WFS signal variance can be optionally executed. The operator has the option to adjust the values of the modal gains (grouped in three sets: tip-tilt, modes from 2 to 36, and higher modes) from the ARGOS control GUI. This procedure is not always required for the modes controlled by the LGSW because the sensitivity of the SH sensor is stable under most operating conditions. However, it is always required for tip and tilt modes because the quad-cell sensor and the pyramid WFS sensitivity depends on PSF size, which in turn depends on seeing.
\end{enumerate}

\subsubsection{Dithering, offsets, and asynchronous interruptions}\label{sec:offsets}

During the execution of a LUCI-ARGOS observing block, there are several circumstances that require pausing the adaptive optics correction for a short period. These include offsetting the field or stopping the laser propagation due to satellite or airplane transit. In the first case, because of the way the LBT handles binocular offsets between the mount and the two telescopes, it is difficult to predict whether the LGS light will stay on target. The easiest solution is to pause the LGS loop by setting the mid- and high-order modal gains to zero. The size of the offset then plays a role in the way the LGS loop is resumed:

\begin{itemize}
\item {\it small offsets}, within the reach of the FLAO field positioning stages ( $3’ \times 2’$ ), the NGS loop is paused, the board stages are then moved to re-centre the star on the tip-tilt sensor and both the NGS and LGS loops are resumed by ramping up in two steps the modal gains to the original value. During the offset execution, the truth sensing is disabled.
\item {\it larger offsets}, where the NGS cannot be re-acquired because it is out of the field reachable by the FLAO stages, the whole adaptive optics loop will simply remain paused. When pointing the telescope mount back after the large offset, the adaptive optics loop will be automatically resumed, with the same procedure as for small offsets.
\end{itemize}

When a satellite or an airplane transit requires laser propagation to be stopped, the control software automatically pauses the LGS loop and then stops the lasers: this permits  the closed-loop observation to quickly resume as soon as the lasers are propagated again on sky without the overhead of a new acquisition. The control software tries to resume the loop automatically. If it fails, it leaves to the ARGOS operator the task of reacquiring the LGSs, which  may have drifted out of the field during the pause. In the current telescope control software scheme there is no way for ARGOS to interact with the LUCI script sequencer, leaving it unaware of these kinds of events and unable to react and optimize the observation. Instead the AO loop status is recorded in the fits header. Since most interruptions are short, the LUCI integration just can continue, with a minor reduction in the data quality.

\subsection{Spectroscopy with ARGOS}\label{sec_spectroscopy}

   \begin{figure}
   \centering
    \includegraphics[width=6cm]{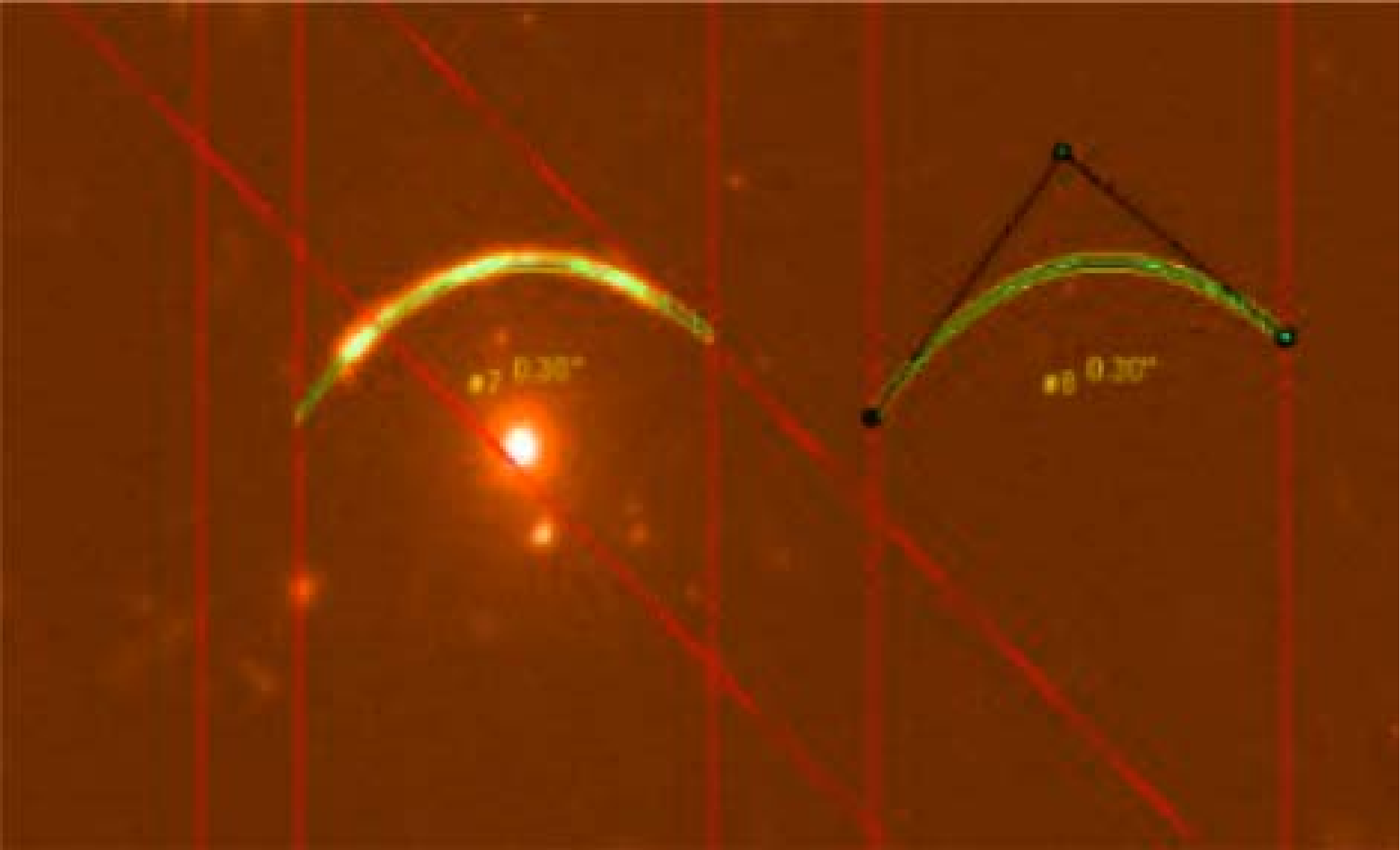}
    \includegraphics[width=6cm]{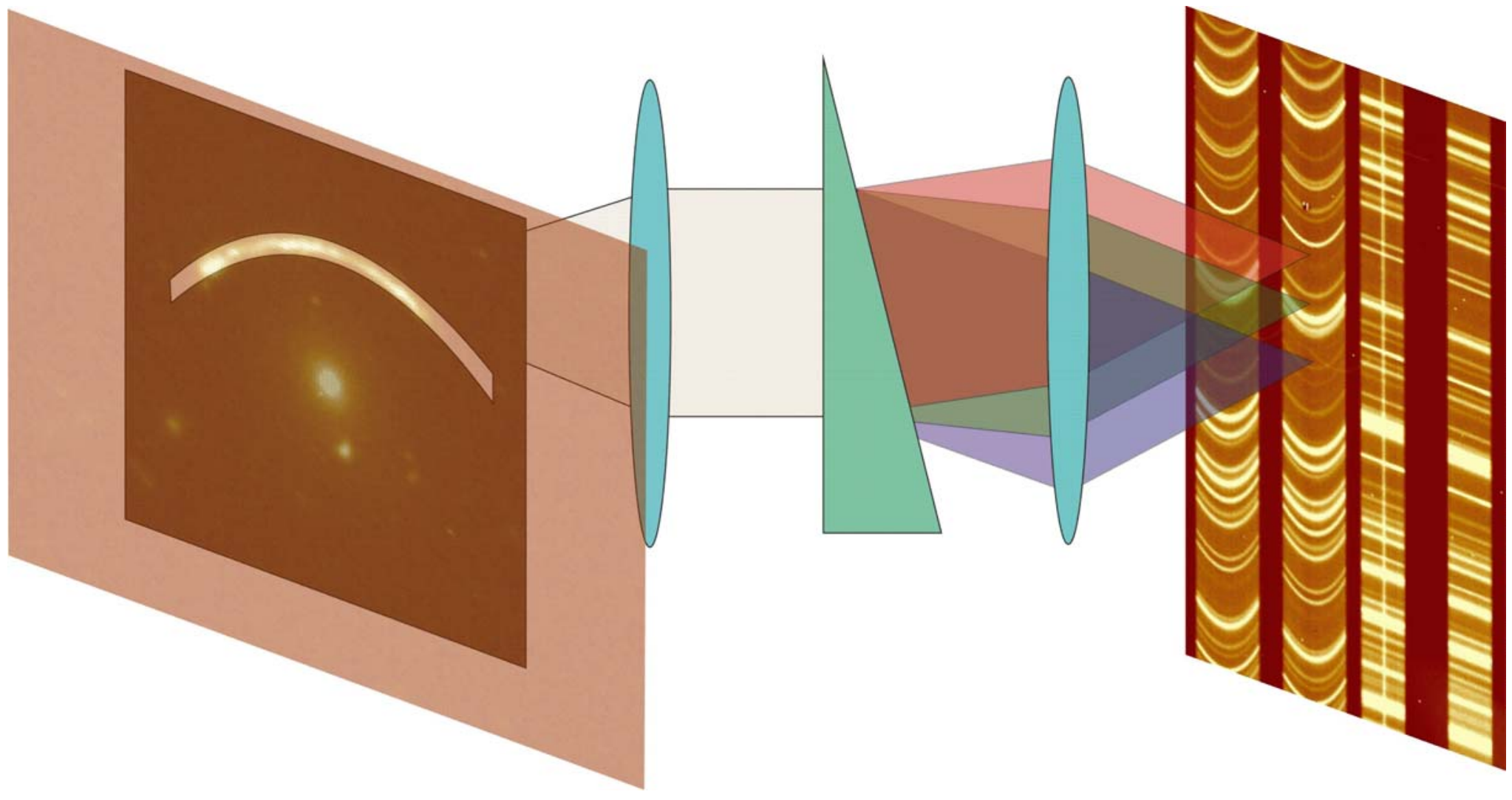}
      \caption{Basic principle of curved slit spectroscopic observations with ARGOS plus LUCI. Top:  Ks-band image of a gravitationally lensed object (the 8 o’clock arc) with the mask design overlaid. From the design, a custom, laser cut mask is made and inserted and stored cryogenically in the LUCI instruments. During the observation only the light from the desired object enters the spectrometer. The ARGOS system allows tiny slits to be used with all light from the object passing through. As a result the spectral resolution is high and the skylines are well resolved. This allows for much better object flux extraction.}
         \label{Fig_spectrograph}
   \end{figure}

One of the really unique capabilities that ARGOS offers is the combination of the GLAO with NIR spectroscopy. To our knowledge LUCI-ARGOS is currently the only facility in the world that can deliver spectra of multiple objects with 0.2 to 0.3$\arcsec$ spatial resolution and spectral resolution R \textasciitilde 10000 at the same time. Additionally the 4x4 arcmin masks for LUCI can be cut to custom shapes matching the object under study, or can contain up to \textasciitilde 70 slits (depending on length) to be placed on individual objects for a high multiplexing advantage. In Figure \ref{Fig_spectrograph} a sketch of such a custom slit observation is shown. Using HST or an ARGOS pre-imaging campaign, we design a mask that matches a lensed arc, and a second identical slit to allow nodding between the two. Additionally, a slit of the same width is always placed on one or more reference stars in the field and its nod position to control alignment during the observation and to have a spatial and spectral reference upon data reduction. Adding alignment boxes on objects over the field finalizes the design step. This design file is sent to the observatory for the laser cutting and cryogenic insertion of the masks for the upcoming semester. Upon observation the mask is grabbed by the robot inside LUCI and inserted in the instrument's focal plane. The ARGOS observation then approximately follows the scheme for seeing-limited MOS observations with some specific additions:

\begin{itemize}
  \item in the observation preparation, and already in the design of the mask, the location of the tilt star needs to be taken into account. All planned dither points must lie within the capture range of the FLAO board;
  \item during pointing of the telescope the set-up follows a two-stage process:  the telescope first needs to set its active optics on a suitable bright star, not too far from the object, and then the telescope and FLAO board position are set to catch the tilt star on the pyramid. This process is handled by the telescope control and does not need the attention of the observer. In parallel the LGSs are launched and the LGS acquisition process is executed, as laid out in Section \ref{sec:operating}.
  \item Once the AO loop is closed, the telescope pointing can be aligned to the MOS mask in the LUCI focal plane. With the ARGOS slits usually being  of the order of 0.3$\arcsec$ wide, this process needs special attention since small misalignments quickly result in light lost at the slits. Therefore, we have developed our own routine that measures and aligns the required offset and rotation on the through slit image and reference objects, preferably some background galaxies, chosen during the mask design.

\end{itemize}

\section{GLAO performance on sky}\label{sec_performance}

\begin{figure}
\includegraphics[width=9cm]{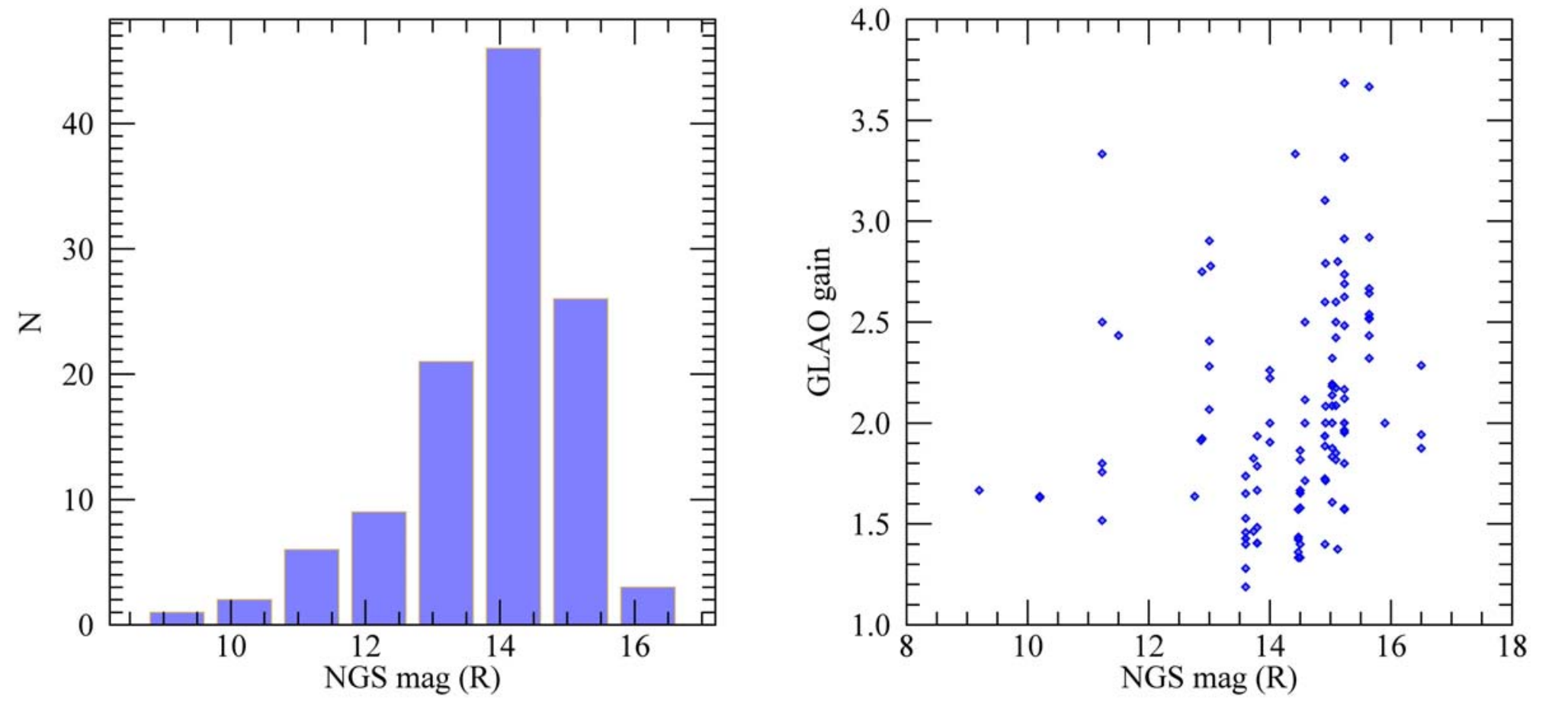}
\caption{Left: Histogram of the mag(R) of the NGS used in 100 science fields during 2016 and 2017 commissioning runs. The values represent the star magnitude taken from the NOMAD Catalog, as used to produce observing script. Right: Distribution of the GLAO performance as a function of the NGS magnitude. The GLAO gain has been evaluated by the ratio between the FWHM measured in open-loop and closed-loop LUCI/ARGOS images. No clear trend is visible in the plot, suggesting a limited impact of the NGS brightness on the GLAO peformance.}
\label{fig:starMagGain}
\end{figure}

From 2015 to 2017, ARGOS has spent approximately 100 nights and many days of commissioning. With multiple systems only being available on site, we had to bring the subsystems together and make them work at the telescope. While usually an extended period in the laboratory as a complete system would be desirable, the pure size of such a test facility and the complexity of the ASM usage did not allow for laboratory system testing. Having seen continuous progress, now the full system in binocular mode is finally available for the community. Commissioning and AO results can be found in several conference proceedings \citep{Busoni2015, Xivry2015, Xivry2016, Rabien2017}. Now, we do see a system that is capable of conducting science operation feeding the two sides of LBT simultaneously with LGS corrected light. The capability of doing full binocular operation with ARGOS and two LUCIs is a real boost for the science outcome. Having two 8.4\,m telescopes available at the same time with a GLAO corrected PSF of 0.2$\arcsec$ to 0.3$\arcsec$ in size will help  some science cases to proceed. Since the PSF size compares well with HST data, ARGOS can complement imaging taken with HST with K-band observations at similar resolution.
In the following sections, we discuss the performance of ARGOS based on several metrics. First, we look at the adaptive optics performance, {i.e.} the wavefront error RMS, and show that it matches our theoretical expectations. Then, in Section~\ref{sec_imaging_performance} we analyse the uniformity of the correction over the full 4$\times$4\,arcmin field of view and conclude that we have marginal observational effects, irrespective of the tip-tilt guide star location in the field. In Section~\ref{sec_time_performance} we look at the performance over an observing period showing that GLAO with ARGOS is also a full width at half maximum (FWHM) stabilizer over time, producing a more consistent FWHM than seeing-limited observations. In Section~\ref{sec_glao_performance}, we summarize the GLAO performance, i.e. the improvement in FWHM with ARGOS with respect to seeing-limited mode, over our commissioning period. Finally, in Section~\ref{sec_psf_performance}, we discuss the GLAO corrected PSF which is well matched by a Moffat analytical profile as expected from theory.
One limitation of our performance analyses is the lack of a dedicated turbulence profiler which would provide the strength of the turbulence as a function of altitude while operating with ARGOS. This is further discussed at the end of Section~\ref{sec_glao_performance}.

\subsection{Adaptive optics performance}

A critical parameter of an AO system based on LGS tomography is the availability of a suitable NGS within a few arcmins of the science field. A statistical analysis of the more than 200 science fields observed during the last 2 years of commissioning shows that $40\%$ of the time the system worked with an NGS fainter than 14${th}$ magnitude, as shown in Figure \ref{fig:starMagGain} and that a few targets were observed with NGS of about 17${th}$ magnitude.

As detailed in section \ref{sec:correcting}, the ARGOS LGS wavefront sensor runs at a fixed framerate of 1\,kHz, while the NGS wavefront sensor framerate can vary between 100\,Hz and 1\,kHz depending on the star magnitude. At the faint end, the typical tip-tilt residual measured by the NGS WFS amounts to 150\,nm rms. However, the pyramid WFS sensitivity strongly depends on the size of the NGS PSF so the tip-tilt residual measured on sky is underestimated. The ratio of the optimal gain applied to the tip-tilt modes on sky (e.g. 2.5) to the value  used in daytime operation when a diffraction-limited light source and 1\,kHz frame rate are used (e.g. 0.75) gives a conversion factor to properly scale the on-sky tip-tilt residual. Considering a faint NGS the residual jitter amounts to $\sim 50$\,mas, so about 1/4 of the closed-loop PSF FWHM obtainable in GLAO assisted images. In conclusion, the NGS brightness has a very limited impact on the GLAO performance (see Figure \ref{fig:starMagGain}, right panel).

\begin{figure}
\includegraphics[width = 9cm, bb = 0 0 400 260]{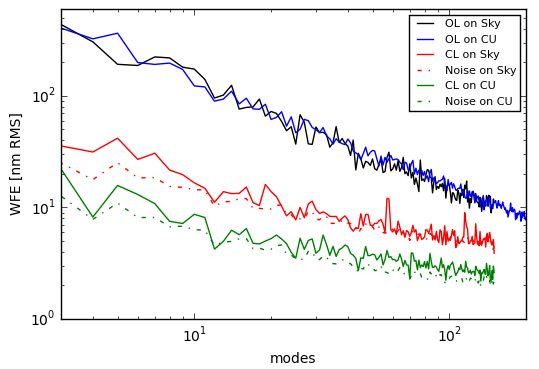}
\caption{Modal decomposition of the wavefront measured by the LGS WFS. The tip-tilt modes have been removed because they are sensed by the NGS WFS. In open loop (OL) the integral of the measured WF amounts to 860\,nm rms in daytime, using the prime focus calibration unit (CU) (blue line), and to 830\,nm rms on sky (black line). When the AO loop is closed the WFE are lowered to 55\,nm rms (green) and 120\,nm rms (red) respectively. The  dashed lines show the modal decomposition of the covariance matrix for the SH WFS measurement error evaluated considering the daytime (green) and on-sky (red) parameters of seeing, flux, and spot dimension.}
\label{fig:confrontoSkyCalSourceNoise}
\end{figure}

Evaluating the performance on sky of the LGS wavefront sensors is difficult because the quality of the tomographic measurement is strongly affected by the vertical distribution of the atmospheric turbulence and because the LGS WFS are only sensitive to the lower layers of the atmosphere. A possible approach is to compare the performance of the LGS WFS on sky with that obtained in daytime, using the prime focus calibration units and emulating the atmospheric turbulence through the adaptive secondaries.

 In daytime the system operates under well-known conditions: the injected disturbance is equivalent to a 0.8$\arcsec$ seeing ($r_0$ = 0.125\,m) with a Kolmogorov spectrum represented by 672 Karhunen-Loeve modes, the daytime LGS flux can be tuned to reproduce the same S/N as obtained on sky ($\sim30$, equivalent to a flux of 650\,$e^-$/subap/ms) and the LGS spots from the calibration source have a FWHM of \textasciitilde~0.9$\arcsec$ with no elongation. The turbulence is only applied to a layer conjugated to the ground where the three LGS WFS  measure up to 150 modes. Neglecting the tip-tilt terms, the typical WFE measured on the higher order modes by the LGS WFS amounts to 860\,nm rms, as shown by the blue line in Figure \ref{fig:confrontoSkyCalSourceNoise}. Closing the AO loop the residual WFE lowers to 55\,nm rms, as given by the green line in the figure. This value can be compared with the covariance matrix $C_N$ obtained by multiplying the noise propagation coefficients by the Shack-Hartman measurement error  reported
in \cite{Cubalchini79},
\begin{equation}\label{eq:noise}
C_N= (\mathcal{I_M^{\mathrm{T}}} \mathcal{I_M})^{-1} \sigma_m^2 \; ,
\end{equation}
where $\mathcal{I_M}$ is the LGS WFS interaction matrix and the measurement error $\sigma_m^2$ can be retrieved from \cite{Hardy1998},
\begin{equation}
\sigma_m = \frac{\pi^2}{4S\!N\!R}
\left[ \left(\frac{3d}{2r_0}\right)^2 + \left(\frac{\theta d}{\lambda}\right)^2
\right]^{1/2} \qquad [\mathrm{rad}] \; ,
\end{equation}
where $d$ = 0.55\,m is the sub-aperture dimension projected on the primary mirror, $\theta$ the LGS spot dimension and $\lambda$ = 532\,nm. Considering the daytime operating conditions described above, the modal decomposition of the LGS WFS measurement error is shown by the dashed green line in Figure \ref{fig:confrontoSkyCalSourceNoise}. Digging into data collected on sky it has been possible to find similar input conditions, where the LGS WFS were measuring an open-loop WFE of about 830\,nm rms (black line) and a residual of 120\,nm rms (red line). However, the LGS spots on sky  typically have a FWHM of $\theta \sim 2$", increasing the LGS WFS measurement error by a factor of 2.2.
Taking into account the increased spot size and the on-sky parameters, the residual WFE measured by the LGS WFS is in agreement with the modal decomposition of the covariance matrix, as can be seen from the red dashed line.

   \begin{figure}
   \centering
    \includegraphics[width=4.4cm]{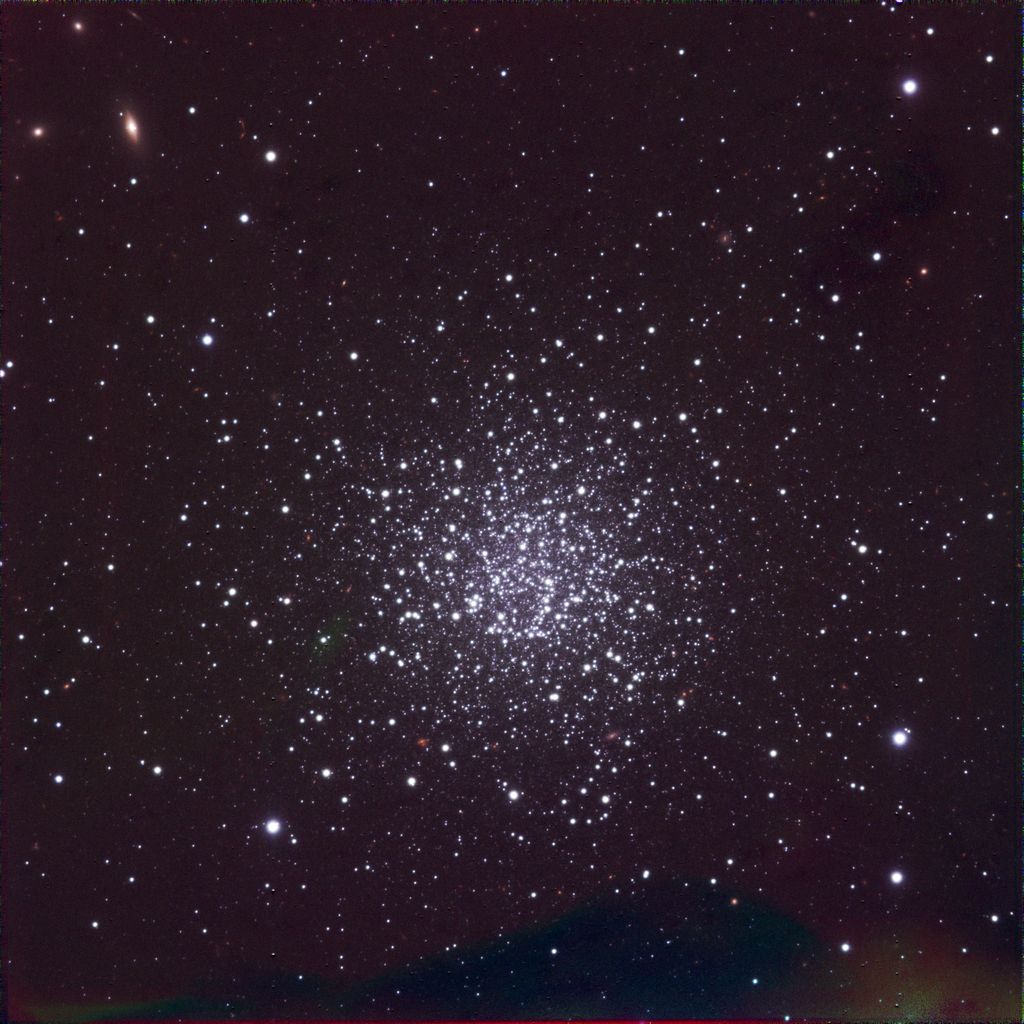}
    \includegraphics[width=4.4cm]{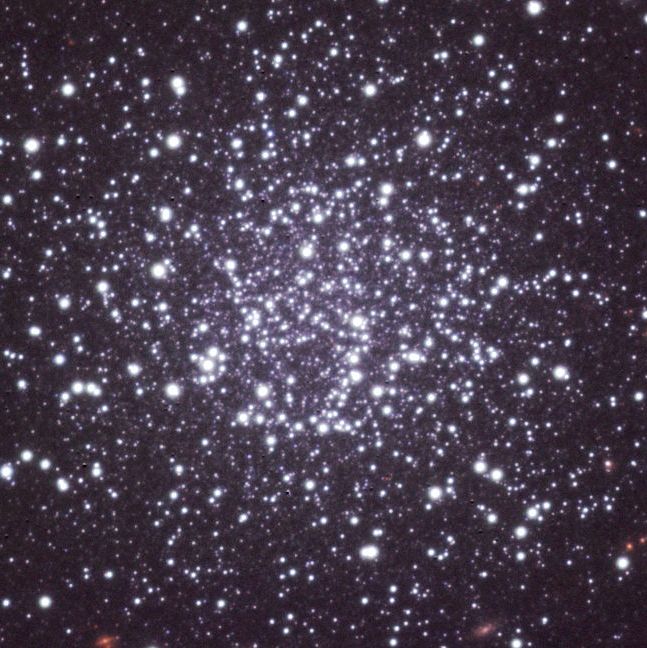}
      \caption{Colour composite image of NGC\,2419 from J, H, and Ks images. On the left side the full 4$\times$4 arcmin field of LUCI is shown, fitting well the whole cluster. To the right a zoom into the cluster core, showing the richness and density. Taken during the commissioning of ARGOS, we can use this example to measure the adaptive optics performance over the field. NGC\,2419 with its large number of stars provides a good map for the uniformity of the FWHM. }
         \label{Fig_2419_img}
   \end{figure}

\subsection{Imaging performance over the field}\label{sec_imaging_performance}

   \begin{figure}
   \centering
   \includegraphics[width=7.25cm]{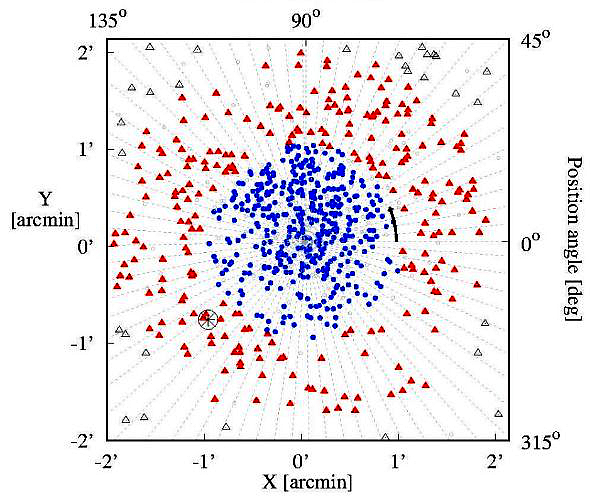}
    \includegraphics[width=6.9cm]{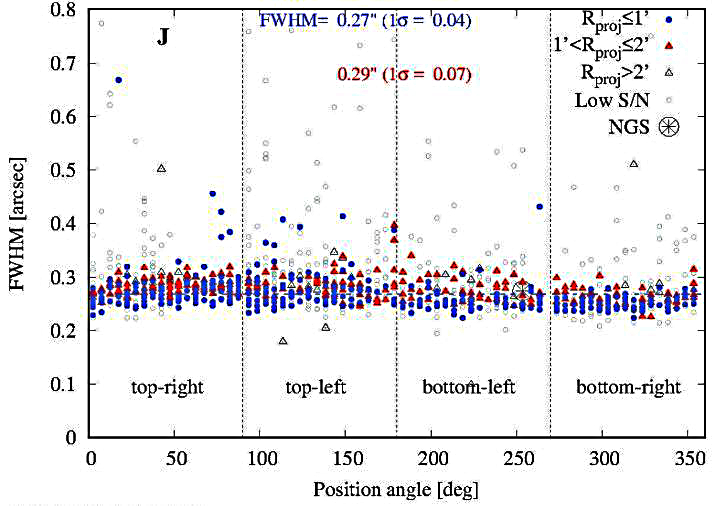}
    \includegraphics[width=6.9cm]{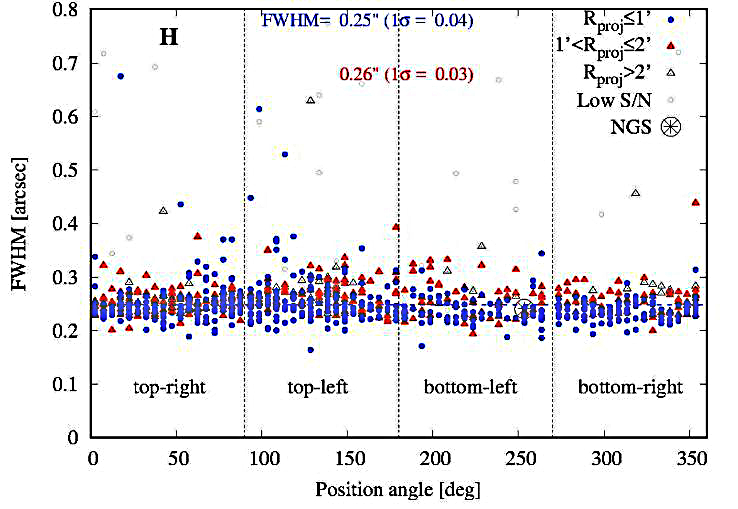}
    \includegraphics[width=6.9cm]{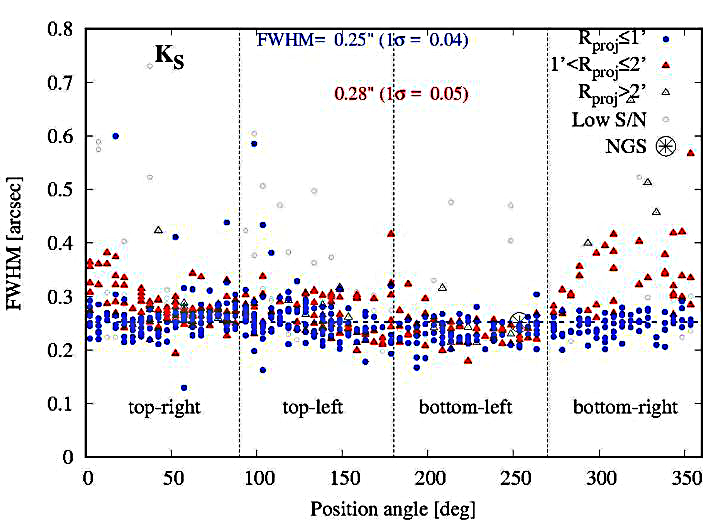}
      \caption{ Analysing the FWHM over the field of the NGC\,2419 data shows the ARGOS performance over the full LUCI2 field of view, measured in 1$\arcsec$ seeing. In the upper panel are indicated the positions of all the stars   used to measure the azimuthal FWHM variation in the field. Colours separate stars at less then 1 arcmin (blue) and 2 arcmin (red) radial distance from the field centre. The grey encircled asterisk marks the location of the NGS at ~225$\deg$, that has been used to track the tilt and low-order modes. A small (<15$\%$) radial increase from 0.25$\arcsec$ to 0.28$\arcsec$ of the FWHM from the centre to the detector edge is possibly seen in this data set. There is one panel shown for each of the J, H, and Ks bands, with the FWHM ordered in quadrants of the detector.}
         \label{Fig_2419_field}
   \end{figure}

Figure \ref{Fig_2419_img} shows the image of NGC\,2419, an old cluster of stars in the halo of the Milky Way. The cluster has been visited during the ARGOS commissioning several times. Apart from the scientific insight into the galactic potential and the cluster colour magnitude diagram (CMD)  outlined in section \ref{sec_N2419}, the cluster also provides us  with nicely distributed stars as PSF probes for the adaptive optics performance.
Judging the correction quality in single-conjugate adaptive optics is usually done by measuring the Strehl ratio, comparing the PSF to the diffraction limit. In GLAO the correction rarely reaches the diffraction limit, with the main focus being on the uniformity and wide-field performance. In analysing the resulting images we selected the FWHM as a reasonable metric for the PSF.

   \begin{figure}
   \centering
   \includegraphics[width=9.3cm]{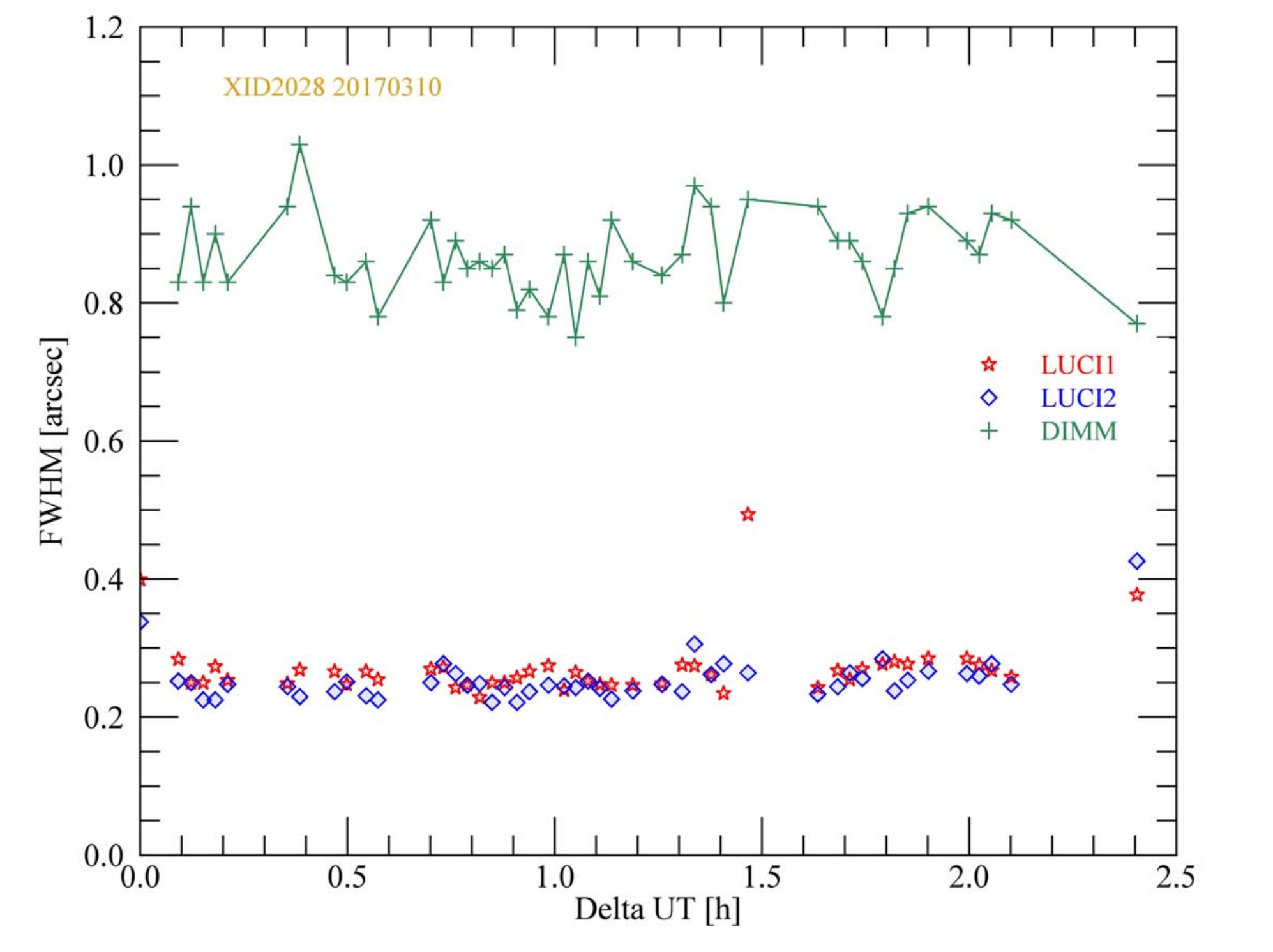}
      \caption{ Temporal evolution of an imaging observation of the XID2028 source. The natural seeing in the visible from the DIMM measurements is shown as a green line and varies between 0.8$\arcsec$ and 1.0$\arcsec$ over this \textasciitilde 2\,h observation. The individual points are the measured FWHM from fitting a Gaussian to a resulting 1min integration of six stacked frames. The red stars indicate the FWHM measured in the LUCI1 frames, the blue diamonds that in LUCI2. With this observation of an extragalactic source, no sky frames are taken, and no open-loop data is  available. The one red star at 0.5$\arcsec$ is a frame marked with `paused',   indicating a laser propagation stop upon aircraft passage while still being closed on the tip-tilt star. The median performance on the LUCI frames over this full 2h operation amounts to 0.26$\arcsec$ with LUCI1, and 0.25$\arcsec$ with LUCI2.}
         \label{Fig_XID_perform}
   \end{figure}

 With the large LUCI field available we can probe the PSF properties over the $4 \times 4$ arcmin fields. Figure \ref{Fig_2419_field} shows the performance over the field as obtained in an observation of NGC\,2419 over the  J, H, and Ks bands. In this  observation under seeing conditions of \textasciitilde~1$\arcsec$, as measured on the differential image motion monitor (DIMM), the measured PSF size in J, H, and Ks amounts to 0.27$\arcsec$$\pm0.04$, 0.25$\arcsec$$\pm0.04$, and 0.25$\arcsec$$\pm0.04$, respectively, in the inner 1 arcmin radius. Towards the outer areas an increase of \textasciitilde 0.02$\arcsec$ in all bands can be seen.
 
 Another inherent property of the GLAO system is the lack of dependence of the PSF shape on the tilt star location. Within the ARGOS and FLAO set-up we are able to chose the location of the tilt star within a $2 \times 3$ arcmin field, asymmetric in the LUCI fields, as given by the travel range of the FLAO stage assembly. Within many observations we have  not seen a  detectable dependence on the tilt star location in the field. For  the many science cases that suffer from sparsely available suitable NGSs, this will be of major importance.

 \subsection{Imaging performance over time}\label{sec_time_performance}

 Many science cases require a stable and well-defined PSF over the observing period. With the natural seeing being often highly variable, the combination of many images over the observation period results either in the combined image being affected badly by the increased seeing moments, or a selection of frames may be required. Getting rid of the ground-layer turbulence contribution does help in that respect. We have seen long exposure periods where the seeing measured on the DIMM had high variations, while in the corrected images a fairly constant PSF size has been achieved. Of course, if the variation takes place in the high layer, the GLAO PSF will follow the seeing variation. In Figure \ref{Fig_XID_perform} the PSF FWHM of an imaging observation under fairly constant conditions is shown. The object of interest in this observation is XID2028 located at z=1.5930, originally discovered in the XMM-COSMOS survey, a star forming QSO, being thought to be in the `feedback phase'. The LUCI observations here complement ALMA data on the same object, resolving extended dust emission in the Ks band coincident with a point source at 0.1'' resolution detected by HST. We  observed this object  for \textasciitilde 2\,h in binocular mode with both LUCIs in Ks band. With the individual integration times being 10\,s long, each data point is already a stack of six images taken in integrated mode. During this 2\,h period the median FWHM on the two LUCIs is 0.26$\arcsec$ and 0.25$\arcsec$  with a standard deviation of 0.014$\arcsec$ and 0.02$\arcsec$, respectively. The data taken in this observation has been published in \citet{Brusa2018}.

\subsection{GLAO performance}\label{sec_glao_performance}

  \begin{figure}
   \centering
    \includegraphics[width=9.5cm]{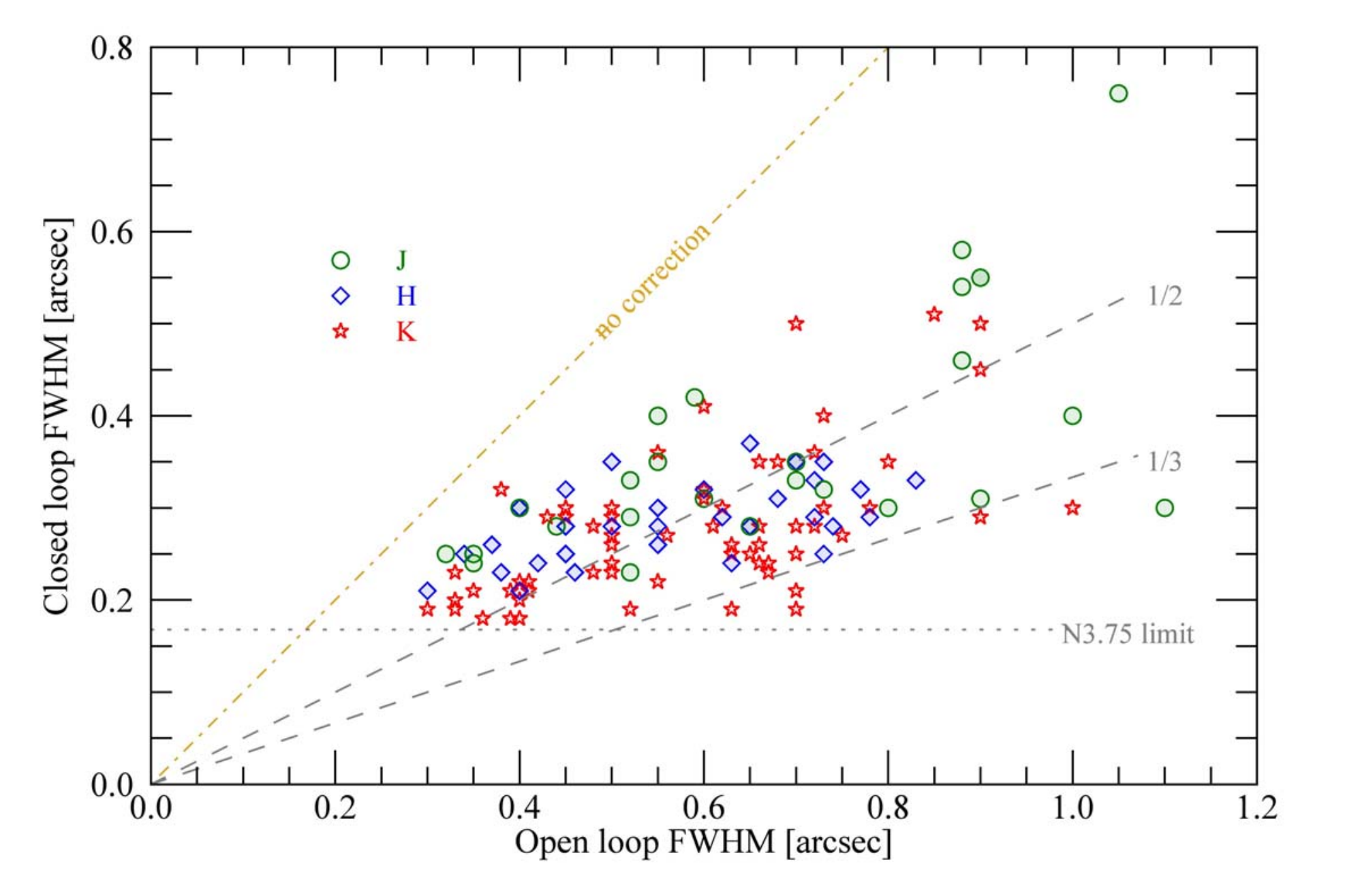}
      \caption{ARGOS commissioning performance plot, summarizing many imaging observations done during commissioning. Taking the open-loop FWHM points as seen in sky frames of an observation and overplotting the achieved GLAO FWHM, we can judge the `improvement' in FWHM. The plot shows 123 data points taken from 43 observations over the J, H, and K bands (green, blue, and red points, respectively). Lines mark the `factor 2' and `factor 3' improvement, and the yellow diagonal of `no correction'. The `N3.75 camera limit' marks the smallest possible size measurement with the specific camera in use, possessing a 0.118\arcsec\ pixel scale.}
         \label{Fig_arg_perform}
   \end{figure}

From early 2015, when the first successful loop closures could be carried out, until late 2017, the technical commissioning was always  accompanied by an attempt to get data on sky in parallel. This strategy  proved to be quite helpful as issues in the whole chain of units--from telescope  to FLAO, ASMs, ARGOS, and finally LUCI as the receiver of the infrared photons--could be found and improved. In the case of nearby galaxies or galactic objects,  sky frames are usually required for background subtraction. Moving to sky with ARGOS is usually done   by opening the high-order and the tilt loops since the sky pointing may be out of the acceptance cone for satellite avoidance (see Section \ref{sec:offsets}). For performance measurements this offers an  opportunity to measure the open-loop PSF size and compare it to the closed-loop performance.  Having seen a  variety of weather and seeing conditions over the commissioning nights, we can now draw  a plot for the open- versus closed-loop PSF size giving the GLAO improvement metric. Figure \ref{Fig_arg_perform} shows the collection of data points out of 43 observations delivering 123 data points in the  J,H, and K bands for this analysis. Observation attempts in very poor conditions, high seeing, or winds with the AO loop crashing have been excluded from this plot since under routine observing programmes one would  expect to switch to a different programme.

 In summary the GLAO system can deliver an improvement of a factor of 1.5--3. The K-band average improvement from these data points is 2.14, with a 0.26$\arcsec$ median PSF size. Nevertheless, this value gives an average over a technically improving system under commissioning conditions. A future routine usage of ARGOS will yield a picture that is statistically more solid. Since LBT had no turbulence profiler in operation over the commissioning period, we could not study the influence of uncorrected high-altitude turbulence layers within this data set. For this reason we  started a SLODAR implementation \citep{Mazzoni2016}, and built a MASS unit \citep{Kohlmann2018} that may be installed in the future as facility devices, helping to decide whether a GLAO observation will yield high gain.

\subsection{Shape of the GLAO corrected PSF}\label{sec_psf_performance}

  \begin{figure}
   \centering
    \includegraphics[width=9cm]{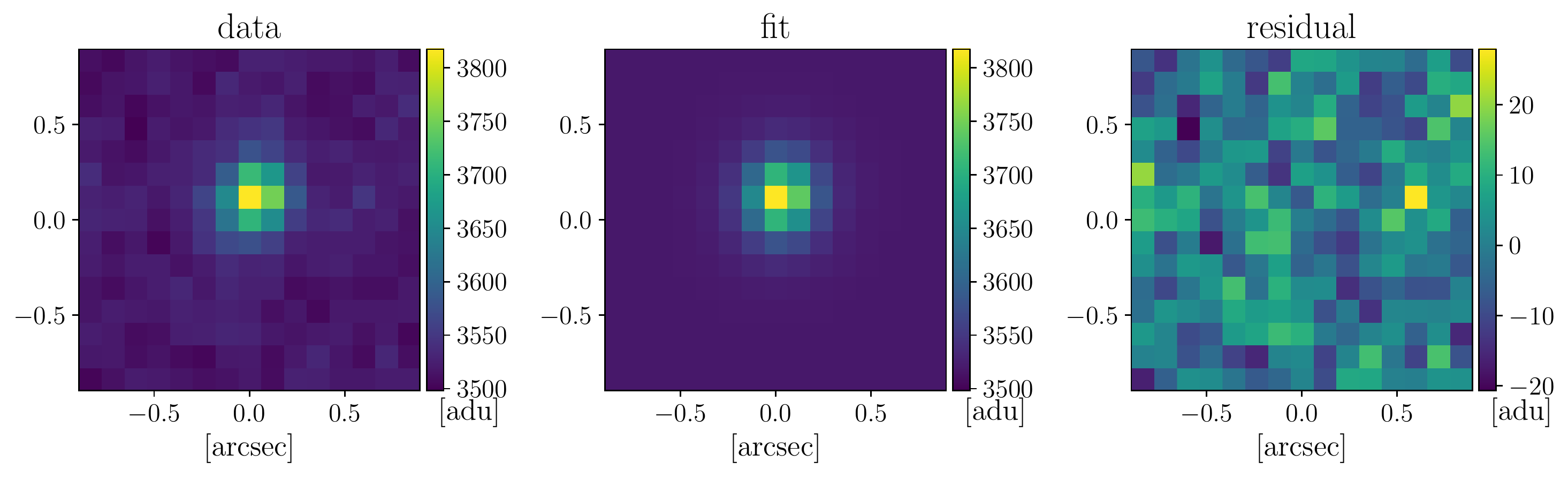}
      \caption{Illustration of the GLAO corrected PSF based on a combined K-band image of a total of 3.8\,min exposure time.
      Left: Point source in the field. Middle and right: Fit by a Moffat profile and its residual after subtracting the model from the data, respectively.
      }
         \label{Fig_arg_moffat_fit}
   \end{figure}

It is well known that the PSF produced by a GLAO system does not exhibit diffraction-limited features,  but is qualitatively very similar to a seeing-limited PSF \citep[see e.g.][]{Andersen2006}. This means the GLAO PSF is well fitted by a Moffat function
\begin{equation}
I(r) =I_0 [1 + (r/R)^2]^{-\beta}
\end{equation}
with $R$ the core width related to the FWHM by $FWHM = 2 R\sqrt{2^{1/\beta} - 1}$, and $\beta$ the power index of the function, with smaller index leading to larger wings.
The seeing-limited PSF under Kolmogorov turbulence is well modelled by $\beta \approx 4.77$, but real PSFs  typically have larger wings or equivalently smaller $\beta$ due to imperfection in the optics \citep{Trujillo2001}. The GLAO PSFs are similar and have a power index between 2.5 and 4.5 \citep{Andersen2006}.

As an illustration of the delivered ARGOS PSF, we analysed J-, H-, and Ks-band images from a typical good and stable night.
The data fit and residual of a single PSF is shown in Figure \ref{Fig_arg_moffat_fit}. In total we fit several hundreds of elliptical Moffat functions to single and combined frames and obtained the following results:

\begin{enumerate}
  \item For the J-band, we have $\beta = 3.4 \pm 1.3$ and $FWHM = 0.34$\arcsec$ \pm 0.04$;
  \item For the H-band, we have $\beta = 3.2 \pm 0.9$ and $FWHM = 0.28$\arcsec$ \pm 0.03$;
  \item For the Ks-band, we have $\beta = 2.5 \pm 0.6$ and $FWHM = 0.21$\arcsec$ \pm 0.02$.
\end{enumerate}

We see that $\beta$ and the FWHM becomes smaller at longer wavelengths which reflects the fact that the PSF is better corrected with a more pronounced core (corrected) and wings (uncorrected). These values also fall in the predicted range obtained by GLAO simulations. Flux residual amounts to less than 1\% and are typically around 0.2\%, which again match the expectation in \citet{Andersen2006}.

\section{Selected science observations with ARGOS}\label{sec_observations}

Many science cases can profit from a GLAO reduced PSF size. From the many objects visited during the ARGOS commissioning which were used as part of the performance testing and science verification, we showcase the use of ARGOS with few selected science cases in this section.  Ordered from near to far, we show the following observations:
\begin{itemize}
\item NGC2419, a distant Galactic globular cluster. The enhanced resolution allows us to overcome crowding effects and measure precise CMD and identify variable stars;
\item NGC\,6384, where the enhanced resolution of ARGOS extends nuclear and star cluster science to a distance of  20\,Mpc;
\item PLCK\,G165.7, a massive lensing cluster, where LUCI-ARGOS K-band data complements HST imaging, contributing to lensed image family identification;
\item Spectroscopic measurements of high-$z$ gravitationally lensed galaxies. Utilizing the custom slit capabilities, we match the shape to the curved arcs and with ARGOS delivering a small PSF size we can probe the objects at high spectral and spatial resolution.
\end{itemize}

\begin{figure}
\includegraphics[width=1.\linewidth]{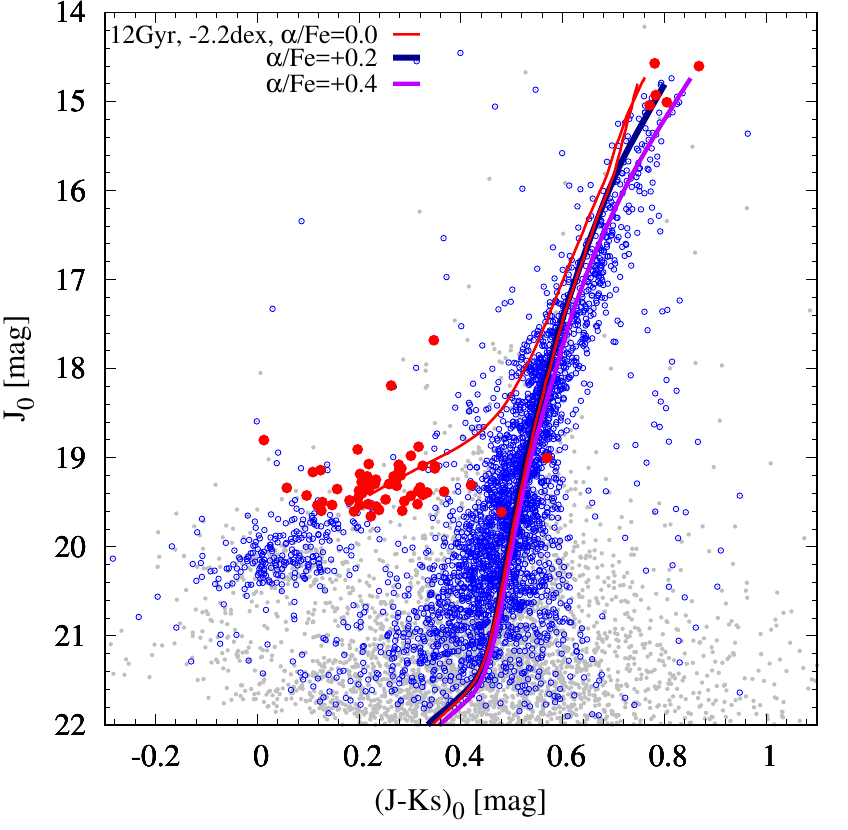}
\includegraphics[width=1.01\linewidth]{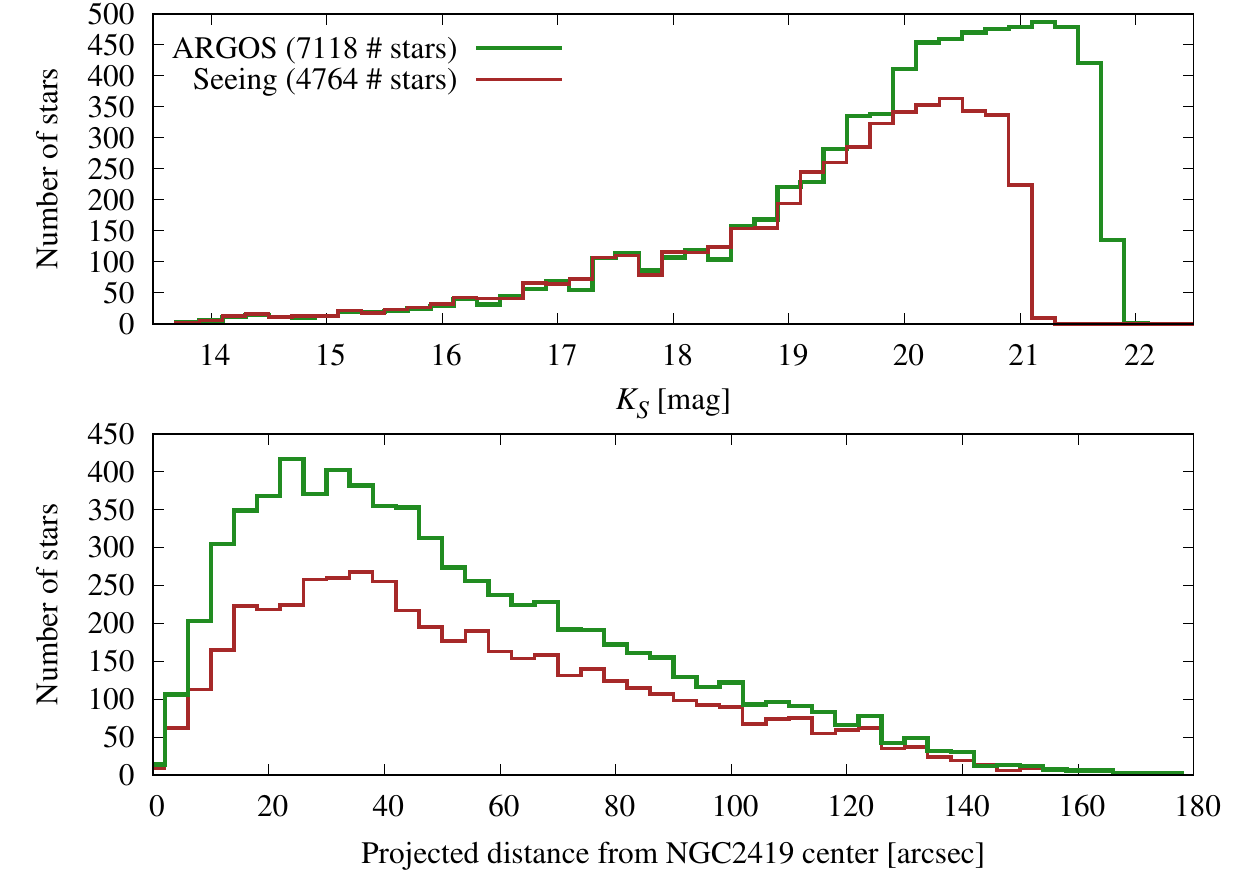}
\caption{Upper panel: NGC\,2419 colour magnitude diagram from data taken during commissioning nights 2016 Oct 19, 23, 24, and 26. The magnitudes are corrected for foreground extinction as indicated by the magnitude subscript. Shown are sources with $S/N_{J,K_S}\!>\!10$ (blue circles), all detections (small grey dots), and the known RR\,Lyrae (and other) variable stars in NGC\,2419 (large solid red dots). For reference, we show with a solid red curve an isochrone adopting the literature values for its age and metallicity and $[\alpha/$Fe$]\!=\!0.0$\,dex. Magenta and blue curves in the background show a comparison to enhanced $[\alpha/$Fe$]\!=\!0.2$ and 0.4\,dex isochrones (see  Sect. \ref{sec_N2419}). Middle and lower panels: Number counts of stars as a function of magnitude  and as a function of distance to the cluster centre. The green histogram shows the ARGOS number counts, while the red line shows the detected stars in non-GLAO data. (Details in Section \ref{sec_N2419}) }\label{Fig_N2419_CMD}

\end{figure}

\begin{figure*}
\centering
\includegraphics[width=0.25\linewidth,bb=5 -150 950 252] {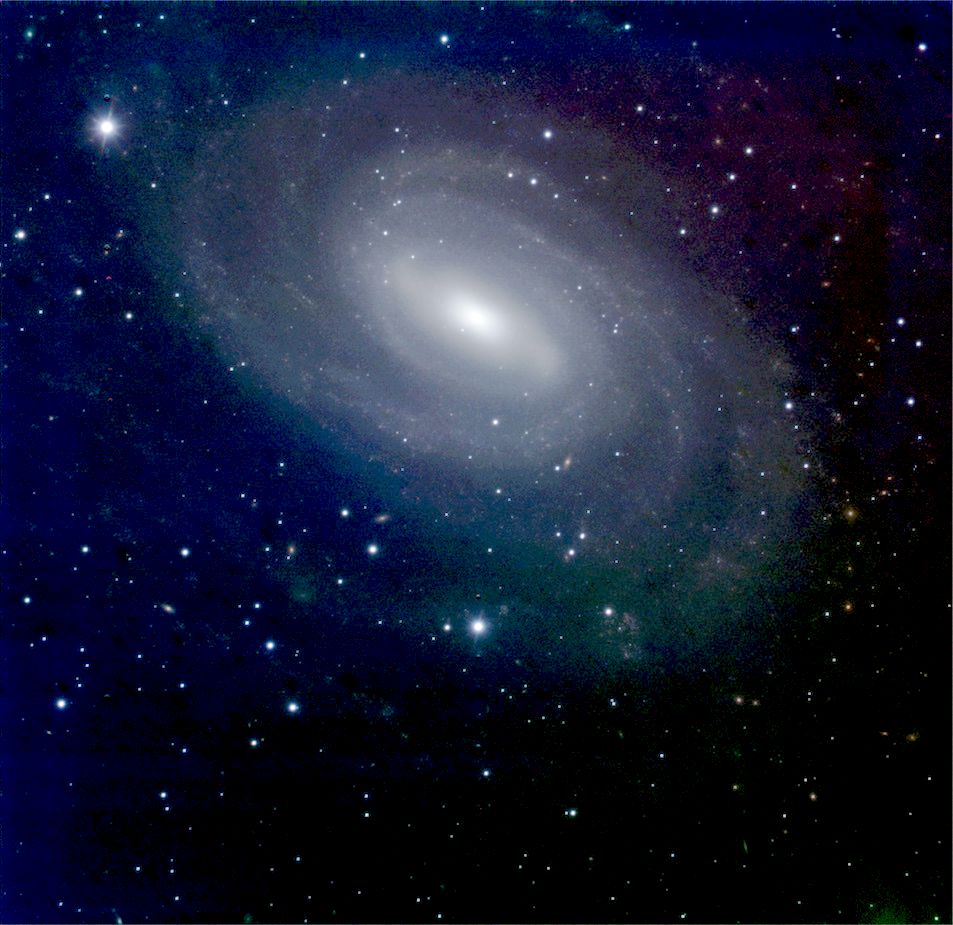}
\includegraphics[width=0.37\linewidth,height=0.31\linewidth,bb=5 -5 258 252] {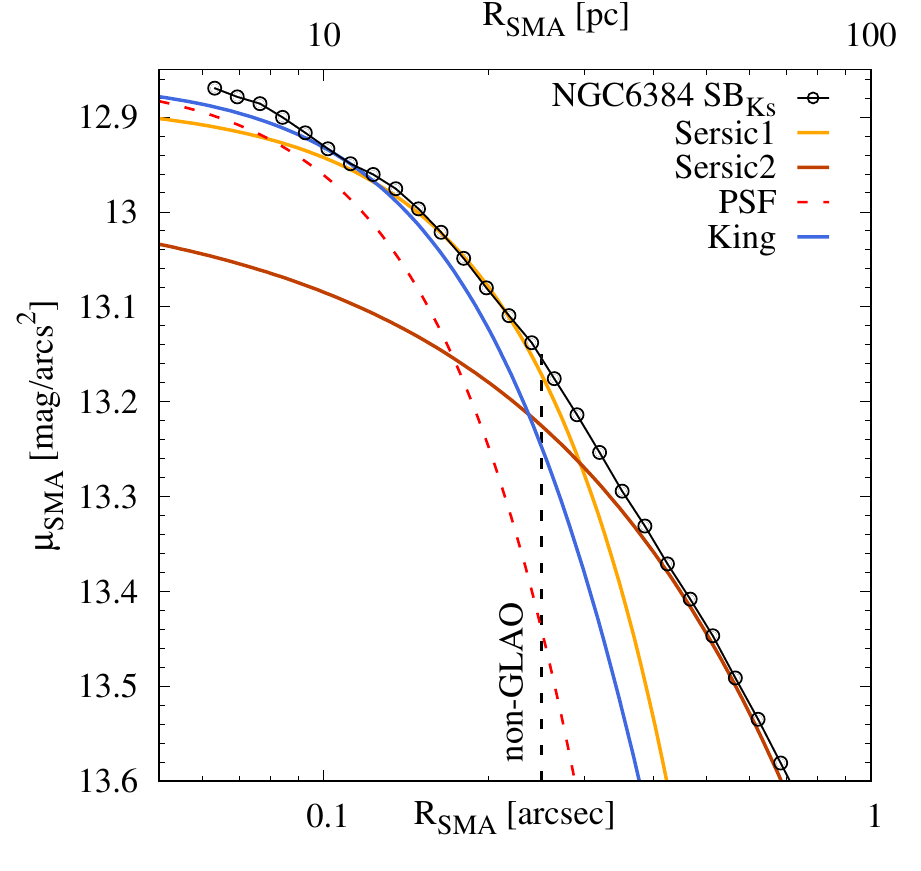}
\includegraphics[width=0.365\linewidth,bb=5 10 275 250] {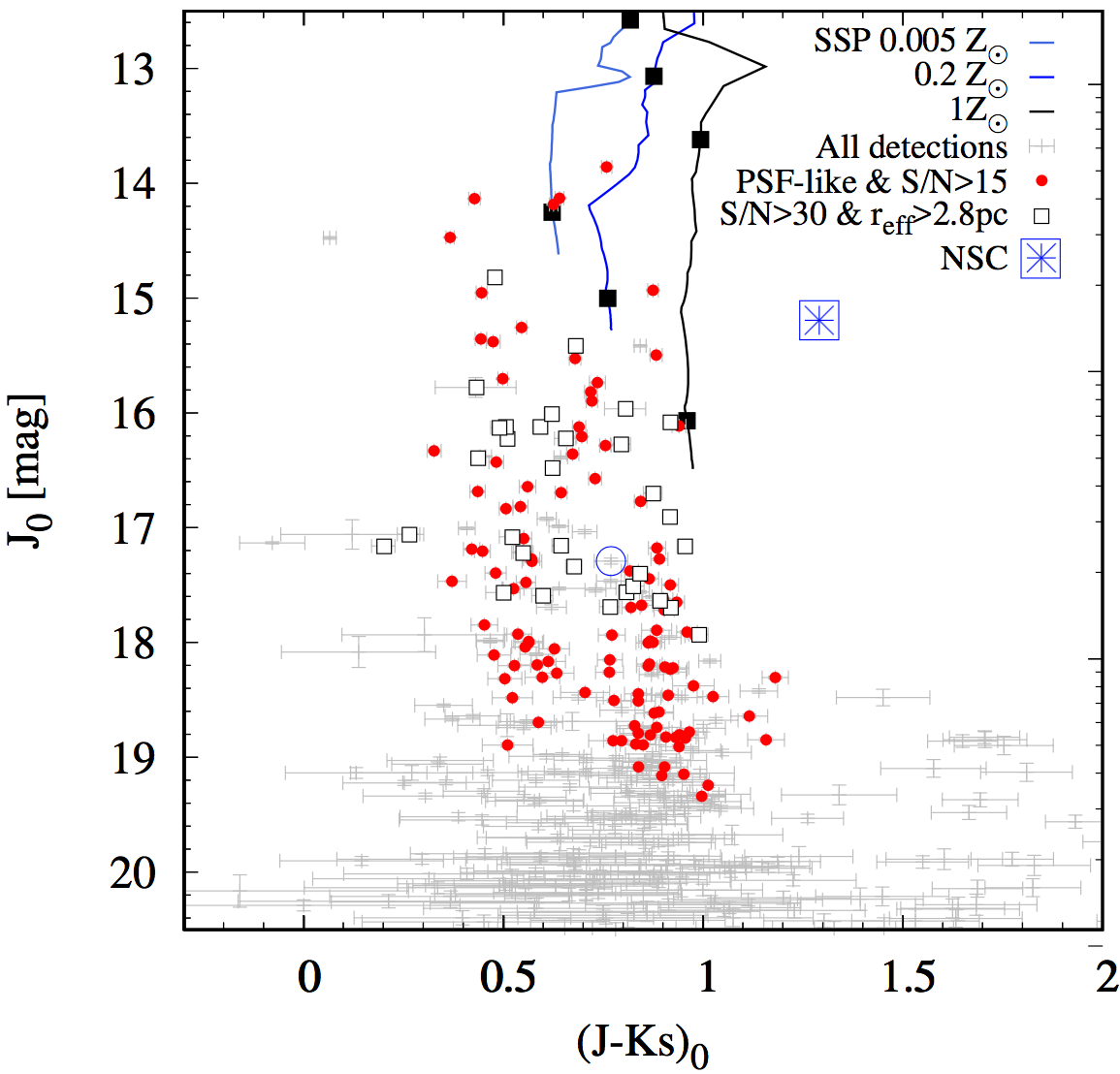}

\caption{Left: Colour composite image of NGC\,6384 from $JHKs$ LUCI1\&2 ARGOS commissioning data. Middle: 1D surface brightness profile decomposition of the central 100\,pc around its nuclear cluster. Lines indicate the various components   obtained via 2D modelling (see  legend). A vertical dashed line indicates the non-GLAO PSF size. Right: Colour-magnitude diagram of PSF sources (solid red dots), extended objects (open squares), and all detections (light grey points) in the field. Figure adapted from \citet{Georgiev2018}.}
\label{Fig_N6384}

\end{figure*}

\subsection{Stellar clusters in the Galaxy: NGC\,2419}\label{sec_N2419}

Galactic globular star clusters (GCs) present one of the best targets for studying the past and present of the Milky Way (MW). To do this it is essential to know precisely their intrinsic properties and global motions around the Galactic Centre. Due to the high stellar densities and large extent of the GCs, their study is often hampered by the small fields of view of SCAO or space-based observations, or the  spatial resolution of seeing-limited observations, which is two times lower. ARGOS provides a solution to these two major problems, and coupled with the binocular capabilities at LBT makes it unparalleled in its efficiency for MW star cluster science.

 As a demonstration of the efficiency and quality of the data, in this section we use the commissioning $JHK_S$ data of NGC\,2419 (see Figure\,\ref{Fig_2419_img}), one of the most massive and remote GCs in the Galactic halo. Knowing the distance and the chemical composition of its stars provides a unique probe of the MW potential and its assembly;  several lines of argument suggest that NGC\,2419 might have been accreted from a dwarf galaxy \citep[see e.g.][]{Massari17,Lee13,Cohen10}.
 NGC\,2419 hosts a large fraction of known RR\,Lyrae stars and other variable stars \citep{Clement01}\footnote{Catalogue: \href{http://www.astro.utoronto.ca/~cclement/cat/listngc.html}{http://www.astro.utoronto.ca/$\sim$cclement/cat/listngc.html}} that are excellent distance indicators, especially in the NIR.
 While precise proper motions may be obtained from HST or perhaps with future Gaia data releases, the most precise distances in the Galactic halo can only be measured for RR\,Lyrae stars. In the Ks band, for example, the period-luminosity relation for RR\,Lyrae stars can provide  distances as precise as 1--2 $\% $.
 The chemical composition sensitivity of the $I\!-\!K_S$ colour index (and in the future combined with optical HST photometry), provides an excellent tool for the  study of the cluster internal composition. We demonstrate this in Figure\,\ref{Fig_N2419_CMD}, which  shows the CMD of NGC\,2419 obtained during four commissioning nights (2016 Oct 19, 23, 24, and 26). From the figure it is evident that with only $\sim6$\,min of effective $J$-band exposure time we reach $S/N\gtrsim10$ at $J\simeq21$\,mag. The quality of the photometric accuracy is such that it already allows us to make a direct comparison with stellar evolutionary isochrones (solid dark and magenta curves in Figure\,\ref{Fig_N2419_CMD}) based on literature estimates of the cluster age, metallicity, and alpha ($\alpha/$Fe) abundances \citep{Kirby08}. As can be seen in Figure\,\ref{Fig_N2419_CMD} the comparison with the isochrones already nicely matches the known chemical variation in the NGC\,2419 stellar population \citep[e.g.][]{Kirby08}. All known variables (solid red circles) are also easily recovered, and coupled with their temporal photometry will provide an excellent distance estimate to NGC\,2419. Due to commissioning task limitations, only four temporal points were observed for this programme; therefore, it will be completed during the upcoming science runs with ARGOS. The results on the distance from the stellar variability as well as detailed analysis of NGC\,2419 CMD content will appear in Testa~et~al. (in prep.). To compare our measurements with seeeing-limited results we have been using the seeing-limited measurement of the PSF in the sky frames, and convolved the NGC2419 data to the corresponding seeing value of the PSF for each filter. We then piped the seeing-limited data through the same routines of iterative star detection and PSF photometry. The result from this experiment is shown in Figure \ref{Fig_N2419_CMD} (bottom panels). The top panel compares the number distributions of stars as a function of $K_S$ magnitude and the bottom panel as a function of projected distance from the cluster centre. Evidently, the total number of stars is nearly twice as high with ARGOS, and the strong radial incompleteness toward the densest regions of the cluster increases by a factor of two as well. In summary, ARGOS with the LUCI cameras and the two 8.4\,m mirrors of LBT provides an uniquely efficient tool for disentangling the properties of distant and complex clusters like NGC\,2419 to gain insights into the assembly of the MW.

\subsection{Galactic nuclei and star clusters: NGC\,6384}

The typical sizes of globular and nuclear star clusters ($\sim\!4$\,pc) in distant galaxies at  20\,Mpc are $\lesssim0.4\arcsec$, which makes them extremely challenging to resolve in seeing-limited observations. This limits these kind of studies to a volume of only $r\sim10$\,Mpc.
In this section we show an example of studying extragalactic nuclei and GCs from the ground out to $\sim21$\,Mpc enabled by the high spatial resolution provided by ARGOS. Being located at that distance, NGC\,6384 showcases this science case, as the factor of two smaller PSF size allows us to spatially distinguish the most extended GCs around NGC\,6348. This galaxy was the first science target during the ARGOS commissioning in 2015, and in February 2018 we were able to complete this observation by obtaining a  $K_S$-band image as deep as those for $J$ and $H$. Our analysis of this ARGOS data with a PSF$_{K_S}\!=\!0.25\arcsec$ obtained within the common area of available HST data ($\sim\!1.5\arcmin\times1.5\arcmin$), shows that the eight background galaxies, falling within the magnitude and colour range of the 27 GC candidates, have a PSF size ${K_S}_{\rm FWHM}\!=\!0.36\arcsec\!-\!0.48\arcsec$. Since their sizes are clearly below the seeing value ${K_S}_{\rm FWHM}\sim\!0.5\arcsec\!-0.6\arcsec$, they would have been selected as GC candidates in non-GLAO data. The twice sharper PSF delivered by ARGOS thus reduces the fraction of background contamination by $\sim\!30$\%, compared to non-GLAO observations. In Figure\,\ref{Fig_N6384}\,left we show a $JHKs$ LUCI1\&2 colour composite image taken in binocular mode with  ARGOS. In Figure\,\ref{Fig_N6384} middle  we show that the 1D decomposition of the radial surface brightness profile of the central 1\arcsec\ resolves the nuclear star cluster (King model), but not  for non-GLAO data (as indicated by the vertical dashed line). For a detailed analysis of the cluster and galaxy structure, see \citet{Georgiev2018}. The different lines illustrate the solutions from fitting the sharpest $K_S$-band images in 2D with {\sc imfit} \citep{Erwin15}. The spatial decomposition of the light profile of the nuclear star cluster in the NIR is important for its accurate mass modelling. Coupled with high spatial and spectral resolution spectroscopy provided by the LUCI longslit and MOS will enable detailed evaluation of its dynamics, and assess whether it could harbour a massive black hole. Similar analysis using a spatially variable PSF model built from the many Milky Way stars in the image was performed for all sources with $S/N>30$ (red solid dots in the right panel of Figure\,\ref{Fig_N6384}) for which a size measurement can be trusted if the source is bigger than $\gtrsim10$\% of the PSF$_{\rm FWHM}$, i.e. $\gtrsim2.8$\,pc at the distance to NGC\,6384. All these `resolved' sources are GCs and young star cluster candidates (open squares in the right panel of Figure\,\ref{Fig_N6384}). From the sheer number of detected clusters, and combined with their photometric stellar population properties, one can efficiently obtain the age and metallicity distributions of the entire cluster population and derive the major star formation history (SFH) of the host galaxy \citep{Georgiev12}. This is possible thanks to the high sensitivity of the $J-K_S$ colour index to metallicity, which is often greater than the measurement uncertainties. We illustrate this by the comparison with single stellar population model tracks \citep{BC03} for three metallicities, as indicated in the legend of Figure\,\ref{Fig_N6384}. As  can be seen, the metallicity resolving power of the $J-K_S$ colour index allows us to clearly distinguish between GCs in a range of metallicities \citep[details in][]{Georgiev2018}. In summary, the high spatial resolution provided by ARGOS over the entire $4\arcmin\times4\arcmin$ and high spectral resolution of the LUCIs will enable efficient investigation of the major star formation history of the host galaxy via the properties of its GCs as well as testing scenarios for the build of galactic nuclei, their nuclear star cluster and whether they coexist with massive black holes.

\subsection{LUCI-ARGOS Data of PLCK G165.7+67.0 (G165)}

High-resolution imaging in the central regions of massive lensing clusters is crucial for identifying the sets of multiple galaxy images which arise from a single galaxy in the background of a massive lens.  Finding such `arclet families' is important as each image multiplicity places a strong constraint on the underlying distribution of the dark matter \citep{Zitrin2009}. While all members of an arclet family have the same redshift, it is the confirmation of their similar colours,  morphologies, spectroscopic redshifts, and model-predicted locations that distinguishes them from being single objects, or two different objects at a similar redshift.  To aid in the search for arclet families, LBT LUCI-ARGOS data is used together with Hubble Space Telescope ({\it HST}) data to extend the wavelength reach (GO-14223, PI:  Frye). Here we describe the data analysis of our  LBT LUCI-ARGOS $K$-band observations in the field of the massive galaxy cluster PLCK G165.7+67.0 (G165; $z$=0.35). Note a detailed description of the data reduction techniques and discussion of the lens appears in \citet{Frye2018}.

This massive lens was discovered as a result of an all-sky census for infrared bright galaxies using {\it Planck/Herschel}. The approach relies on the search for intensely star forming and dusty galaxy sources which produce strong thermal emission by warm dust ($T \approx 40$\,K). This rest-frame far-infrared emission peak is detected in the observed {\it Planck} High Frequency Instrument sub-millimeter bands for redshifts of $z$\,=\,2\,-\,4. The brightest 228 sources detected by {\it Planck} that are consistent with being infrared-bright galaxies are followed up using Herschel Space Observatory ({\it Herschel}). At the higher resolution, the vast majority of sources separate out into several submillimeter bright sources, as expected of galaxy over-dense
structures at $z$\,=\,2\,-\,3 \citep{2015A&A...582A..30P, Martinache2018}. At the same time, a small minority of sources remain compact at higher resolution as expected for strong lensing \citep{Canameras2015}. These 11 compact sources turn out to be individual infrared bright galaxies whose brightness is boosted by lensing amplification.  This discussion focuses on an investigation of one of the eleven strongly lensed galaxies which we designate as G165\_DSFG\_1.

\begin{figure}[t]
\includegraphics[width=\linewidth]{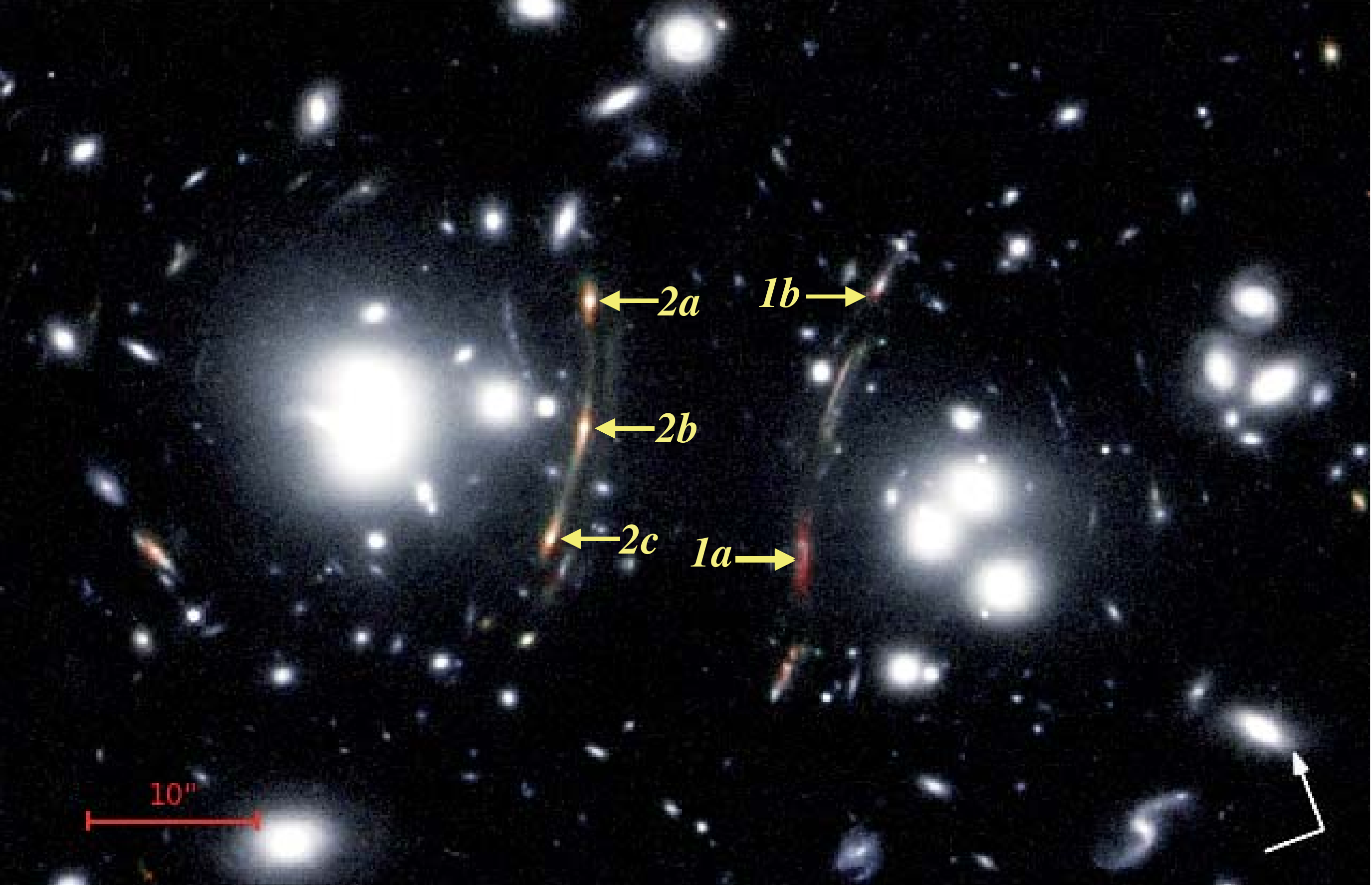}
\caption{Colour composite image for the central region  of the massive lensing cluster G165, using LBT/ARGOS $K$-band for the 2016 Dec 15 run (red), + HST/WFC3 $F160W$ (green), and $F110W$ (blue) data.   In total, 11 arclet families are discovered in this field as a result of the combined HST + $K$-band analysis, demonstrating the value of LUCI-ARGOS to provide high-resolution imaging that enables side-by-side comparisons \citep{Frye2018}. In addition to detecting the spatially resolved giant arc `1a'  with its high estimated magnification factor of \textasciitilde 30, the $K$-band data also unveil another image of this same background galaxy `1b', that falls at its model predicted location.  The secure identification of this doubly-imaged family enables us to construct a robust lens model. The brightest objects in our $K$ image belong to the triply-imaged {\it Arcs 2a, 2b} and {\it 2c}, as labeled. North is in the direction of the compass arrow and east is to the left. A 10$\arcsec$ scale bar is shown in the bottom left corner for reference. }
\label{figG165}
\end{figure}

We acquired imaging of G165 in $K$-band using LBT LUCI-ARGOS during instrument commissioning time on two separate nights: 46\,min of observation using LUCI2 on 2016 December 9 and 42\,min using LUCI1 on 2016 December 15. We custom-built much of the reduction software to ensure high flatness across the chip and to maximize the signal-to-noise of the data.  Briefly, after subtracting the dark frames from all object frames, we find the best estimate of the background.  Our approach is to construct a running boxcar in 10\,min intervals to cope with the rapidly varying background. Following the background-subtraction, we divide through by the flat-field which further improves image flatness and removes image artifacts. Finally, we stack the  sky-subtracted, flat-fielded object frames to produce the image product as shown in Figure \ref{figG165}.  Our final mean $K$-band FWHM is 0.53$\arcsec$ for the 2016 Dec 9 run (LUCI2), and 0.29$\arcsec$ for the 2016 Dex 15 run (LUCI1). In all, we reach 10$\sigma$ magnitudes for point sources inside an aperture of 4 $\times$ FWHM of 22.63 (AB) mags and 23.50 (AB) mags for the 2016 Dec 9 and 15 runs, respectively.

Our high-resolution LBT LUCI-ARGOS imaging of the G165 field shows dozens of lensed galaxies.  The main objective of this initial analysis is to establish which of these arcs belong to sets of arclet families. As DSFG\_1a (`1' in Figure \ref{figG165}) already has a spectroscopic redshift \citep{Harrington2016}, the most important task is to search its counter-image produced as a result of strong lensing.
We readily detect the NIR counterpart of the {\it Planck/Herschel} source as a giant arc with an angular extent of $\sim$5$\arcsec$ that is merging with the critical curve ({\it G165\_DSFG\_1a}). Our lens model also predicts  there to be a counter-image ({\it G165\_DSFG\_1b}), although the {\it HST} data are too blue and shallow to detect this fainter lensed source. In our longer wavelength  LBT LUCI-ARGOS data and {\it Spitzer/IRAC} data, we detect G165\_DSFG\_1b at the model-predicted location as shown in Figure \ref{figG165}. We note that the brightest arcs in our $K$-band image are the triple-imaged arclet family designated  {\it Arcs 2a, 2b,} and {\it 2c} in Figure \ref{figG165}.  Each of the three arcs are extremely bright, with $K_{AB}$-band magnitudes of $\approx$18.5 mag, yet the redshifts are unknown. These are excellent sources for follow-up spectroscopy.
In sum, we identify 11 sets of arclet families using the LUCI-ARGOS $K$ imaging combined with {\it HST}, from which we construct our strong lensing model for the cluster \citep{Frye2018}. From our model, we estimate that {\it G165\_DSFG\_1a} has a minimum magnification of a factor of 30.  Interestingly, {\it G165\_DSFG\_1a} is merging with the critical curve, from which we infer that there may be potentially large transverse velocities relative to the critical curve, which may result in caustic crossing events which can lead to still higher magnifications \citep[e.g.][]{Diego2018, Windhorst2018}.

   \begin{figure}
   \centering
    \includegraphics[width=9cm]{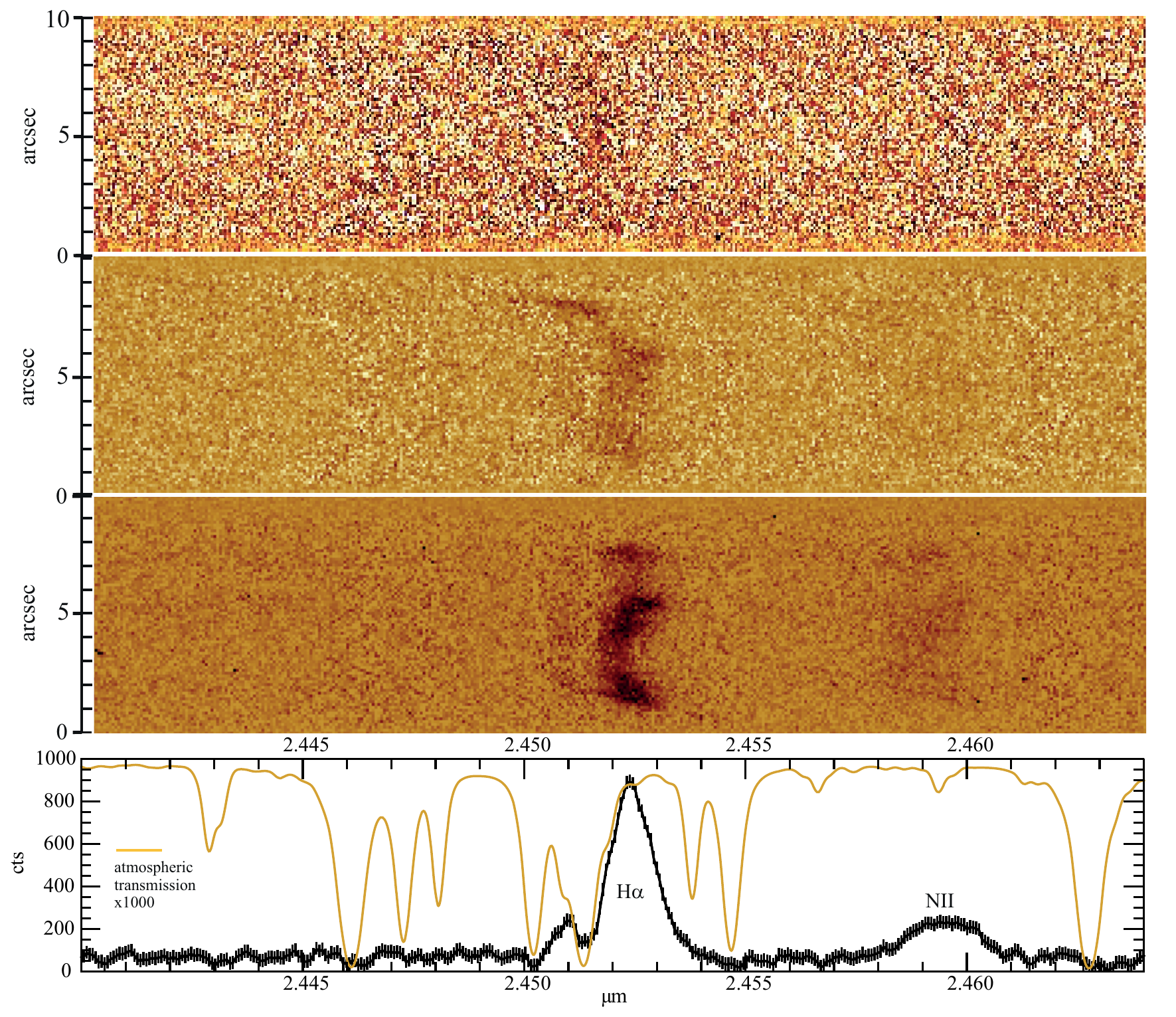}
      \caption{Emission lines H$\alpha$ and [NII] detected from the 8 o'clock arc. From top to bottom: A 5\,min integration with the ARGOS loop open; the emission is barely visible, due to the small slit width.  A 5\,min integration with the adaptive optics loop on; all the flux from the object is nicely squeezed into the curved slit. Combining  multiple nodded integrations for an hour of observation and removing the slit curvature; the velocity distribution and a richness of details can be seen already. In the bottom panel a 1d spectrum is shown, formed by  collapsing the 2d spectrum from above along the spatial axis. While this averages out all the velocity distribution that is seen above, it compares directly to data taken by others. The shown spectrum is not corrected for telluric absorption in the atmosphere, which is strong at that wavelength. To guide the readers eye a calculated absorption spectrum is overplotted in yellow, being responsible for some features in the spectrum, and contributes to the invisibility of the weaker second [NII] line at 2.4462$\mu$m.
              }
         \label{Fig_8oclock}
   \end{figure}

\subsection{ARGOS LUCI observation of gravitationally lensed arcs}

As demanding spectroscopy targets, we have been observing gravitationally lensed high-$z$ galaxies during ARGOS commissioning campaigns. Gravitational lensing offers a great opportunity to study objects at high redshift, due to the flux enhancing effect of the lens, which makes  dim objects appear brighter and sometimes makes them observable within a reasonable time. Due to the mass distribution of the lensing clusters or massive galaxies, many lensed objects appear as small,  extended arcs on sky. In  some cases these arcs extend over tens of arcseconds in length, but only over 0.2 -- 0.3$\arcsec$ in width. Due to the length of the  objects, small field AO IFU studies  can only look at parts of the object, while straight slit spectrographs do not match the curvature. In that respect LUCI-ARGOS offers an ideal facility for detailed studies of those arcs. With LUCI we can make curved matched shape slits over the full extent of the arcs, while with ARGOS we sharpen the object, such that it concentrates all the light through the narrow 0.3$\arcsec$ slit to resolve velocities spatially unsmeared at R \textasciitilde 10000. 

In the following, as illustration, we  show two objects  that have been observed through commissioning: SDSSJ0022+1431‚ the 8 o‘clock arc, and SDSSJ1038+4849, the `Cheshire cat' or `SMILE'. These data have been reduced using {\it flame}, a data reduction pipeline developed at MPE \citep{Belli2018}.  As we have targeted several lensed objects over the commissioning period, a study on the emission line properties has been carried out by \citet{Perna2018}. For all the gravitationally lensed arcs we have designed and used custom cut slit masks inserted into the LUCI spectrographs, extending over the full length of the arc and following its curvature.

\subsubsection{SDSSJ002240+1431: the 8 o'clock arc}

   \begin{figure}
   \centering
    \includegraphics[width=9.2cm]{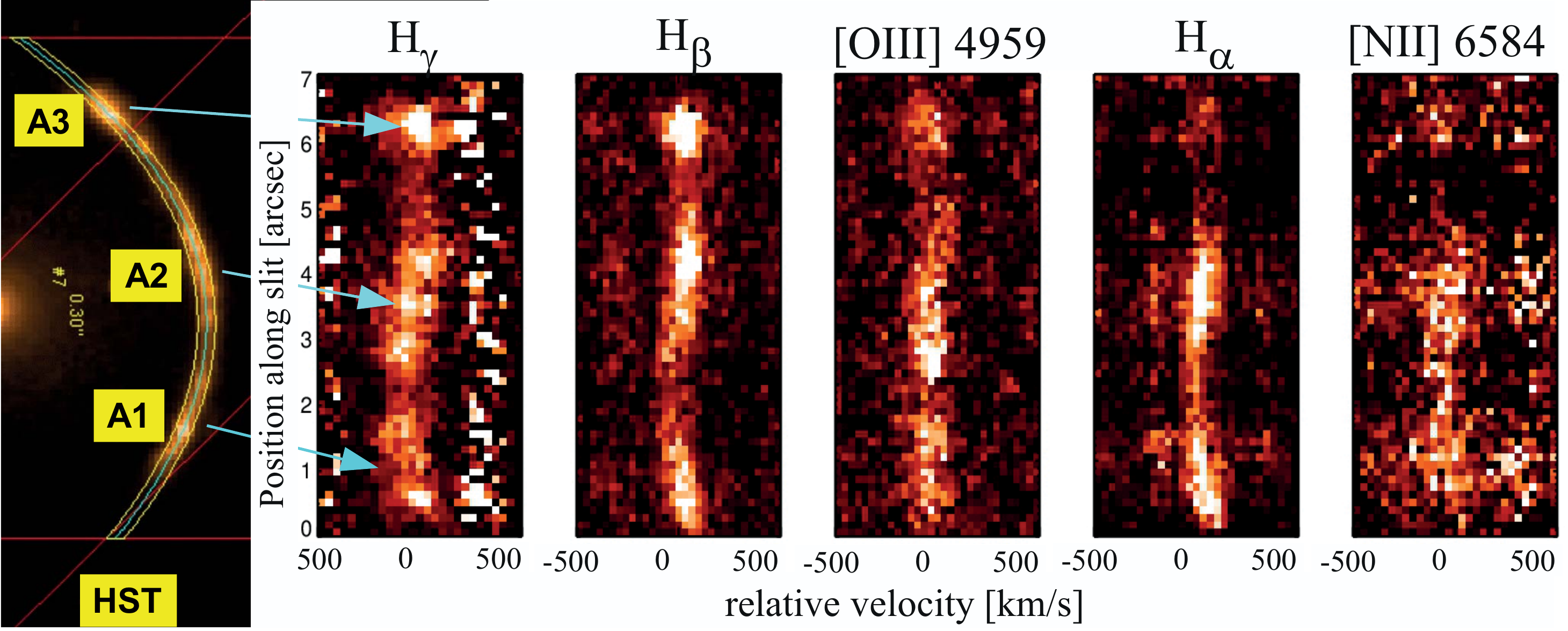}
    \caption{Left: Curved slit used for the LUCI-ARGOS observations on a HST infrared image of the 8 o'clock arc covering
      the three  multiple lensed images of the galaxy at $z$ = 2.73. Right: 2D spectra of all lines observed in the H and K bands at 55 {\rm km s$^{-1}$}
      spectral resolution. The integration times and spatial resolution
      of each band are  t$_{int}$(K)=2 h, PSF(K)=0.35$\arcsec$ and t$_{int}$(H)=3.9 h, PSF(H)=0.35$\arcsec$.}
         \label{Fig_8oclock_HST}
   \end{figure}

As one of the spectroscopy test cases we have targeted SDSSJ0022+1431, a Lyman break galaxy named the 8 o‘clock arc \citep{Allam2007}. This object is located at  $z=2.73$ being strongly lensed by a $z=0.38$ luminous red galaxy SDSS J002240.91+143110.4. It has been observed spectroscopically with SINFONI \citep{Shirazi14}, NIRI \citep{Finkelstein09}, and Xshooter \citep{Dessauges11}, so the ARGOS observations can be compared with the previous work. At that redshift this object is especially demanding to observe in H$\alpha$ emission, since it falls at 2.45\,$\mu$m outside the K band where the thermal background is high. With the ability of LUCI-ARGOS to keep the slits small, concentrate the object's flux into the slit, and have a high spectral resolution, we can very easily extract the arc's signal  out of the massive thermal background. We have been targeting this object since October 2016. Because  it is one of the early science cases, many first-time issues had to be solved: drifting  of the object due to systematic movements between the tilt sensor and the focal plane, spectral focusing, and wavelength drifts in LUCI. The spectrum as detected with LUCI-ARGOS of the 8 o'clock arc H$\alpha$ emission is shown in Figure \ref{Fig_8oclock}. Without the GLAO correction the emission is barely visible in a 5\,min integration time,  partly due to the small slit width of 0.3$\arcsec$. With the adaptive optics on, the emission can be seen in a single 5 min integration. Summing up several integrations,  removing the slit curvature, and performing wavelength calibration results in the final spectrum (see  Figure~\ref{Fig_8oclock}). The spectrum shows details of the velocity distribution and dispersion. This level of detail has not been seen in previous observations due to the lack of spectral resolution and signal-to-noise ratio. The resulting average 1D spectrum of the 8 o’clock  arc is shown in the lowest panel of Figure~\ref{Fig_8oclock}, reaching a H$\alpha$ peak flux-to-noise ratio of \textasciitilde 50 within 1h of observation. Compared to the NIRI observations, the lines are much better resolved and compared to the X-Shooter observations, the S/N is a magnitude higher. Our curved slit  covered the whole arc at once, resolving the multiple images  A1, A2, and A3 (Figure~\ref{Fig_8oclock_HST}). We have  observed this arc in the clear filter targeting the H$\alpha$ line at 2.45\,$\mu$m  at the maximum spectral resolution ($\sigma$ $\sim$20\,{\rm km s$^{-1}$}) and in H band  at a lower spectral resolution ($\sigma$ $\sim$55\,{\rm km s$^{-1}$}). We combined only the best seeing data that result in a final spatial resolution equal to $\sim$ 0.35$\arcsec$ for all data. We show the resulting 2D spectra of H$\alpha$ (rebinned to the lower spectral resolution used for H-band observation), [NII]$\lambda$6583, H$\gamma$, H$\beta$, and [OIII]$\lambda$4959 in  Figure \ref{Fig_8oclock_HST}. The H$\alpha$ 2D spectrum at full spectral resolution of A2 is shown in the  bottom panel of Figure \ref{Clump}. The integrated spectra of this source are presented in \citet{Perna2018}.

   \begin{figure}
   \centering
    \includegraphics[width=9.3cm]{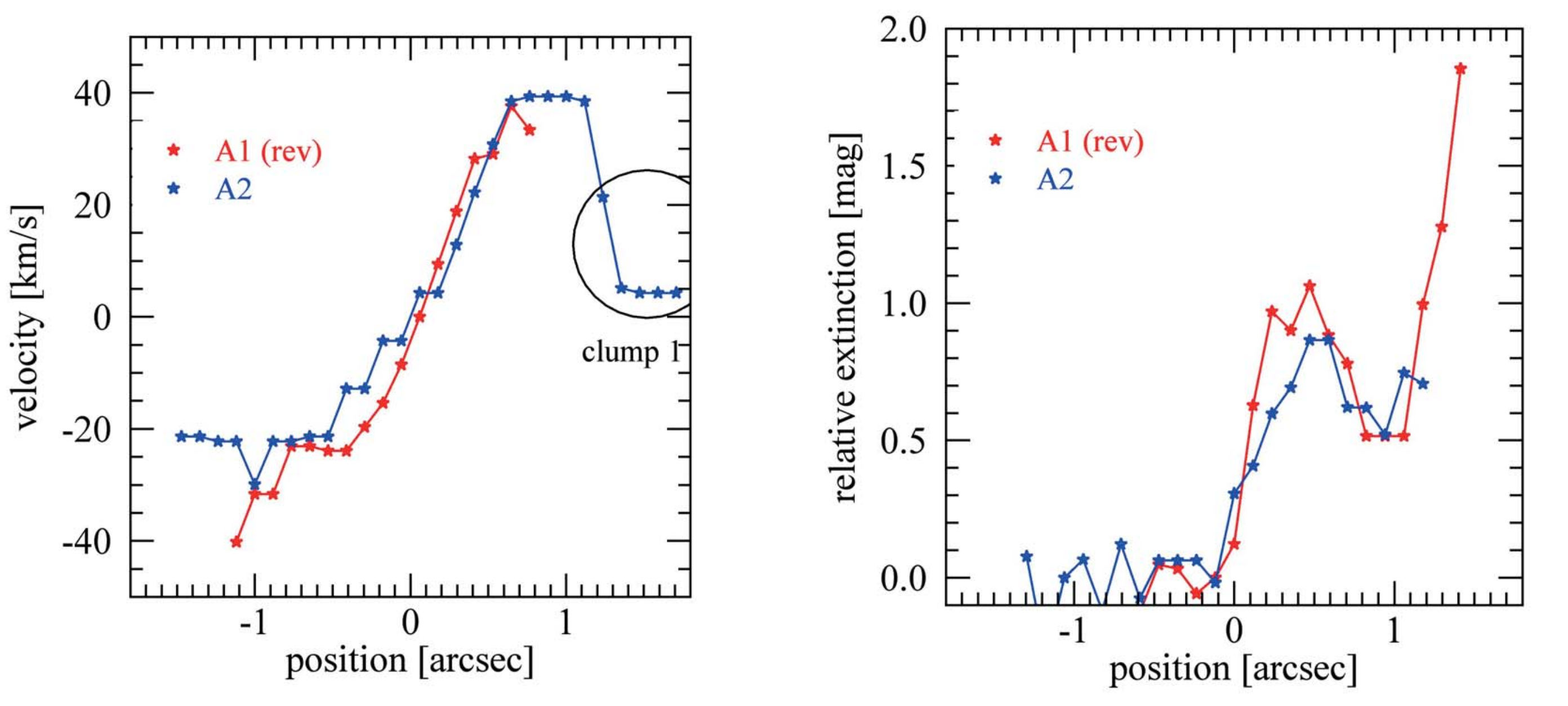}
      \caption{ Left panel: Rotation curve of the A2 and A1 (reversed) images derived from H$\alpha$ of the 8 o'clock arc. Right panel: {Relative} internal extinction E(B-V) profile of A1 and A2 obtained by binning the data and by assuming that E(B-V) at the edge of the images are zero.    }
         \label{Fig_8oclock_results}
   \end{figure}
  Previous studies of the 8 o'clock arc  established that  A1, A2 (the reversed image of A1), and A3 are multiple images of a galaxy with $Log M_* = 10.3 M_{\odot}$ \citep{Shirazi14}. The rotation curves of A2 and A3  (left panel of Figure \ref{Fig_8oclock_results}) extracted from our data confirm  this hypothesis, showing that the two rotation curves  agree and that this  lensed galaxy has a regular rotating disc.
  
  The high spectral and spatial  resolution of the ARGOS observations  allows us to resolve several clumps in each image, especially in the A2 H$\alpha$ emission (see lower left panel of Figure \ref{Clump}), that were not resolved in previous studies \citep{Shirazi14}.
  Assuming that the flux ratio between Balmer emission lines is determined by the case B recombination \citep[see e.g.][]{Osterbrock2006} and is therefore identical throughout the arc, we interpret any variation in the observed H$\gamma$-to-H$\beta$ ratio as being due to different amounts of dust extinction. The lack of a spectrophotometric standard star in this observation prevents us from measuring the \emph{absolute} line ratio and therefore the absolute dust attenuation. We therefore assume that the E(B-V) value at the bottom (top) edge of A1 (A2) is zero, and derive the \emph{relative} E(B-V) profile along the spatial dimension of A1 and A2. Since these two images have different gravitational stretches, we align them to the spatial position where the H$\gamma$-to-H$\beta$ ratio is similar.
  
  The result is shown in the right panel of  Figure \ref{Fig_8oclock_results} for A2 and A1  (reversed), where we binned the data in order to enhance the signal. We note that the derivation of the absolute clump extinction E(B-V), which will be possible in ARGOS routine phase, will allow the calculation of the clump SFR, which is an important  parameter for the  study of the role of these clumps in the evolution of high-$z$ galaxy discs.
  
  A similar profile derivation for the [NII]/H$\alpha$ ratio, tracing the metallicity, is too uncertain given the weakness of [NII]. Our derived global [NII]/H$\alpha$ ratio of A2 is $\sim$ 0.22. Taking into account  the uncertainty introduced by the subtraction of the continuum, this value is compatible with that expected for a galaxy of similar stellar mass and redshift  of 0.15, as given in \citet{Wuyts2016} and corresponds to a metallicity of 12+$log$ $(O/H)$ = 8.52 (assuming \citet{Pettini2004}) similar to the value measured for the same source by \citet{Dessauges11} adopting  the same method.

\subsubsection{SDSS1038+4849: the SMILE arc}

   \begin{figure}
   \centering
    \includegraphics[width=9.2cm]{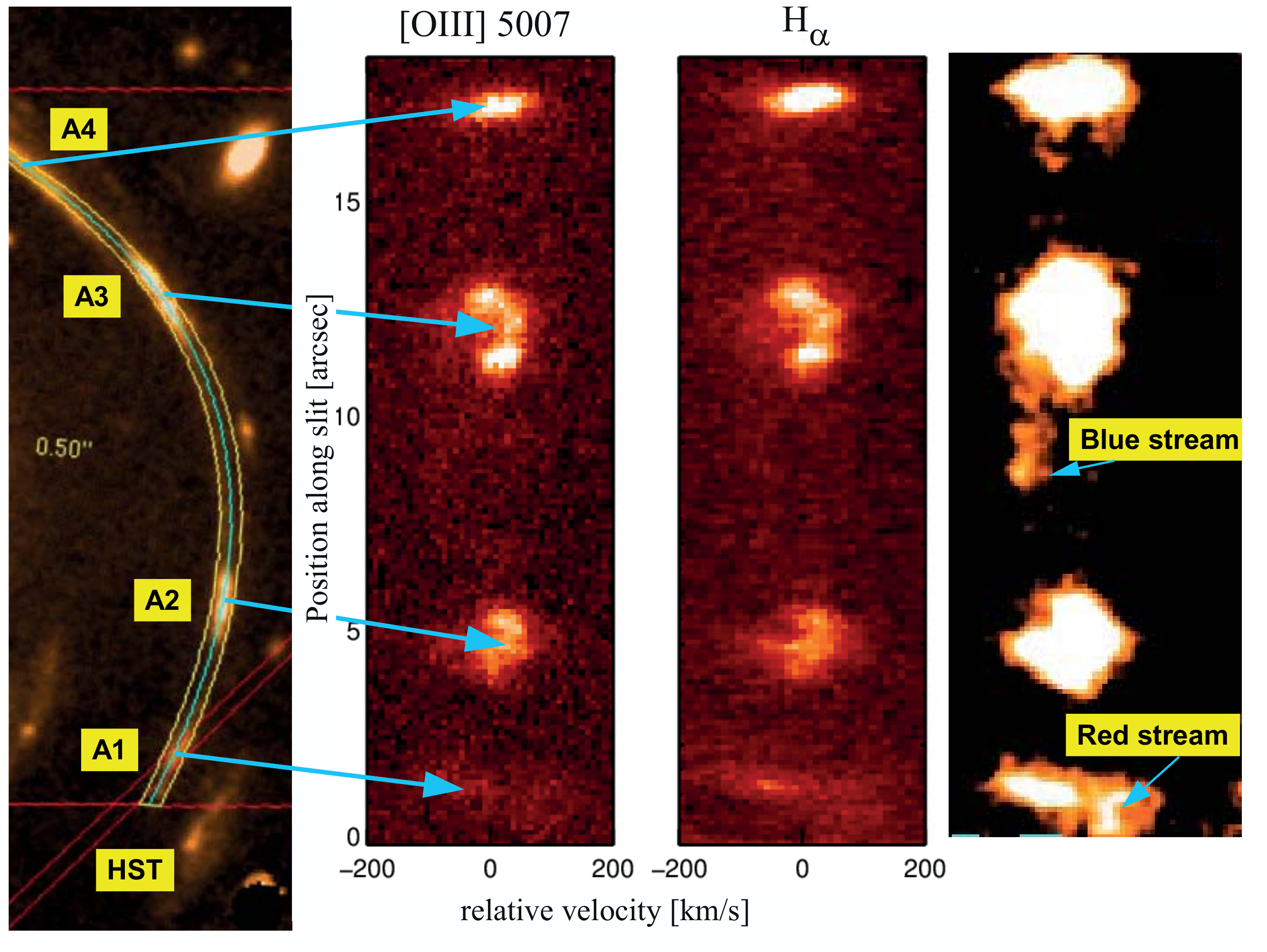}
      \caption{Left panel: Curved slit used for the LUCI-ARGOS observation of the SDSS1038+4849 (SMILE) arc at $z=2.197$ on an HST infrared image that covers
      the multiple images of the same galaxy (A2, A3, and A4) and of the companion galaxy A1.
      Middle panels: 2D spectra (t$_{int}$=4 {\rm hrs}) of the [OIII$\lambda$5007] line at 0.55$\arcsec$ resolution  and of  the H$\alpha$ line ( t$_{int}$=4
      {\rm hrs}) at 0.35$\arcsec$ resolution. Right panel: Smoothed  [OIII$\lambda$5007] 2D spectrum to enhance the blue and the red streams.}
     \label{SMILE_HST}
   \end{figure}

\begin{figure}
   \centering
    \includegraphics[width=9.3cm]{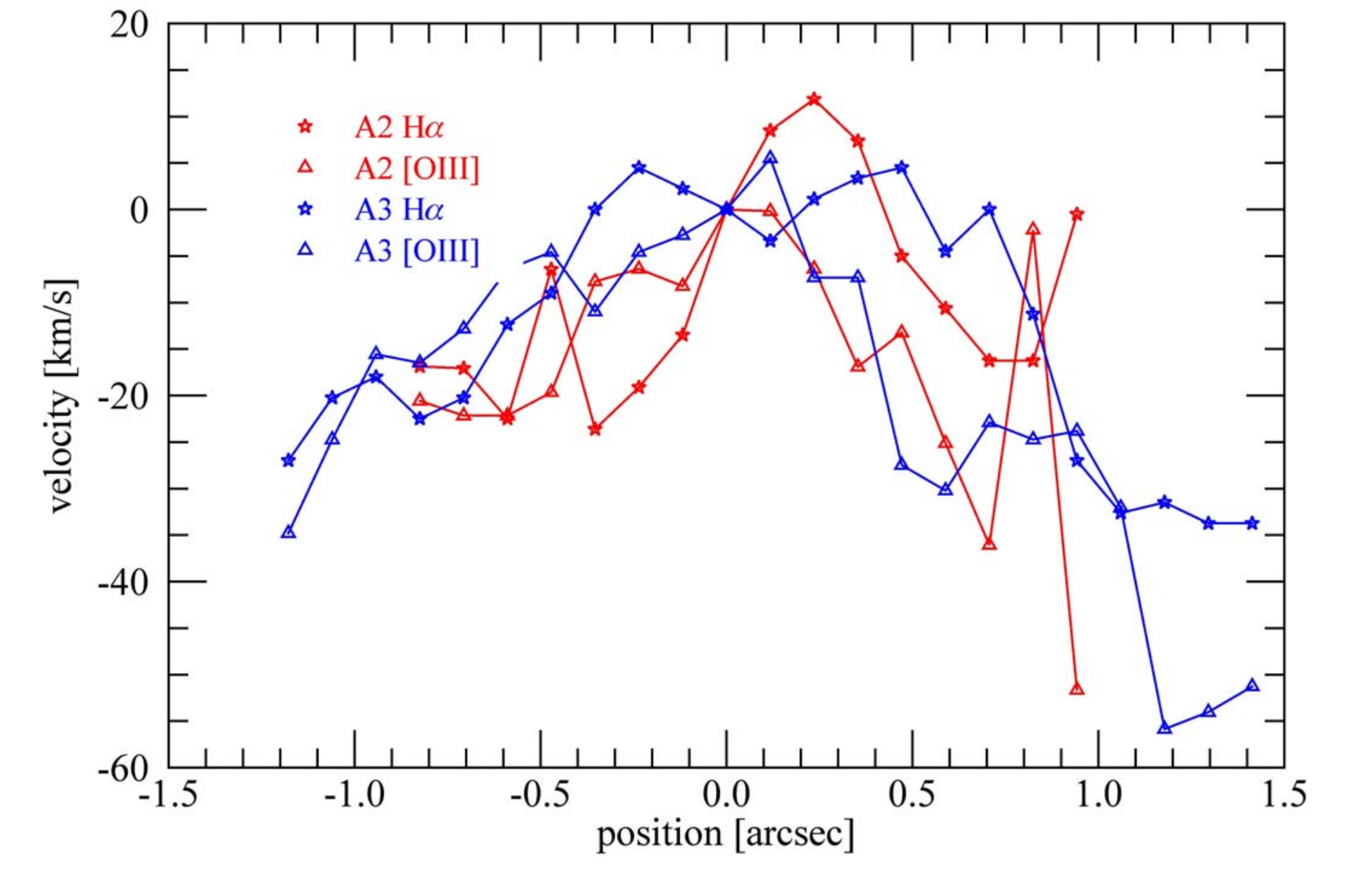}
    \caption{Kinematics of the two images A3 (blue) and A2 (red, reversed) in SDSS\,1038+4849 derived from the H$\alpha$ observations (stars) and from the [OIII$\lambda$5007] observations (triangles). The resulting curves of A2 and A3 have the typical reversed U shape of interacting systems, suggesting that the A2 and A3 images are composed of at least three clumps that are interacting with each other.}
       \label{Fig_SMILE_result}
   \end{figure}

SDSSJ1038+4849 is a lensed system at $z = 2.197$. The arc  represents multiple images of an interacting system  composed of two galaxies of  $Log M_*=9.1 M_{\odot}$ (images A2, A3, and A4) and $Log M_*=9.9 M_{\odot}$ (A1, \citet{Jones2013}) that were all included in the curved slit (Figure \ref{SMILE_HST}). We observed this arc  in the K and H bands with the highest spectral resolution ($\sigma$ $\sim$ 20\,km/s). In that observation we could already make use of the full binocular operation of ARGOS, with one LUCI observing OIII, and the second simultaneously in H$\alpha$. The best seeing (0.35$\arcsec$ in K band and 0.55$\arcsec$ in H band) final position-velocity diagrams (PVD) of H$\alpha$, and   [OIII]$\lambda$5007 are shown in Figure \ref{SMILE_HST}. No [NII] was detected, suggesting that these are low-metallicity systems. Again, thanks to the superior ARGOS spectral and spatial resolution, we can resolve at least three clumps in A3 in all detected lines. What is striking in Figure \ref{SMILE_HST} is the peculiar velocity pattern of A2 and A3 (one is the reverse image of the other) that does not resemble the typical curve of a rotating disc. We have verified that the peculiar velocity pattern is not due to a further multiple reversed image in A2 and in A3. In fact, the [OIII]/H$\alpha$ ratio profile of the bottom and top portion of A3 and A2  are very different indicating that these are indeed different clumps. We have extracted the spatial kinematic profile of all images (Figure \ref{Fig_SMILE_result}). The resulting  curves of A2 and A3, have the typical reversed U shape of interacting systems \citep{Rafanelli1993} and the PVD morphology resembles that of the Antenna galaxy, a local prototype of an interacting system \citep{Oestlin2015}. Therefore, we suggest that the A2 and A3 images are composed of at least three clumps that are interacting with each other, while the whole system is interacting with  A1. Tidal streams detected in the PVD (Figure \ref{SMILE_HST}) further support this interpretation: a blue-shifted stream from A3 and a red-shifted stream from A1. Each of these streams is likely moving  towards the interacting partner.

\subsubsection{Clump analysis}

 In the last decade, there has been  increasing evidence that high-redshift rotationally supported galaxies are clumpier than those in the local Universe \citep{Elmegreen2005, Genzel2011, Grogin2011, Guo2015, FoersterSchreiber2011, Wuyts2012}. Some of these are giant clumps that host extreme star formation and eject outflows in the disc \citep{Newman2012a}. Theoretical studies \citep{Dekel2009, Krumholz2010, Ceverino2012, Mandelker2017} also indicate  that the fate of these clumps is important for the evolution of the host galaxy. For example,  if they   migrate toward the galaxy centre, they can  be the building blocks for the formation of the  the galaxy bulge \citep{Elmegreen2008, Elmegreen2009}. Whether the clumps evaporate or survive (and hence can migrate) depends on some  physical properties of the clumps,  such as the timescale necessary to accrete new gas with respect to that required to expel  material  via outflow and/or to consume it by forming stars.
 
 The results reported in the previous sections on the ARGOS observations of gravitationally lensed high-redshift galaxies illustrate  how  very high spectral and spatial resolution enables  detailed analyses of the physical conditions of these clumps (such as extinction, kinematics, metallicity). This information can  greatly help in understanding the disc stability and thus the evolutionary pattern of galaxies. The highest spatial resolution observation of clumps in high-redshift galaxies existing so far  reach 0.2$\arcsec$ spatial resolution, comparable to  what is   achievable with ARGOS. On the other hand, thanks to the possibility of using small slit widths, LUCI-ARGOS can reach a spectral resolution in the NIR that is up to three times better.
 
\begin{figure}
   \centering
    \includegraphics[width=9cm]{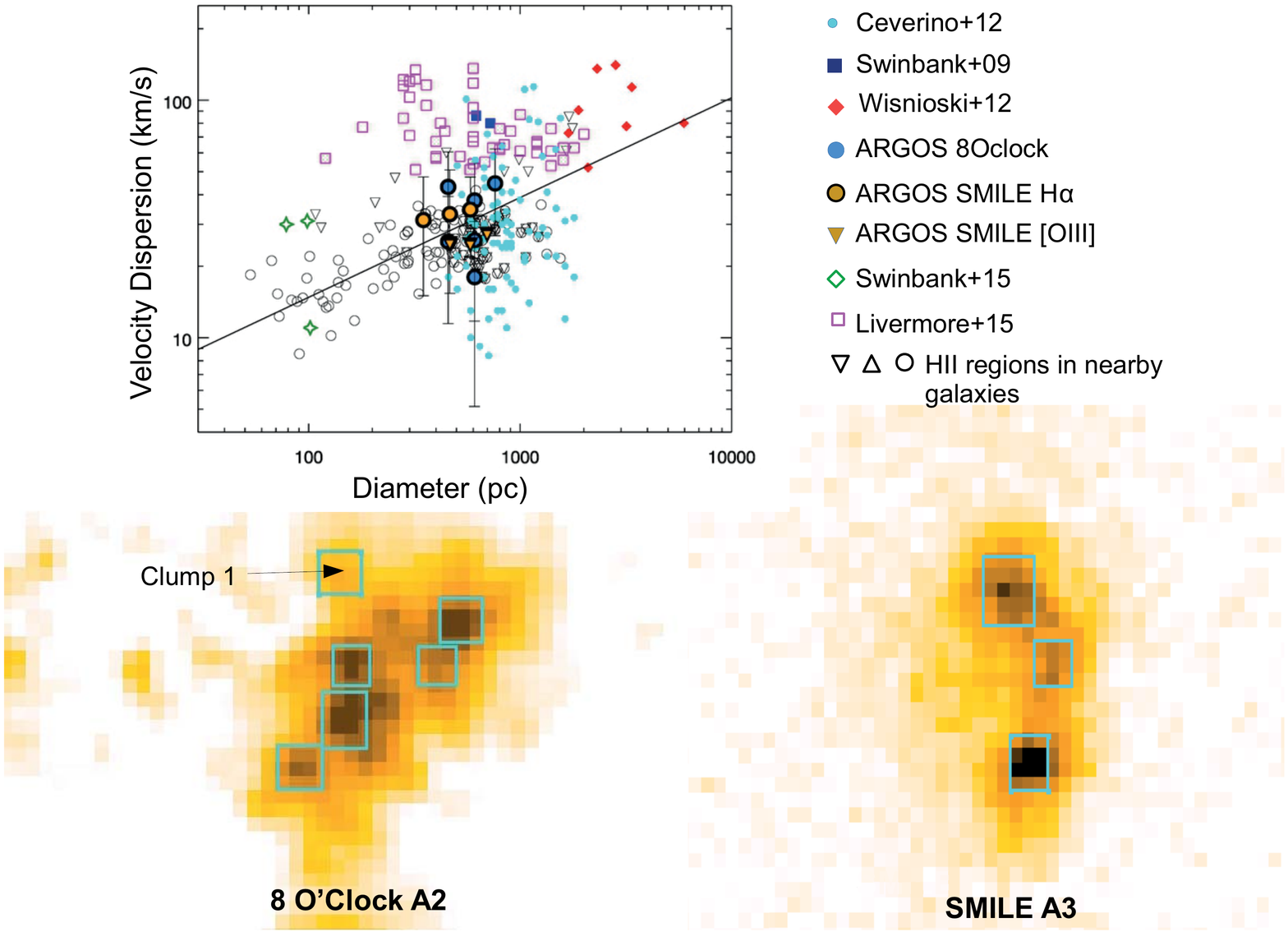}
      \caption{ The $\sigma$-$diameters$ relationship of clumps in high-redshift galaxies and in HII regions in nearby galaxies (upper panel).
      The position of the clumps selected on the A2 2D spectrum  of the 8 o'clock arc (bottom left) and on the A3 2D spectrum
      of the SMILE arc (bottom right) fall on the relation obtained by \cite{Wisnioski2012} (solid black line in the upper panel), but covers a range of clump sizes and velocity dispersion smaller than those reported in other high-redshift surveys and close to those followed by giant HII regions in the local Universe. Uncertainties for the velocity dispersion are overplotted denoting the Gaussian fit error in combination with the spectral resolution of the observation, {i.e.} 20\,km/s.}
         \label{Clump}
   \end{figure}

 We have identified clumps in both the 8 o'clock and the SMILE best resolved images (A2 and A3 respectively), both spatially and spectrally, as shown in Figure \ref{Clump}. We have extracted the H$\alpha$ (and [OIII] in the case of the SMILE) clump spectra and measured their dispersion, which we then corrected for the intrinsic spectral resolution ($\sim$ 20\,km s$^{-1}$). We compare our results with those of other authors on the velocity dispersion-diameter (corrected for  magnification) relation (Figure \ref{Clump}).

 The clumps studied with ARGOS fall on the  relation predicted by \citet{Wisnioski2012} and we   probe properties of clumps with sizes and velocity dispersion similar to those of the giant HII regions in  nearby local galaxies, and significantly smaller than those reached from other rest-frame optical wavelength observations of high-redshift galaxies.
 For example, a study of lensed galaxies conducted by \citet{Livermore2015} with a spatial resolution $\sim$0.2$\arcsec$, but spectral resolution three times worse than that of the ARGOS data, reports higher velocity dispersion  for clump radii similar to ours. Only CO ALMA observations of lensed galaxies, which have a spectral  resolution of $\sim$10\,km/s,  reach a velocity dispersion as low as those analysed in this work at comparable clump sizes \citep{Swinbank2015}.
 
 Thus, we conclude that in order to have a reliable census of the clump physical properties, and possibly kinematics, in high-redshift  galaxies, one needs the  spectral and spatial resolution delivered by ARGOS. These data are complementary to studies at  sub-mm wavelengths where similar resolution is reachable, but probing a different gas phase.

\section{Conclusions}

The ARGOS system corrects the ground-layer atmospheric distortions for both 8.4~m eyes of the LBT, improving the imaging resolution by a factor of 2 -- 3 over a 4$\times$4\,arcmin field of view. It provides a uniform correction over the field of view and a much more stable and well-defined PSF over a given observing time than is provided by natural seeing. We have repeatedly demonstrated that ARGOS can deliver a PSF with a FWHM of $\sim0.25-0.3\arcsec$ in the $J$, $H$, and $Ks$ bands.
As detailed in this paper, ARGOS is a constellation LGS-based adaptive optics instrument and has just entered the era of routine scientific operation. During the extent of the commissioning period a variety of scientific observations were successfully tested and have largely benefited from the enhanced resolution:
\begin{itemize}
    \item by overcoming the crowding effects in the globular cluster NGC2419 we were able to measure more precisely the CMD and identify variable stars, which in turn will provide  better distance estimates and can help the study of the past and present evolution of the Milky Way;
    \item by enhancing the spatial resolution, ARGOS enables an investigation of the inner part of galaxies at a greater distance, as shown by the study of the nucleus and the stellar clusters of NGC6384 at 20~Mpc;
    \item by providing high-resolution imaging at wavelengths inaccessible to HST, we identified the counter-image of the red DSFG with an estimated magnification factor of $>30$;
    \item on combining ARGOS with the multi-object spectrograph and exploiting its custom slit capabilities, we can match the slits to curved gravitational arcs and obtain high spatial and spectral resolution 2D spectra. In particular, we can uniquely study the clump physical properties of high-redshift (lensed) galaxies at spectral and spatial resolution that is otherwise only available at sub-mm wavelengths, tracing a different gas phase.
\end{itemize}
The first scientific data with LUCI+ARGOS show the great potential of the GLAO enhanced data. On the technical side, despite its complexity, the ARGOS system is operating robustly, its Rayleigh laser system is working flawlessly, and the adaptive optics loop can be closed in minutes. This makes ARGOS manageable and maximizes its scientific impact. In the light of these technical and early scientific achievements, we consider the concept and our implementation of the LGS ground-layer adaptive optics system a great step forward for the LBT.

\begin{acknowledgements}
Over the extent of the project many people and institutions have helped to make ARGOS work. In no particular order, we would like to thank all the members from the workshops, design departments, industry partners, and students for building, installing, and commissioning the many pieces that form a complex system. Many thanks to Matthias Honsberg, David Huber, Stephan Czempiel, Martin Deuter, Simon Krämer, Luca Carbonaro, Marcus Haug, Markus Thiel, Stefan Kellner, Reinhard Lederer, Nancy Ageorges, Christina Loose, Peter Mayer, Gerhard Hölzl, Thomas Ertl, Franz Soller, Stefan Huber, Felix Dressler, Florian Gorgeon, Sebastian Ihle, Robert Hartmann, Thomas Blümchen, Michael Lehmitz, Monica Ebert, Klaus Meixner, Norbert M\"unch, Thomas Hahn, and in memoriam to Srikrishna Kanneganti.
We would also like to thank the LBTO engineering, mountain crew, and telescope operators for excellent support over all the years.
Very special thanks to Jasmin Zanker-Smith for flawlessly organizing all the many meetings, travels, and commissioning trips to the mountain.
Great thanks also to the enthusiastic spotting crew standing outside in the wind and snow watching out for aircraft during all the commissioning nights.

The LBT is an international collaboration among institutions in the United States, Italy, and Germany. LBT Corporation partners are: The University of Arizona on behalf of the Arizona Board of Regents; Istituto Nazionale di Astrofisica, Italy; LBT Beteiligungsgesellschaft, Germany, representing the Max-Planck Society, The Leibniz Institute for Astrophysics Potsdam, and Heidelberg University; The Ohio State University and The Research Corporation on behalf of The University of Notre Dame, University of Minnesota, and University of Virginia.

This work has been supported amongst others by funds from the Max Planck Society,
the European Union through the Optical Infrared Co-ordination Network
for Astronomy, Grant Agreement 226604, and by the National Science Foundation under award AST-1040559.
\end{acknowledgements}

   \bibliographystyle{aa}
   \bibliography{argos_bibfile3}

\end{document}